\documentclass[letterpaper,12pt]{book}

\usepackage[letterpaper, 
    left=1.5in,
    right=1.5in,
    top=1.15in,
    bottom=1in,
    heightrounded,
    headheight=15pt,
    includehead,
    includefoot,
    centering
]{geometry}
\usepackage{setspace}

\usepackage[T1]{fontenc}
\usepackage{lmodern}
\usepackage{ragged2e}
\usepackage{titling}
\usepackage{layout}
\usepackage[utf8]{inputenc}
\usepackage{enumerate}
\usepackage[figuresright]{rotating}
\usepackage[section]{placeins}
\usepackage{bm}
\usepackage{color}
\usepackage[small, it]{caption}
\usepackage{fancyhdr}
\usepackage{fix-cm}
\usepackage{titlesec}
\usepackage{latexsym}
\usepackage[switch]{lineno}
\usepackage{mathtools}
\usepackage{comment}
\usepackage{float}
\usepackage{relsize}
\usepackage{lmodern}
\usepackage[dvipsnames]{xcolor}
\usepackage{amssymb}
\usepackage{upgreek}
\usepackage{lipsum}
\usepackage{soul}
\usepackage{tabularx}
\usepackage{booktabs}
\usepackage{subfig}
\usepackage{amsmath}
\usepackage{subcaption}
\usepackage{xspace}
\usepackage{graphicx}
\usepackage{hyperref}
\usepackage{url}
\usepackage{listings}

\usepackage[
    backend=biber, 
    style=numeric, 
    sorting=none,    
    maxbibnames=3,   
    minbibnames=3    
]{biblatex}
\title{Detector Designs for Frontier Measurements in Neutrino and Collider Physics in the 21st Century}
\author{Wonyong Chung}
\date{May 2, 2025}

\titleformat{\chapter}[display]
    {\centering\normalfont\huge\bfseries\singlespacing}
    {\vspace*{0.0\textheight}\thechapter} 
    {20pt}                                
    {\LARGE}
\titlespacing*{\chapter}{0pt}{-50pt}{30pt} 

\newcommand{\chapteropeningtext}{
  \vspace*{0.0\textheight} 
}

\titleformat{\section}
  {\normalfont\large\bfseries} 
  {\thesection} 
  {1em} 
  {}

\titleformat{\subsection}
  {\normalfont\normalsize\bfseries} 
  {\thesubsection} 
  {1em} 
  {}

\newcommand{\EBz}{Z_\text{B}}
\newcommand{\Rin}{R_\text{inner}}
\newcommand{\PHISEGMENTS}{N_\Phi}
\newcommand{\nomfw}{C_\text{fw}}

\newcommand{\Fdz}{F_\text{dz}}
\newcommand{\Rdz}{R_\text{dz}}

\newcommand{\xbotyinc}{x_\text{c}^\text{0}}
\newcommand{\xbotyinl}{x_\text{L}^\text{0}}
\newcommand{\xbotyinr}{x_\text{R}^\text{0}}
\newcommand{\xtopyinc}{x_\text{c}^\text{1}}
\newcommand{\xtopyinl}{x_\text{L}^\text{1}}
\newcommand{\xtopyinr}{x_\text{R}^\text{1}}
\newcommand{\DPHIGLOBAL}{d\Phi}

\newcommand{\THETASIZEENDCAP}{\Theta_E}

\newcommand{\DTHETABARREL}{d\theta_B}

\newcommand{\NPHIBARRELCRYSTAL}{N\phi_B}

\newcommand{\DPHIBARRELCRYSTAL}{d\phi_B}

\newcommand{\thC}{\theta_\text{c}}
\newcommand{\gamm}{\phi_\text{offset}}
\newcommand{\rinner}{r_\text{in}}
\newcommand{\router}{r_\text{out}}
\newcommand{\yinner}{y}
\newcommand{\rotZ}{\text{Rot}_Z}
\newcommand{\rSlice}{r_\Phi}
\newcommand{\dispSlice}{P_\Phi}
\newcommand{\rlocal}{r_\text{c}}
\newcommand{\dispLocal}{P_\text{offset}}
\newcommand{\dispGlobal}{P_\text{global}}

\newcommand{\gPhi}{\Phi_c}

\begin{document}
\pagenumbering{gobble}


\begin{titlepage}
\begin{center}

  \scshape

    {\raggedright \scshape \Huge Detector Designs for \\ Frontier Measurements in \\ Neutrino and Collider Physics \\ in the 21st Century \par}

    \vspace*{0.6\textheight}
    {\raggedright \scshape \Large Wonyong Chung \par}
    {\raggedright \scshape \Large Princeton University \par}

\end{center}
\end{titlepage}

\begin{titlepage}
\begin{center}

    \vspace*{0.165\textheight}  
    {\large\textcopyright\;Copyright by Wonyong Chung, 2025.\\
    All rights reserved.\par}

\end{center}
\end{titlepage}

\begin{titlepage}

  \begin{center}
{\bfseries\Large Abstract\par}
\end{center}
\noindent The last energy-frontier lepton collider, LEP, established several limits that still hold today.
A key one is the counting of three light neutrino species from the invisible decay width of the Z boson.
From a collider calorimetry standpoint, the missing energy is an invitation to design an experiment to directly measure the neutrino mass.
We present a new type of EM spectrometer which leverages the first adiabatic invariant in magnetic gradient drift to achieve exponentially bounded resolution in a highly compact and scalable format, enabling the PTOLEMY experiment to not only measure the neutrino mass at the tritium endpoint, but one day directly detect the Cosmic Neutrino Background.
Meanwhile, the next lepton collider promises to expose the Higgs self-coupling and complete the accounting of lepton universality.
We present a dual-readout, segmented crystal calorimeter for future collider detectors, combining new hardware capabilities with novel AI/ML reconstruction techniques towards realizing a detector that must definitively and unambiguously surpass its predecessors.
Together, these studies confront the most pressing challenges for 21st century particle physics experiments to achieve the sensitivities needed to bridge the gap between the largest and smallest scales of reality.

\end{titlepage}

\begin{titlepage}
\begin{center}

  \vspace*{0.165\textheight}  

  \textit{\small for gowoony}

\end{center}
\end{titlepage}

\cleardoublepage

\pagenumbering{arabic}
\setcounter{page}{1}

\tableofcontents


\chapter[The Neutrino Mass and the Cosmic Neutrino Background]{The Neutrino Mass and\\ the Cosmic Neutrino Background \\ \large PTOLEMY}
\chapteropeningtext

\noindent In 1962 Weinberg examined how a relic massless, degenerate sea of neutrinos would Pauli-block phase-space in tritium $\beta$-decay, giving rise to a narrow capture peak just above the spectral endpoint~\cite{weinberg1962}. In 1965 the Cosmic Microwave Background was discovered, with the theoretical and detection papers appearing back-to-back in the Astrophysical Journal~\cite{cmb1965,cmb_discovery1965}. Based on the measured temperature excess of 3.5 Kelvin, Dicke, Peebles, Roll and Wilkinson laid out a possible thermal history of the universe, from which it could be inferrred that a sea of relic neutrinos which decoupled from the rest of the universe roughly 1 second after the Big Bang would have a flux in the present day of around 300 per cubic centimeter~\cite{cmb1965}.

These relic neutrinos are collectively known as the Cosmic Neutrino Background and would be the earliest observation of the universe possible based on the current theory of fundamental particles and interactions. Relic neutrinos would capture on tritium to produce a measureable line roughly $2m_\nu$ above the endpoint of the beta decay spectrum. In the forward beta-decay, a non-zero neutrino mass would decrease the total energy available to the electron and shift the cutoff of the spectrum down by an amont equal to the neutrino mass. These predictions make studying the tritium endpoint an attractive endeavor. After the discovery of neutrino oscillations in 1998~\cite{neutrino_oscillations1998}, Cocco, Mangano, and Messina re-examined the Cosmic Neutrino Background and its experimental consequences in view of a finite neutrino mass, laying out a path toward direct detection~\cite{Cocco_2007}.

The Mainz and Troitsk tritium $\beta$-decay experiments ran almost ten years before the observation of neutrino oscillations, using molecular gaseous diatomic tritium and a MAC-E filter. The KATRIN experiment, using the same scheme and representing the maximum scale of this type of apparatus with a main spectrometer nearly 70 meters long~\cite{katrindesign2005}, has since placed an upper limit on the electron neutrino mass of 0.45\,eV at 90\% CL~\cite{katrin2025}.

Meanwhile, breathtaking direct observations in the EM have cemented the arrival of observational cosmology. Hubble, WMAP, Planck, and JWST have shown the beauty and mystery of our universe to us in truly humbling ways, with cosmological models closing in on the available parameter space for neutrino masses: Planck constrains the sum of neutrino masses to be under 0.12\,eV~\cite{Planck2018}, and results from oscillation experiments expect the one of the mass eigenstates to be at least 0.05\,eV~\cite{pdg2024}.




To close in on the absolute mass scale in a tritium $\beta$-decay experiment, one first has to drain the kinetic energy of an endpoint electron precisely enough to bring it to a near stop, then measure the rise in resistance that registers when the electron is absorbed by a metal-superconducting bilayer near the phase transition and the heat of its kinetic energy breaks the superconductivity. To achieve the highest precisions in making this mesasurement, the Pontecorvo (n\'ee Princeton) Tritium Observatory for Light, Early-universe Massive-neutrino Yield advances research in condensed matter theory, nanomaterial synthesis, cyclotron radiation emission spectroscopy, particle transport, TES microcalorimetry, and EM spectroscopy. 



The contribution of this work is the spectrometer, magnet design, and global transport mechanism and transmission of electrons from the tritium target to the detector.

The work in this chapter has been published in~\cite{filter_paper1, filter_paper2, fusion_bookchapter}.
\section{Endpoint structure of the tritium beta decay}
The most direct and model-independent method for determining the absolute mass scale of neutrinos relies on the precise measurement of the kinematics of weak decays. The beta decay of tritium ($^3$H) offers a particularly advantageous system due to its low energy release (Q-value) and simple nuclear structure:
\begin{equation}
^3\text{H} \rightarrow ^3\text{He} + e^- + \bar{\nu}_e .
\end{equation}
The energy released in the decay is shared among the electron, the antineutrino, and the recoiling helium ion. The electron energy spectrum is continuous, extending up to a maximum value known as the endpoint.

The kinematic endpoint of the electron energy spectrum is directly sensitive to the mass of the emitted electron antineutrino $m_{\bar{\nu}_e}$. By energy conservation, the maximum kinetic energy available to the electron is the Q-value minus the rest mass energy of the antineutrino:
\begin{equation}
T_{e,max} = Q - m_{\bar{\nu}_e} c^ .
\label{eq:endpoint}
\end{equation}
If the neutrino were massless, the electron spectrum would extend precisely up to $T_e = Q$. However, a non-zero neutrino mass truncates the spectrum, shifting the endpoint to a lower energy by an amount equal to the neutrino mass.

The shape of the electron energy spectrum near this endpoint is dictated by the available phase space for the final state particles. Neglecting recoil and Coulomb corrections for simplicity, the differential decay rate $dN/dE_e$ is proportional to the product of the electron and neutrino phase space factors:
\begin{equation}
\frac{dN}{dE_e} \propto p_e E_e \cdot p_\nu E_\nu
\end{equation}
where $p_e, E_e$ and $p_\nu, E_\nu$ are the momentum and total energy of the electron and neutrino, respectively.

Using the energy conservation relation $E_\nu = Q - E_e$ and the relativistic energy-momentum relation $E_\nu^2 = p_\nu^2 c^2 + m_{\bar{\nu}_e}^2 c^4$, the neutrino momentum can be expressed as 
\begin{equation}
p_\nu c = \sqrt{(Q - E_e)^2 - (m_{\bar{\nu}_e} c^2)^2}.
\label{eq:mass_energy}
\end{equation}
Substituting these into the phase space factor and considering the region very close to the endpoint where $E_e \approx Q$ and $p_e, E_e$ are approximately constant, the spectral shape becomes
\begin{equation}
\frac{dN}{dE_e} \propto (Q - E_e) \sqrt{(Q - E_e)^2 - (m_{\bar{\nu}_e} c^2)^2} \quad \text{for } E_e \lesssim Q .
\label{eq:spectrum}
\end{equation}
This formula reveals the characteristic signature of a non-zero neutrino mass. For $m_{\bar{\nu}_e} = 0$, the spectrum $dN/dE_e \propto (Q - E_e)^2$, approaching zero tangentially at $E_e = Q$. For $m_{\bar{\nu}_e} > 0$, the spectrum terminates sharply at $E_e = Q - m_{\bar{\nu}_e} c^2$, and importantly, the derivative $d/dE_e(dN/dE_e)$ diverges at this point, meaning the spectrum approaches the endpoint with a vertical tangent.

The Kurie plot linearizes the spectrum under the assumption of zero neutrino mass. The Kurie function $K(T_e)$ is defined as:
\begin{equation}
K(T_e) = \sqrt{\frac{dN/dT_e}{F(Z, E_e) p_e E_e}} \propto \sqrt{(Q - T_e)^2 - (m_{\bar{\nu}_e} c^2)^2}
\end{equation}
where $F(Z, E_e)$ is the Fermi function accounting for Coulomb interactions between the outgoing electron and the daughter nucleus. If $m_{\bar{\nu}_e} = 0$, $K(T_e)$ is proportional to $(Q - T_e)$, resulting in a straight line intersecting the energy axis at $T_e = Q$. If $m_{\bar{\nu}_e} > 0$, the plot curves downwards near the endpoint, intersecting the axis at $T_e = Q - m_{\bar{\nu}_e} c^2$. Precise measurement of this deviation from linearity provides the experimental handle on the neutrino mass.

It is important to note that beta decay experiments are sensitive to an effective electron antineutrino mass, $m_\beta$, due to the fact that the electron neutrino flavor state $\nu_e$ is a superposition of mass eigenstates $\nu_i$ with masses $m_i$, described by the Pontecorvo-Maki-Nakagawa-Sakata (PMNS) mixing matrix elements $U_{ei}$. Since current experiments cannot resolve the individual mass eigenstates, the measured quantity is an incoherent sum
\begin{equation}
m_\beta^2 = \sum_{i=1}^{3} |U_{ei}|^2 m_i^2 .
\end{equation}
The presented spectral formula (Eq. \ref{eq:spectrum}) is derived under several approximations. More rigorous treatments include relativistic corrections to the kinematics and nuclear matrix elements, as well as radiative corrections. For tritium decay, these corrections are generally small compared to other uncertainties but must be considered for achieving the highest precision.

The primary experimental technique used by the Mainz, Troitsk, and KATRIN experiments is the MAC-E filter (Magnetic Adiabatic Collimation with Electrostatic Filter). The MAC-E filter uses a strong, guiding magnetic field that decreases adiabatically from the source region to a minimum in the center of the spectrometer. Under the first adiabatic invariant, the orbital magnetic moment $\mu$, the transverse momentum of the electrons are collimated into longitudinal momentum. A high-pass electrostatic potential barrier is applied in the low magnetic field region, allowing only electrons with sufficient longitudinal kinetic energy to overcome the potential barrier to be transmitted to the detector. By scanning the retarding potential near the Q-value, the integrated spectrum near the endpoint can be measured with high precision.

The Mainz experiment in its later phase utilized a tritium source formed by quench-condensing molecular tritium ($T_2$) onto a cold graphite substrate. Their final analysis yielded an upper limit of $m_\nu < 2.3$ eV/c$^2$ (95\% C.L.)~\cite{Kraus:2004zw}. A significant portion of the systematic uncertainty stemmed from the molecular nature of the source. The decay occurs within a $T_2$ molecule, leading to a distribution of final states for the daughter ($^3$HeT$^+$) ion, each with slightly different energy releases. This effectively smears the endpoint energy.

The Troitsk experiment ran contemporaneously with Mainz and also used a MAC-E filter but opted for a Windowless Gaseous Tritium Source (WGTS). This approach avoided some issues related to solid sources but still contended with the molecular final state problem. Troitsk reported a persistent anomaly: a step-like excess of counts appearing a few eV below the endpoint, the position of which varied periodically over time. The origin of this anomaly remains unexplained. After empirically correcting for this feature, Troitsk derived an upper limit of $m_\nu < 2.05$ eV/c$^2$ (95\% C.L.)~\cite{Lobashev:1999tp}. The need to model and subtract this anomaly introduced a significant systematic uncertainty specific to the Troitsk results. The Mainz experiment did not observe a comparable anomaly.

Building upon the experience of Mainz and Troitsk, the KATRIN experiment was designed to reach sub-eV sensitivity. It features a large-scale WGTS providing a high luminosity of $10^{11}$ decays/s and a very high-resolution MAC-E filter spectrometer, spanning a total length of 70 meters. KATRIN began taking data in 2018 and has progressively improved the limit on the neutrino mass. The latest result, based on 259 days of data collected between 2019 and 2021, sets the most stringent direct limit to date: $m_\nu < 0.45$ eV/c$^2$ (90\% C.L.). KATRIN continues data taking until the end of 2025, aiming for a final sensitivity of $\sim 0.3$ eV (90\% C.L.). Despite its scale and precision, KATRIN still utilizes molecular tritium and must carefully account for the complex distribution of final molecular states, which remains a significant systematic uncertainty limiting its ultimate reach.

The PTOLEMY experiment proposes to use atomic tritium loaded on monolayer graphene to circumvent the smearing from the molecular form. This approach comes with its own tradeoffs. The most favorable scenario would be a Mossbauer-like decay in which the recoil momentum exactly hits a lattice eigenmode of the substrate. However this decay is exponentially suppressed. The electronic effects from neighboring tritium atoms and the vibrational modes of the lattice introduce new systematics which are being carefully studied~\cite{angelo2022}. It is hypothesized that because the helium ion enters a bound or unbound state due to the neighboring potentials, the degeneracy of these states could manifest at the endpoint as a step-like structure. With enough experimental precision, this step-structure could be used as a handle on the neutrino mass.



\section{The Transverse Drift Filter}
The transverse drift filter is a new method of EM spectroscopy which enables precision analysis of the energy spectrum of electrons near the tritium $\beta$-decay endpoint and is invented as a compact and scalable replacement for the MAC-E filter, achieving comparable or better performance within the space of 1 meter. This section details the physics of its spectroscopic capability and principles of operation. A geometric aperture and theoretical acceptance is calculated using the analytical field representations. A conceptual magnet design to produce the required magnetic field profile is presented and is the design upon which the LNGS demonstrator magnet is based. The filter is realized in full simulation and methods for implementation and optimization are presented. An efficient method for static end-to-end electron transport from the tritium target, through the filter, to the microcalorimeter is devised and a static transmission efficiency of 10\% is achieved.  

The challenges of low-energy electron transport in cases of low field are discussed, such as the growth of the cyclotron radius with decreasing magnetic field, which puts a ceiling on filter performance relative to fixed filter dimensions. Additionally, low pitch angle trajectories are dominated by motion parallel to the magnetic field lines and introduce non-adiabatic conditions and curvature drift. These effects can be first be minimized by collimating the energy at emission with a miniature MAC-E like setup, then inside the filter, with a three-potential-well design to simultaneously drain the parallel and transverse kinetic energies throughout the length of the filter. These optimizations are implemented in simulation to achieve low-energy electron transport from a 1\,T iron core (or 3\,T superconducting) starting field with initial kinetic energy of 18.6\,keV drained to $\approx$ 10\,eV ($\approx$ 1\,eV) in about 40\,cm (60\,cm). 
\subsection{Filter mechanism}

An electron moving perpendicular to a magnetic field will, in general, undergo cyclotron motion from the Lorentz force. The 
central axis of its trajectory with respect to a magnetic field can be described by the guiding center system (GCS) variables formed by taking the average across one cyclotron orbit of all forces acting on the particle.  
The GCS is a non-inertial reference frame in which the transverse plane is oriented orthogonal to the magnetic field direction~\cite{roederer2014particle}.
The direction of the GCS trajectory deviates from the direction of the magnetic field lines in the presence of four fundamental drift terms:
\begin{equation}
\label{eq:drifts}
\bm{V}_D = \bm{V}_\perp = \left( q\bm{E} + \bm{F} - \mu \bm{\nabla} B - m \frac{d\bm{V}}{dt} \right) \times \frac{\bm{B}}{qB^2} 
\end{equation}
where $q$ and $m$ are the electron charge and mass, respectively, and $\bm{V} = \bm{V}_\perp + \bm{V}_\parallel$ is the total phase-averaged velocity of the GCS trajectory with perpendicular and parallel components with respect to the magnetic field line. The perpendicular component is referred to as the drift velocity, $\bm{V}_D$. In equation~(\ref{eq:drifts}), the four drift terms, from left to right, are given by (1) the $\bm{E} \times \bm{B}$ drift; (2) the external force drift (such as gravity); (3) the gradient-$B$ drift; and (4) the inertial force drift.

The GCS description is valid in the limit that the $E$ and $B$ fields vary slowly spatially relative to the cyclotron radius, $\rho_c$, and slowly in time, through the motion of the particle, compared to the cyclotron
period, $\tau_c$, namely:
\begin{align} \label{eq:adinv1}
\rho_c & \ll \left| \frac{B}{\nabla{B}} \right|, \ \left| \frac{E}{\nabla{E}} \right| \ ; \ \mathrm{and} \\
\tau_c & \ll \left| \frac{B}{dB/dt}\right|, \ \left|\frac{E}{dE/dt}\right| \ ; \label{eq:adinv2}
\end{align}
where the total variation per unit time seen by the particle comes from the variation in time at a fixed point in space and the variation due to the displacement while the field is fixed in time: $d/dt = \partial/\partial t + \bm{V \cdot \nabla}$.
These conditions, if satisfied, allow the motion of the electron to be accurately described by adiabatic invariants, and, in particular, the first adiabatic invariant.  The derivation of the first adiabatic invariant
is found in these references~\cite{Alfven1940,cary2009hamiltonian}
and follows from the action-angle variable description of the Hamiltonian in 
terms of the gyroaction $J \equiv (mc/q)\mu$ canonically conjugate to the cyclotron phase angle, where $\bm{\mu}$, with magnitude $\mu$, is the
orbital magnetic moment of the electron with respect to a magnetic field $\bm{B}$.
Starting with a non-relativistic treatment, $\mu$ in the GCS frame is given by
\begin{equation}
\mu = \frac{m v_\perp^{*2}}{2 B}
\end{equation}
where $\bm{v}_\perp^*$ is the instantaneous velocity of the electron
perpendicular to the magnetic field line in the GCS frame (starred quantities) and are related to the inertial frame 
instantaneous velocity $\bm{v} = \bm{v}_\perp + \bm{v}_\parallel$ by
$\bm{v}_\perp^* = \bm{v}_\perp - \bm{V}_D$ and
$\bm{v}_\parallel^* = \bm{v}_\parallel - \bm{V}_\parallel \approx 0$.
The angle, $\alpha$, between $\bm{v}$ and $\bm{B}$, also equal to
\begin{equation}
\alpha = \arccos \frac{v_\parallel}{v} \ ,
\end{equation}
is the {\it pitch angle} of the electron.

In the presence of a non-uniform magnetic field, the Hamiltonian term 
$U = - \bm{ \mu \cdot B}$ gives rise to a total net force given by
\begin{equation}
\bm{f} = - \bm{\nabla} U = - \mu \bm{\nabla} B \ .
\end{equation}
The parallel component, $f_\parallel$,
is the well-known {\it mirror force} responsible for magnetic adiabatic 
collimation and the magnetic bottle effect for trapping charged particles in
non-uniform magnetic fields.
The perpendicular component, $f_\perp$, is the source of the gradient-$B$ drift.
This drift is particularly interesting for a filter since only non-electric
drifts can lead to a change in total kinetic energy.  Drifts due to externally applied electric fields are always perpendicular to $\bm{E}$ by construction and therefore cannot do any work because it is always possible to boost into a frame of reference in which the drift is zero -- electrons under $\bm{E} \times \bm{B}$ drift follow surfaces of constant voltage.

In contrast, the gradient-$B$ drift can do work because the presence of an non-uniform magnetic field implies that in the stationary reference frame of the electron, there is a time-dependent magnetic field which induces an electric field
\begin{equation}
\nabla\!\times\!\mathbf E \;=\; -\,\frac{\partial \mathbf B}{\partial t}\;\neq\;0 .
\end{equation}

This induced electric field can do work, provided there is something to do work against.
More precisely, when both $\bm{E} \times \bm{B}$ and gradient-$B$ drifts are present, the $\bm{E} \times \bm{B}$ drift acts as a transport drift to move the electron to regions of varying $B$. 
The induced $E$ from the gradient-$B$ drift can then do work against the very potential which provides the $\bm{E} \times \bm{B}$ drift, reducing the internal kinetic energy of gyromotion, or the transverse kinetic energy, for a corresponding increase in voltage potential.
This is described by, inserting terms from equation~(\ref{eq:drifts}), 
\begin{equation} \label{eq:energycon}
\frac{d T_\perp}{dt} = - q \bm{E \cdot V}_D = - q \bm{E \cdot}
\left( q\bm{E} - \mu \bm{\nabla} B \right) \times \frac{\bm{B}}{qB^2}  
= \frac{\mu}{B^2} \bm{E \cdot} ( \bm{\nabla} B \times \bm{B} ) 
\end{equation}
where $T_\perp$ is the internal kinetic energy of gyromotion in the GCS frame.

The principle underpinning this energy draining mechanism is the adiabatic invariant $\mu$.
Since $\mu=T_\perp/B$ by definition, as the electron moves into a region of smaller $B$, the transverse kinetic energy $T_\perp$ is also decreased by a proportional amount.
Physically, a charged particle undergoing cyclotron motion is akin to a loop of current, with a certain amount of magnetic flux enclosed inside the loop. In the absence of a $\bm{E} \times \bm{B}$ to transport the electron to regions of different $B$, nature's tendency to want to conserve flux dictates the direction of gradient-$B$ drift along flux-conserving drift shells. Once an $\bm{E} \times \bm{B}$ drift is added and the electron is transported to regions of different flux, the adiabatic invariant dictates the transverse energy draining.

These drift motions and the associated energy draining can be realized into an electromagnetic filter, or spectrometer, by a special configuration of field components such that the trajectories and amount of energy draining are known and controllable, resulting in a geometrical aperature and acceptance. A physical realization can then be instrumented to generate the required fields along the local region of the envisioned trajectories.

\subsection{Analytical formulation}
To implement this mechanism into a design for an electromagnetic filter, or spectrometer, we first set the arbitrary but convenient requirement that the net transverse trajectory of the electron, i.e. the GCS trajectory, is a straight line in the transverse direction. We formulate the field and drift conditions that would produce such a trajectory, then build a filter geometry around the trajectory with the necessary voltage settings to produce the fields. The trajectory is referred to as the center line of the filter.
\begin{figure}[h!]
    \centering
    
    \includegraphics[width=0.75\textwidth]{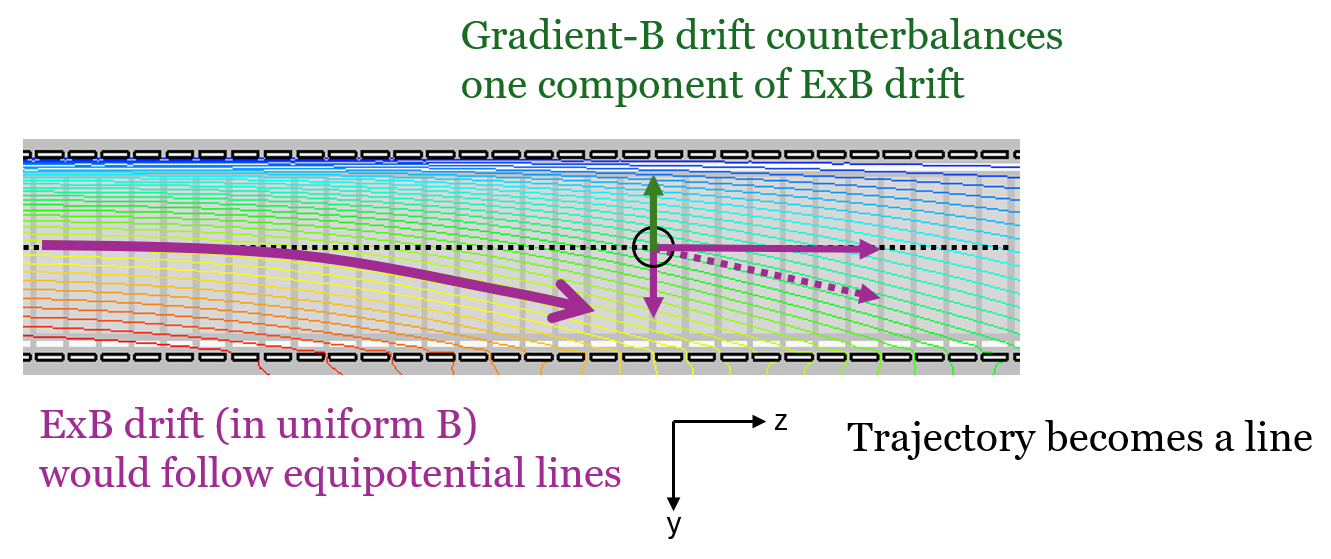}
    
    \caption{Schematic illustrating the realization of a filter concept by balancing drift components along a desired GCS particle trajectory. The magnetic field points out of the page and decreases in magnitude from left to right. The equipotential lines for the potential are shown. The rectangular elements along the top and bottom are electrodes that generate the field. The gradient-$B$ drift is in green, the $E\times B$ drift is in purple.
    }
    
    \label{fig:drift_balancing_schematic}
\end{figure}

Consider the following field configurations:
\begin{eqnarray}
V(x,y,z) &=& V_0\ \sin \left( \frac{y}{\lambda} \right) e^{-z/\lambda} \\
B_x &=& B_0 \cos \left( \frac{x}{\lambda} \right) e^{-z/\lambda} \\
B_y &=& 0 \\
B_z &=& - B_0 \sin \left( \frac{x}{\lambda} \right) e^{-z/\lambda}
\end{eqnarray}
with the potential as defined giving the electric fields
\begin{eqnarray}
E_x &=& 0 \\
E_y &=& \frac{V_0}{\lambda} \cos \left( \frac{y}{\lambda} \right) e^{-z/\lambda} \\
E_z &=& - \frac{V_0}{\lambda} \sin \left( \frac{y}{\lambda} \right) e^{-z/\lambda} .
\end{eqnarray}

The sinusoidal dependence is constructed to satisfy the curl expressions of Maxwell's laws. The ansatz exponential in $z$ sets the magnitude of the fields and, crucially, allows for the derivatives in $z$ to cancel out with the ansatz, leaving only a $-1/\lambda$ term. Additionally, the $E$ and $B$ fields share the same characteristic decay factor $\lambda$ in the exponential, which is a requisite condition to generate a straight trajectory. Along an arbitrary GCS trajectory, the drifts are:
\begin{eqnarray}
\mathbf{V}_{\nabla B} &=& - \frac{\mathbf{\mu} \times {\nabla_\perp B}}{q B} \\
\mathbf{V}_{E \times B} &=& \frac{\mathbf{E} \times \mathbf{B}}{B^2} = \frac{E_\perp}{B} .
\end{eqnarray}

With the fields as defined above, consider only the motion along the $y-z$ plane at $x=0$ such that
\begin{eqnarray}
\mathbf{V}_{\nabla B}(z)  &=& - \frac{\mathbf{\mu} \times {\nabla_\perp B}}{q B}
= - \frac{\mathbf{\mu} \times \mathbf{\nabla_\perp B(z)}}{q B(z)}
= - \frac{\mu}{qB_x} \frac{d B_x}{dz} {\mathbf{\hat y}} \\
\mathbf{V}_{E \times B}^{y}(z)  &=&\frac{\mathbf{E} \times \mathbf{B}}{B_x^2} = \frac{E_z B_x {\mathbf{\hat y}}}{B_x^2} = \frac{E_z}{B_x}{\mathbf{\hat y}} .
\end{eqnarray}
Then enforce the condition $\mathbf{V}_{E \times B}^{y}(z) = \mathbf{V}_{\nabla B}(z)$, giving
\begin{eqnarray}
- \frac{V_0}{B_0 \lambda} \sin \left( \frac{y}{\lambda} \right) = \frac{\mu}{q \lambda}.
\end{eqnarray}

Solving for $V_0$ and rewriting the first adiabatic invariant $\mu = T_{\perp 0}/B_0 e^{-z_0/\lambda}$, with $T_{\perp 0}$ being the actual initial transverse kinetic energy of an electron released at $(y_0,z_0)$,
\begin{eqnarray}
V_0 = - \frac{T_{\perp 0} e^{z_0/\lambda}}{q \sin \left( \frac{y_0}{\lambda} \right) } =  \frac{T_{\perp 0}e^{z_0/\lambda}}{ \sin \left( \frac{y_0}{\lambda} \right) } \ \text{[eV]} .
\end{eqnarray}
Rearranging,
\begin{eqnarray}
\label{eq:v0condition}
T_{\perp 0} = V_0 \sin \left( \frac{y_0}{\lambda} \right) e^{-z_0/\lambda} ,
\end{eqnarray}
which makes clear that the initial $T_{\perp 0}$ needed for a straight trajectory is a function of $V_0$ and $(y_0,z_0)$.

Additionally, because a net forward drift in $z$ is needed to drive the filtering mechanism, this limits $y_0$ to the region where $(E_y/E_z > 1)$,
\begin{eqnarray}
0 < \frac{y_0}{\lambda} < \frac{\pi}{4} \ .
\end{eqnarray}

\subsection{Geometrical aperture and acceptance}

In general, for electrons with a $T_{\perp 0} \neq V_0$, let 
\begin{eqnarray}
\Delta T_\perp = T_{\perp 0} - V_0 \sin \left( \frac{y_0}{\lambda} \right) e^{-z_0/\lambda}.
\end{eqnarray}
Then, expressing $\Delta T_\perp$ along the trajectory in $z$ as
\begin{eqnarray}
\Delta T_\perp = V_0\ e^{-z/\lambda} \left[ \sin \left( \frac{y}{\lambda} \right) - \sin \left( \frac{y_0}{\lambda} \right) \right] \ ,
\end{eqnarray}
the $y$-position of the trajectory can be written as a function of $z$,
\begin{eqnarray}
y(z) = y_0 + \lambda \arcsin \left[ \sin \left( \frac{y_0}{\lambda} \right) + \frac{ \Delta T_{\perp}}{V_0} e^{z/\lambda} \right] \ ,
\label{eq:geo_acceptance}
\end{eqnarray}
which tells us the window of $\Delta T_\perp$ needed to maintain a trajectory that is within a desired $y'$ of $y_0$ at a given $z$. In other words, this defines the analytical aperture and acceptance of the spectrometer along $z$.

\begin{figure}[h!]
\centering

\includegraphics[width=1\textwidth]{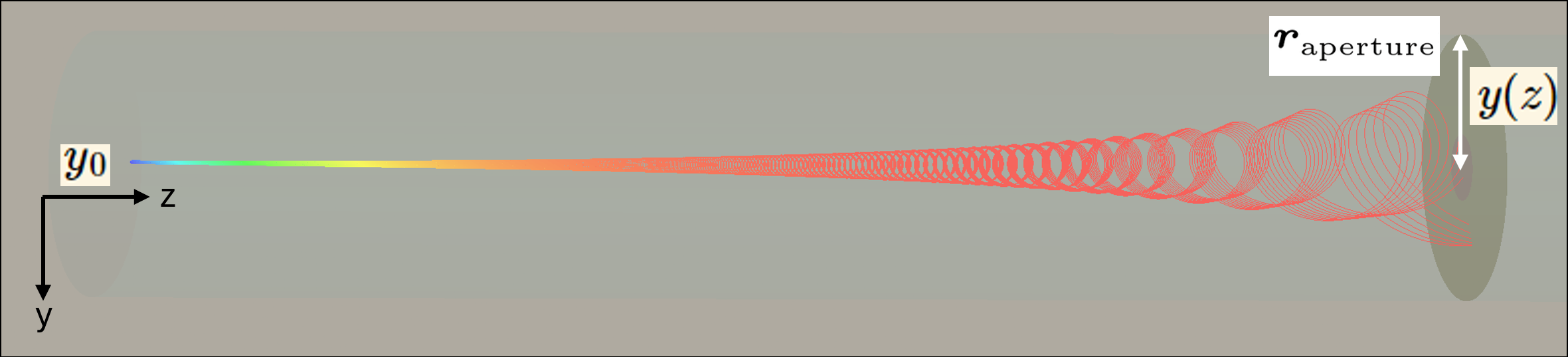}

\par\bigskip

\resizebox{\textwidth}{!}{
\includegraphics[height=4cm, keepaspectratio]{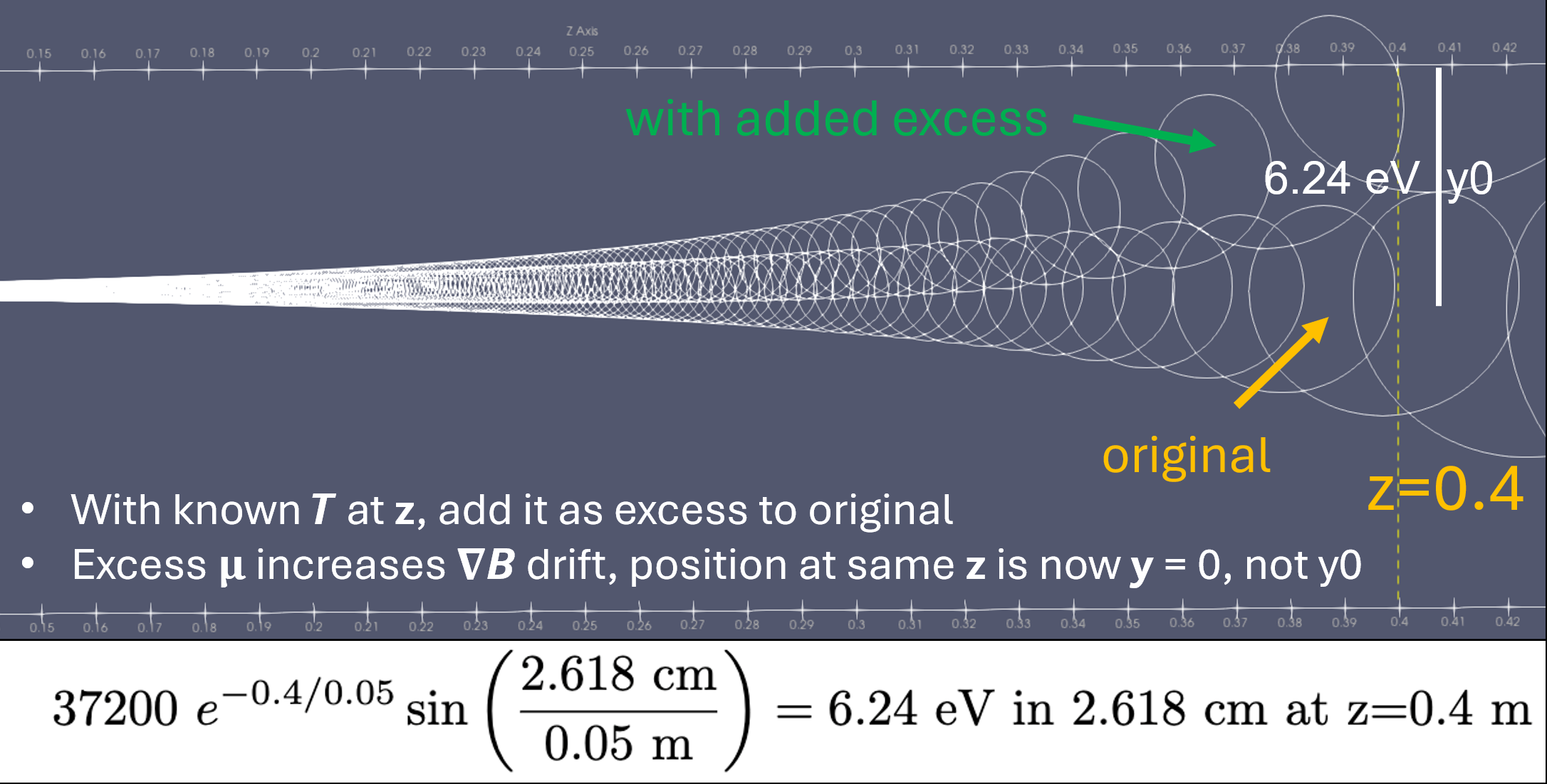}
\hspace{1em}
\includegraphics[height=4cm, keepaspectratio]{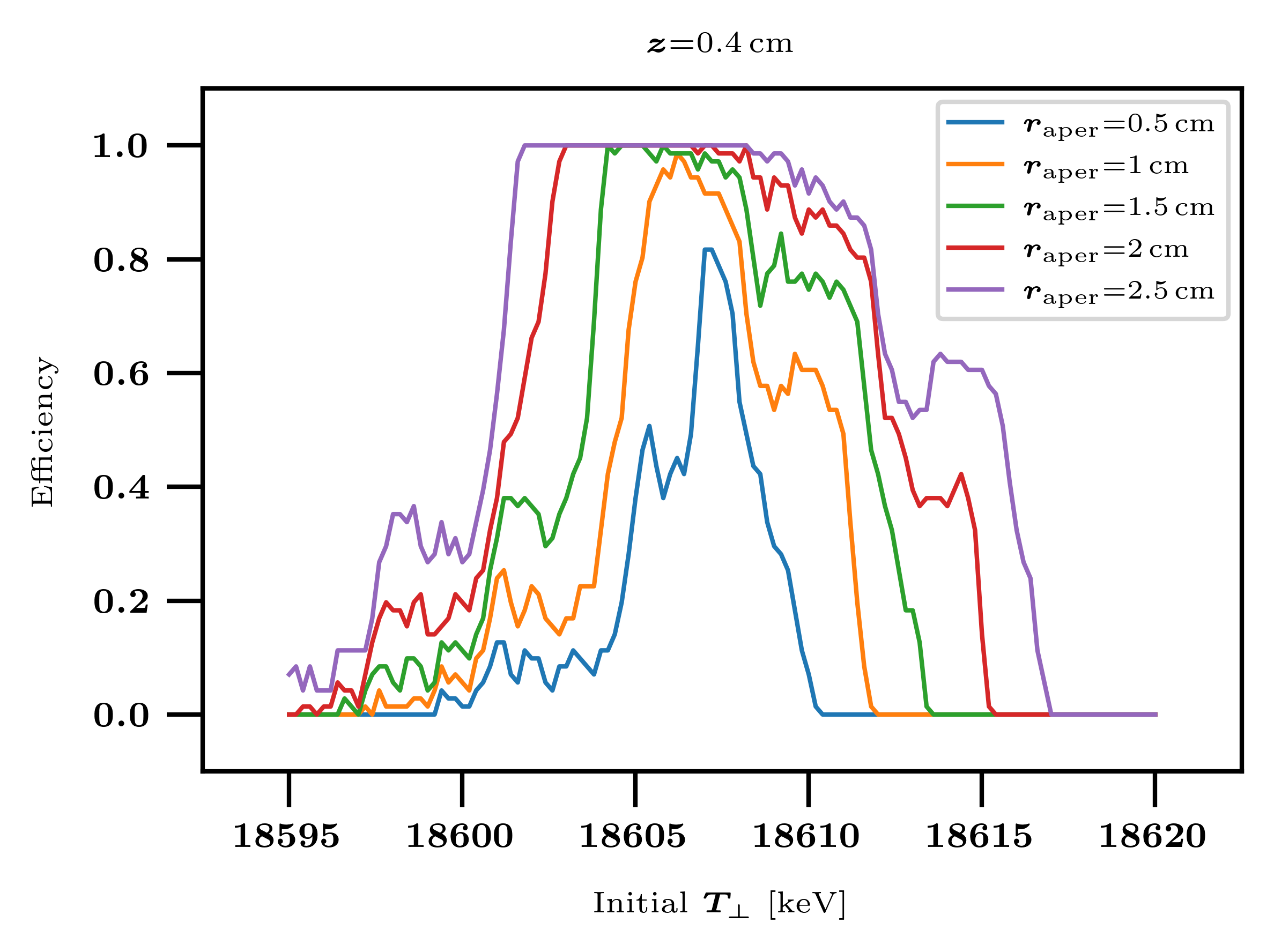}
}

\caption{Geometrical aperture with example acceptance calculation according to~\ref{eq:geo_acceptance}
}

\label{fig:aperture_all}
\end{figure}

\begin{figure}[h!]
\centering

\includegraphics[width=1\textwidth]{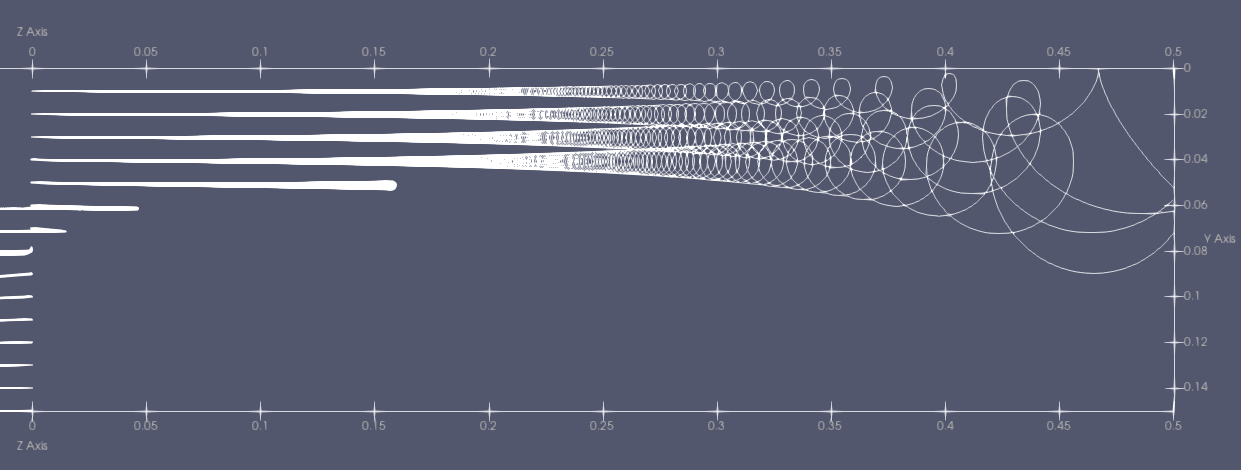}
\par\bigskip
\resizebox{\textwidth}{!}{

\includegraphics[height=4cm, keepaspectratio]{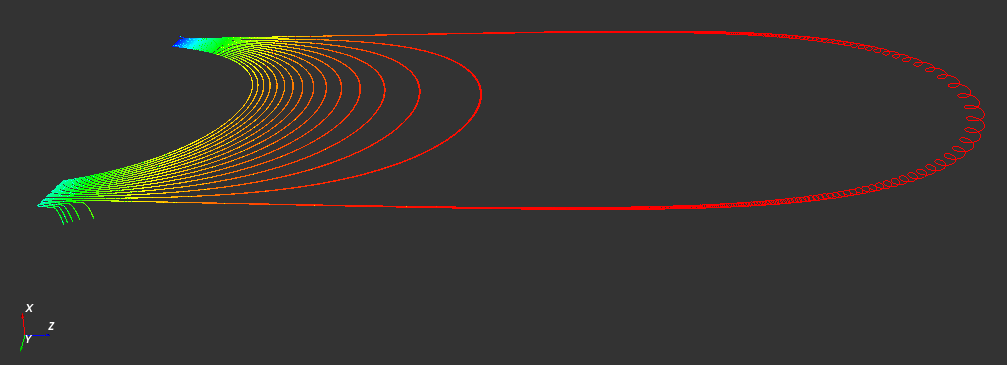}
\hspace{1em}
\includegraphics[height=4cm, keepaspectratio]{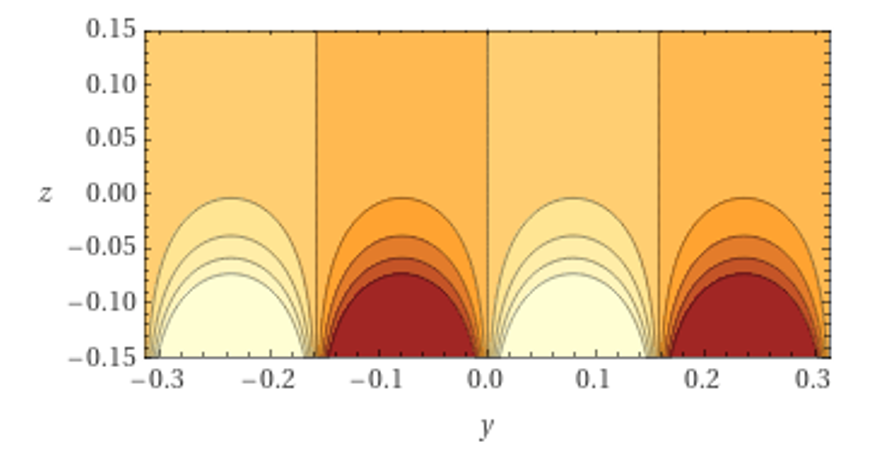}
}

\caption{Top panel: Pitch 90 electrons released at various points along the $y$-axis, with each transverse kinetic energy set to satisfy~\ref{eq:v0condition}, resulting in straight GCS trajectories in $z$. Bottom left: Pitch 90 electrons with uniform energy, released along the $y$-axis, resulting in curved GCS trajectories that sketchout the contours of the analytical field, pictured bottom right. }

\label{fig:aperture_untuned}
\end{figure}
    
\begin{figure}[h!]
\centering

\includegraphics[width=1\textwidth]{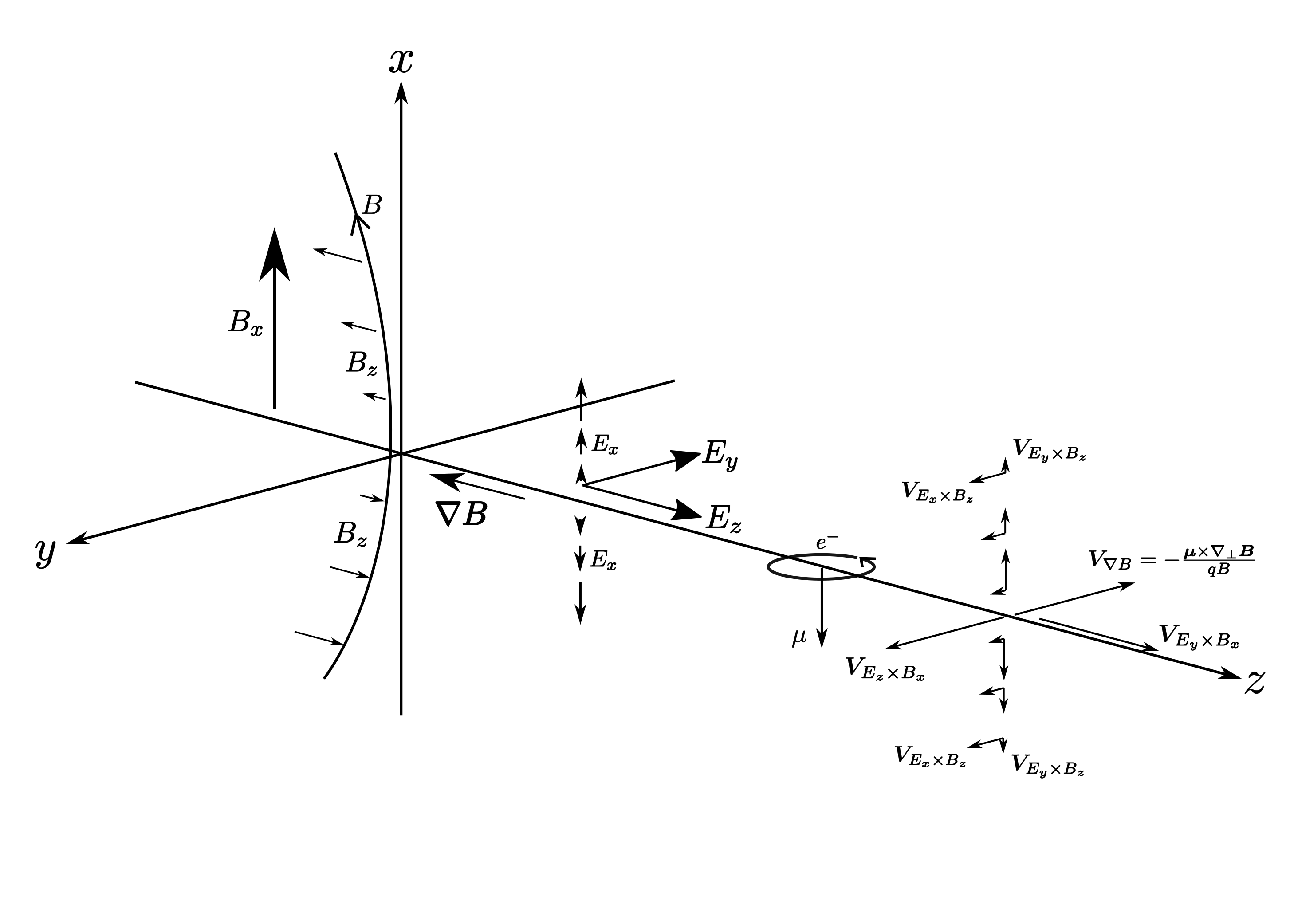}

\caption{Field and drift components for an electron in the filter. The relevant drift terms are the zeroth-order $\bm{E} \times \bm{B}$ drift, which drives transverse transport, and the first-order non-electric gradient-$B$ drift, which does work against the increasing potential along the trajectory. The gradient-$B$ drift is proportional to $\nabla_\perp \bm{B} / B$; this term is constant if $B$ is an exponential in the transverse direction $z$ with a characteristic decay parameter $\lambda$ that is also the radius of curvature of the magnetic field lines. The magnitude of $\bm{E} \times \bm{B}$ drift is proportional to $E/B$, so if $\bm{E}$ is also an exponential with the same $\lambda$ as $B$, then exact canceling is achieved between one of the $\bm{E} \times \bm{B}$ components and the gradient-$B$ drift, and the other $\bm{E} \times \bm{B}$ components are also constant.
}

\label{fig:component_diagram}
\end{figure}

A schematic diagram of the field and drift components is shown in Figure~\ref{fig:component_diagram}. 

\subsection{Energy resolution}

The amount of energy drained in the filter is proportional to the decrease in the magnetic field strength, as dictated by the adiabatic invariant:
\begin{eqnarray}
\mu &=& \frac{T_\perp}{B} .
\end{eqnarray}
In the $z$-direction, since $B_z = - B_0 \sin \left( \frac{x}{\lambda} \right) e^{-z/\lambda}$, the energy resolution is therefore exponentially bounded to arbitrary precision.

\begin{figure}[h!]
    \centering

    \resizebox{\textwidth}{!}{

    \includegraphics[height=6cm, keepaspectratio]{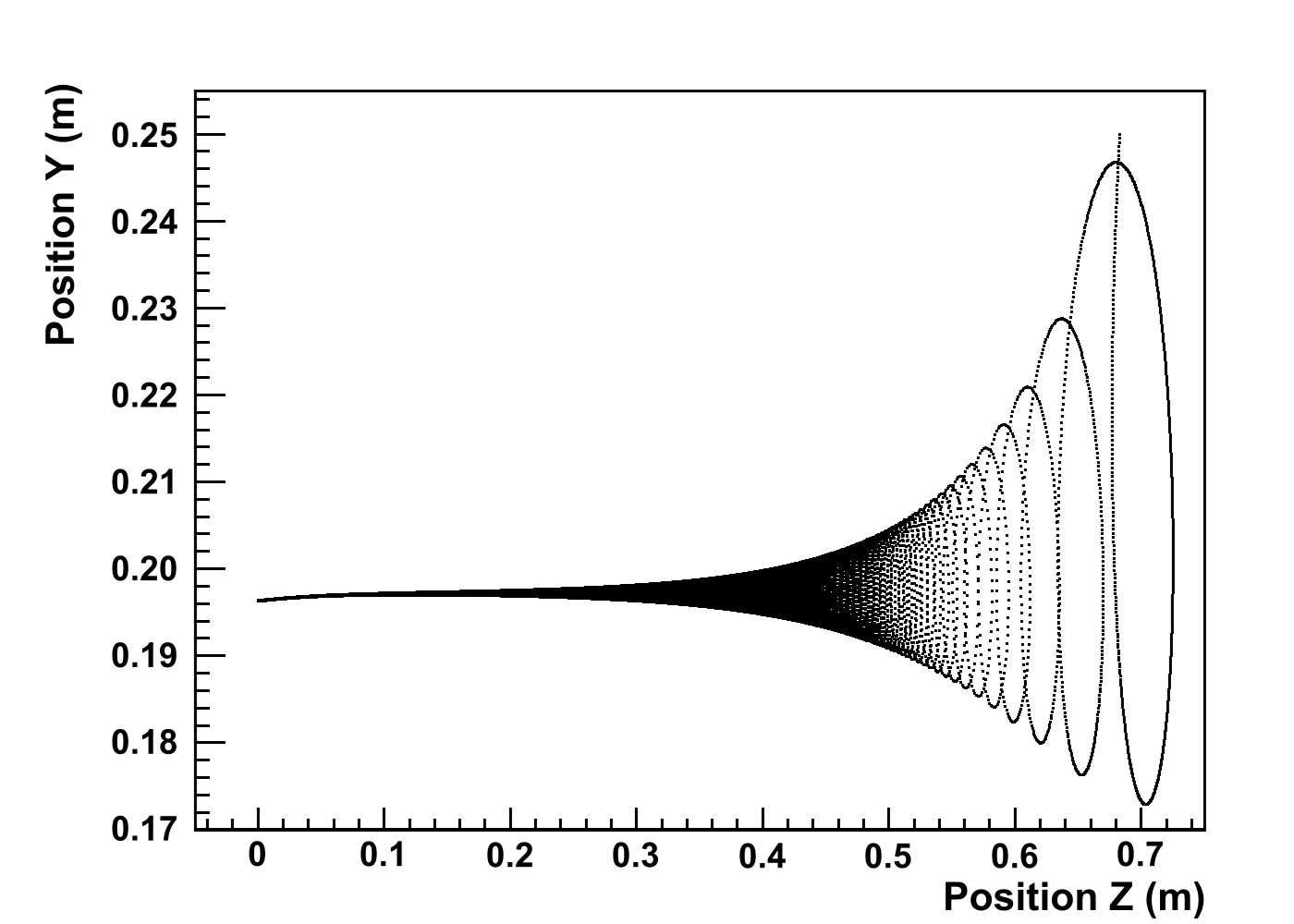}
    \hspace{1em}
    \includegraphics[height=6cm, keepaspectratio]{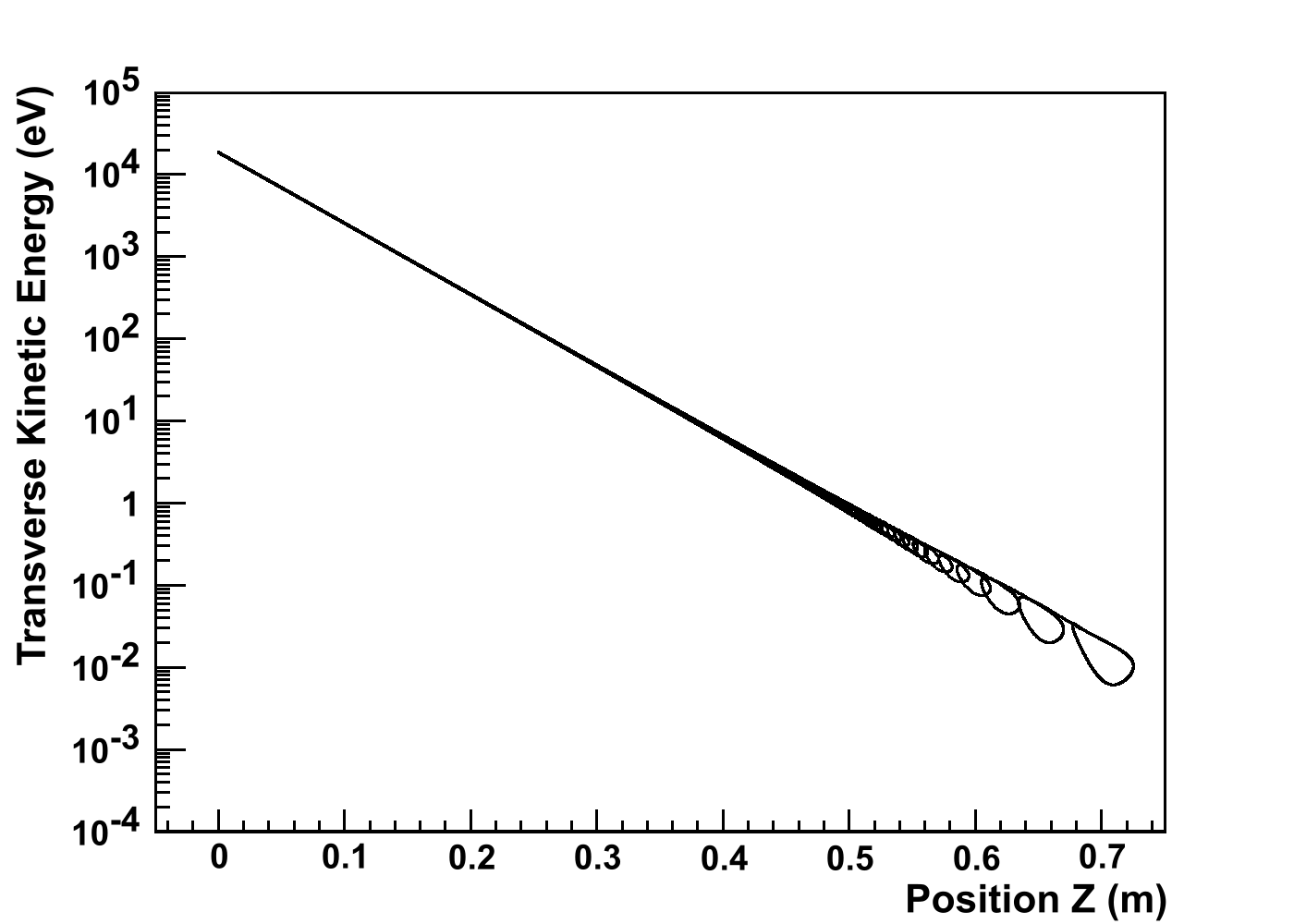}
    \hspace{1em}
    \includegraphics[height=6cm, keepaspectratio]{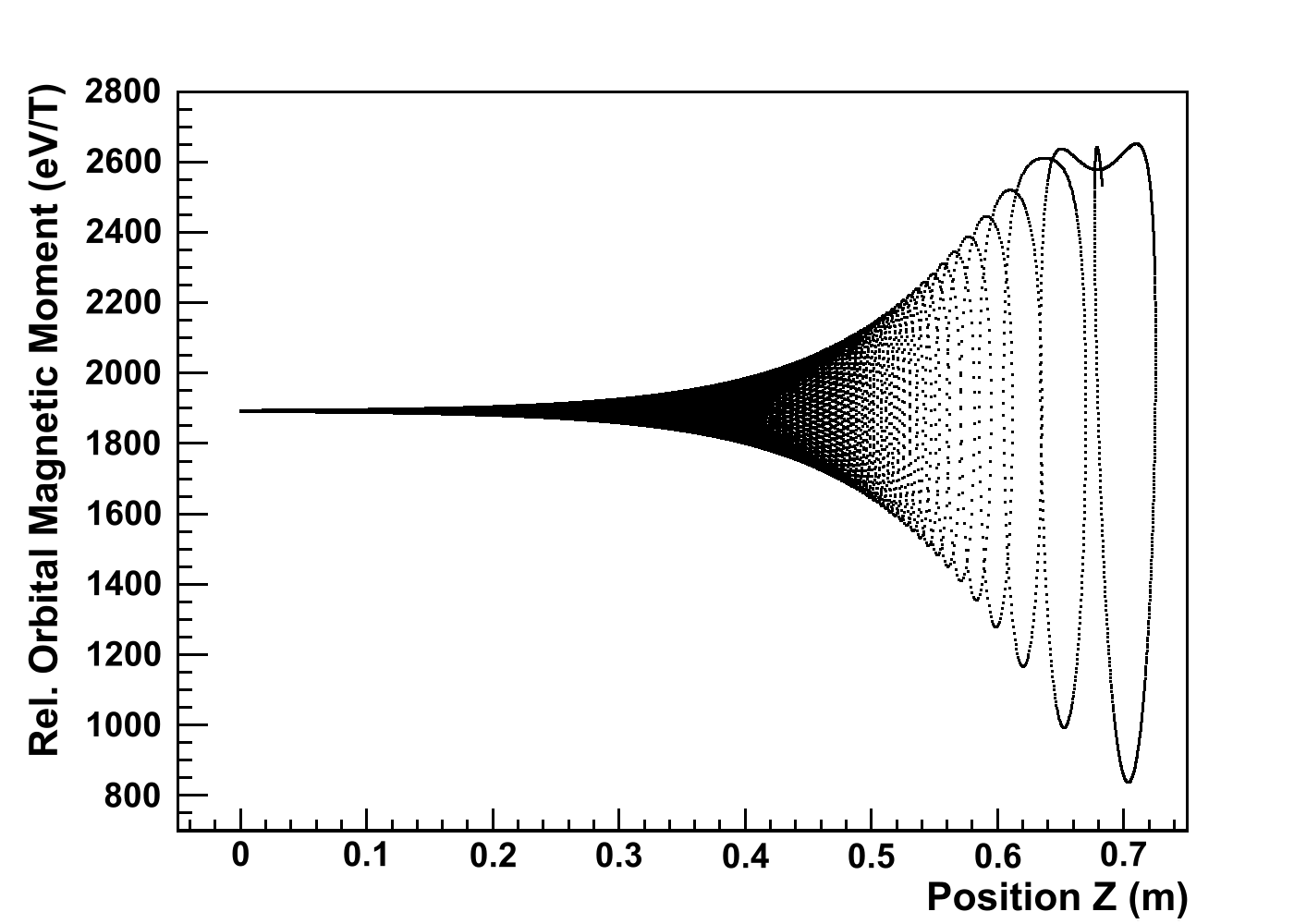}
    }

    \caption{Analytical simulations showing exponentially bounded energy resolution. }

    \label{fig:energy_resolution}
\end{figure}
\subsection{MAC-E vs. transverse drift filters}

\begin{figure}[h!]
\centering
\includegraphics[width=0.85\textwidth]{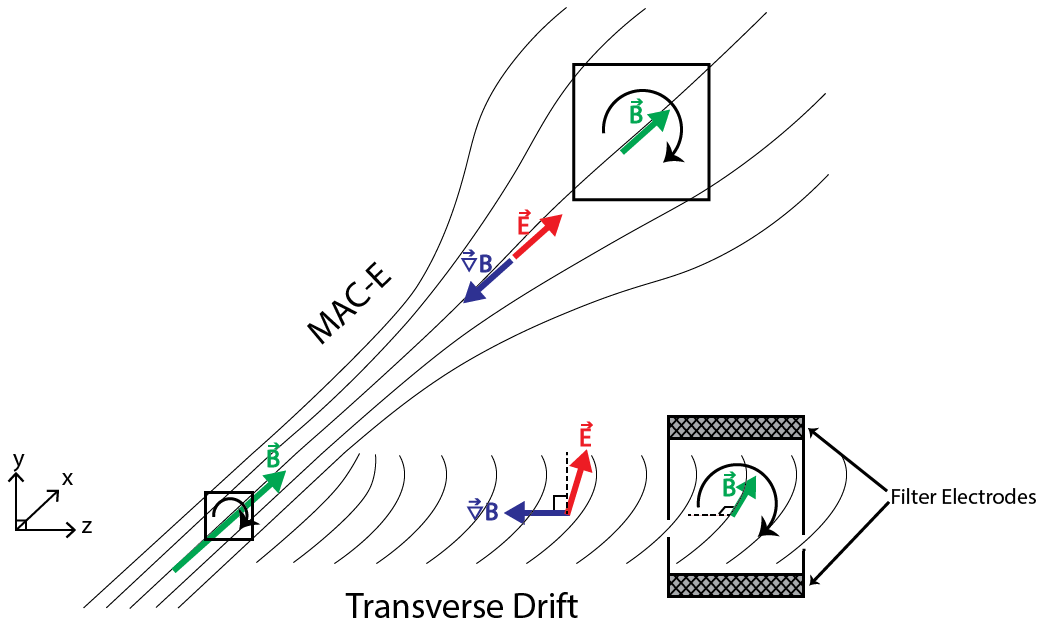}
\caption{Comparison of the magnetic field, magnetic field gradient and electric fields of the MAC-E and transverse drift filters.  A snapshot of the position space in the high magnetic field region of the spectrometer shows how position space expands within the aperture of the MAC-E filter as the momenta are collimated under the action of the magnetic field gradient.  For an identical snapshot in a transverse drift filter, a narrow window of transverse momenta remain within the aperture of the filter while all other momenta are pushed into the top or bottom electrodes.
}
\label{fig:MACEcomp}
\end{figure}

The distinguishing characteristic of a transverse drift filter in comparison with a MAC-E filter is the orientation of the magnetic field gradient with respect to the field direction~\cite{betti2019design, Beamson1980Collimating}.  The two configurations are orthogonal to one another in this respect, as shown schematically in Figure~\ref{fig:MACEcomp}.

The MAC-E filter gradient is along the field direction, hence the effect on the electron momentum is to rotate the transverse components along the field as the electron moves from a high field region to low field.  The rotation speed is conveniently computed in terms of the first adiabatic invariant, the orbital magnetic moment $\mu$, under adiabatic conditions, which are a function of the electron's parallel velocity and the magnitude of the gradient.  The electrostatic filtering is achieved with an electric field parallel to the direction of the magnetic field gradient in such a way that the electron climbs a potential barrier at a rate following the filter design where the parallel momentum component is preferably dominant to the transverse momentum component and hits zero when a turning point is reached.

In contrast, for a transverse drift filter the magnetic field gradient is orthogonal to the magnetic field direction.  The electric field is also orthogonal to the magnetic field direction, but tilted with respect to the magnetic field gradient so that net drift of the electron motion is against the direction of the magnetic field gradient.  This motion pushes the electron from a high magnetic field region to a low field.  Here again, under adiabatic conditions, the rate of reduction in the electron transverse momentum is computed using the first adiabatic invariant.  Unlike the MAC-E filter, there is no collimation effect, the electron parallel momentum is nominally unaffected by the transverse drift filter in this respect.  The work done by the magnetic field gradient term goes directly into the reduction of the electron transverse momentum.

The advantages of the transverse drift filter are in the compactness of the filter dimensions and the direct transition to a zero magnetic field region at the end of the filter. The compact size allows for many filter elements to operate simultaneously, scaling up the effective tritium target.  The zero field region at the end of the filter is ideal for installing a transition-edge sensor (TES) microcalorimeter~\cite{rajteri2020tes}.  The microcalorimeter resolution has an energy measurement resolution on order of 0.05\,eV, providing an additional two orders of magnitude greater sensitivity to the endpoint measurement from the filter alone.

\subsection{Magnet design}
To realize the filter experimentally, the magnetic field must first be realized as it sets the overall scale of the experimental apparatus and the transport between the different modules, i.e. from the tritium target, through the RF, to the filter, and finally to the microcalorimeter. In the discussion above we stipulated that the magnetic field magnitude in $x$ along the center line of the filter, $B_x(z)$, should be a decaying exponential in order for the gradient-$B$ drift to be constant along the center line. Here we present a magnet design that can produce such a field.

\begin{figure}[h!]
\centering
\includegraphics[width=0.5\textwidth]{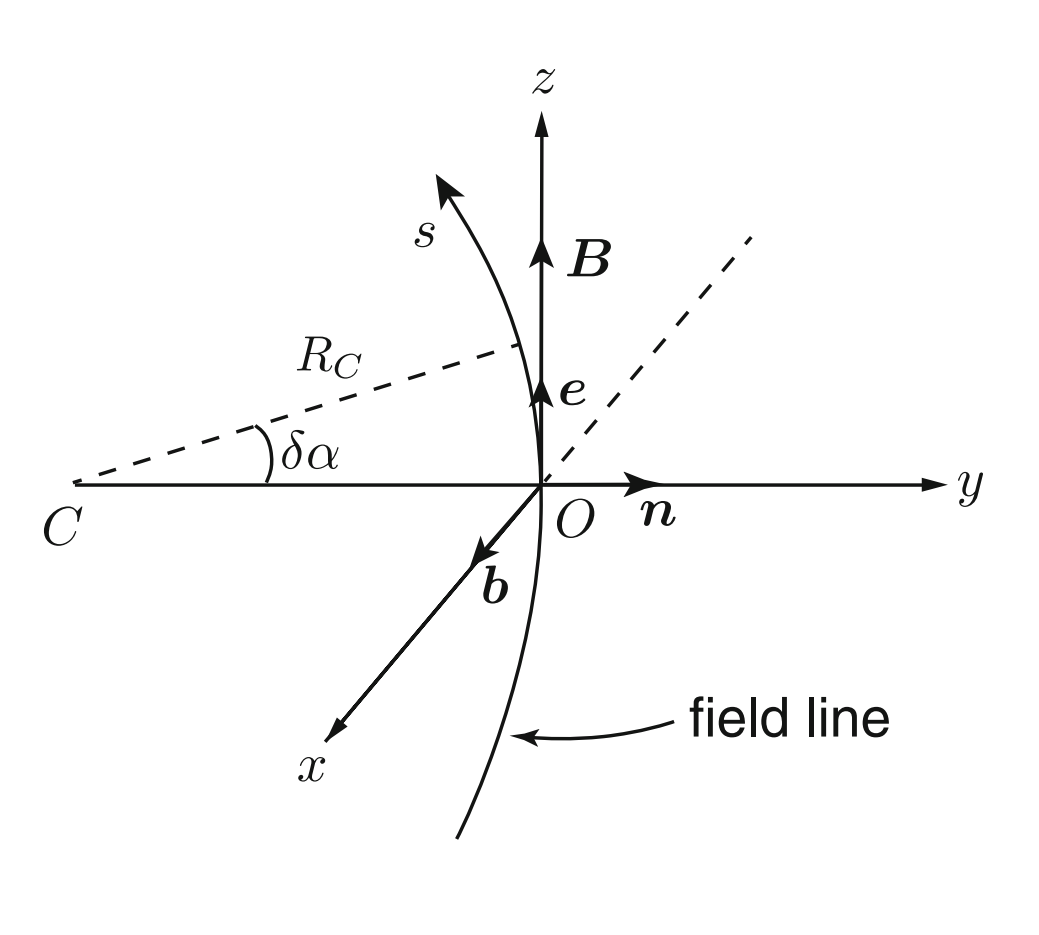}
\caption{The local natural coordinate system of magnetic field lines~\cite{roederer2014particle}.
}
\label{fig:b_coords}
\end{figure}

It can be shown~\cite{roederer2014particle} that in the absence of local currents, $\bm{\nabla \times B} = 0$, the transverse gradient of the magnetic field can be written as
\begin{equation}
\bm{\nabla_\perp} B =  \left( \bm{B \cdot \nabla} \right) \left( \frac{\bm{B}}{B} \right) - \left( \bm{\nabla} \times \bm{B} \right) \times \left( \frac{\bm{B}}{B} \right) = - \frac{B}{R_c} \bm{\hat{n}}
\end{equation}
where $\bm{\hat{n}}$ is the instantaneous unit vector normal to the magnetic field line and $R_c$ is the instantaneous radius of curvature of the field line (Figure~\ref{fig:b_coords}). In the limit where $B_y$ and $B_z$ go to zero, the transverse gradient $\bm{\nabla_\perp} B(z)$ can be expressed as $d B_x/dz = -B_x(z)/R_c$.
This is satisfied if $B_x(z)$ is an exponential, 
\begin{equation}
B_x(z) = B_0 \ e^{-z/\lambda}
\end{equation}
where $\lambda = R_c$, showing that the exponential decay parameter $\lambda$ is equal to the radius of curvature of the field line.
Knowing this, from the uniqueness theorem we know that if we observe a pattern of magnetic field lines that have an apparently constant radius of curvature along one dimension, then the field magnitude in that region must be decreasing exponential with a decay parameter $\lambda$ equal to the radius of curvature of the field lines.

A standard dipole magnet typically produces a region of uniform field in the gap between the two pole faces, and away from the gap, the field is roughly dipole in character and the magnitude decreases as approximately $1/z^3$, where as above we take the direction of the field to be positive-$x$, and $z$ to point transversely outward from the gap. This relation can be observed visually through the increasing radius of curvature of the field lines away from the gap, as all of the flux exiting one pole face must eventually return to the other pole face.

\begin{figure}[h!]
\centering
\includegraphics[width=1\textwidth]{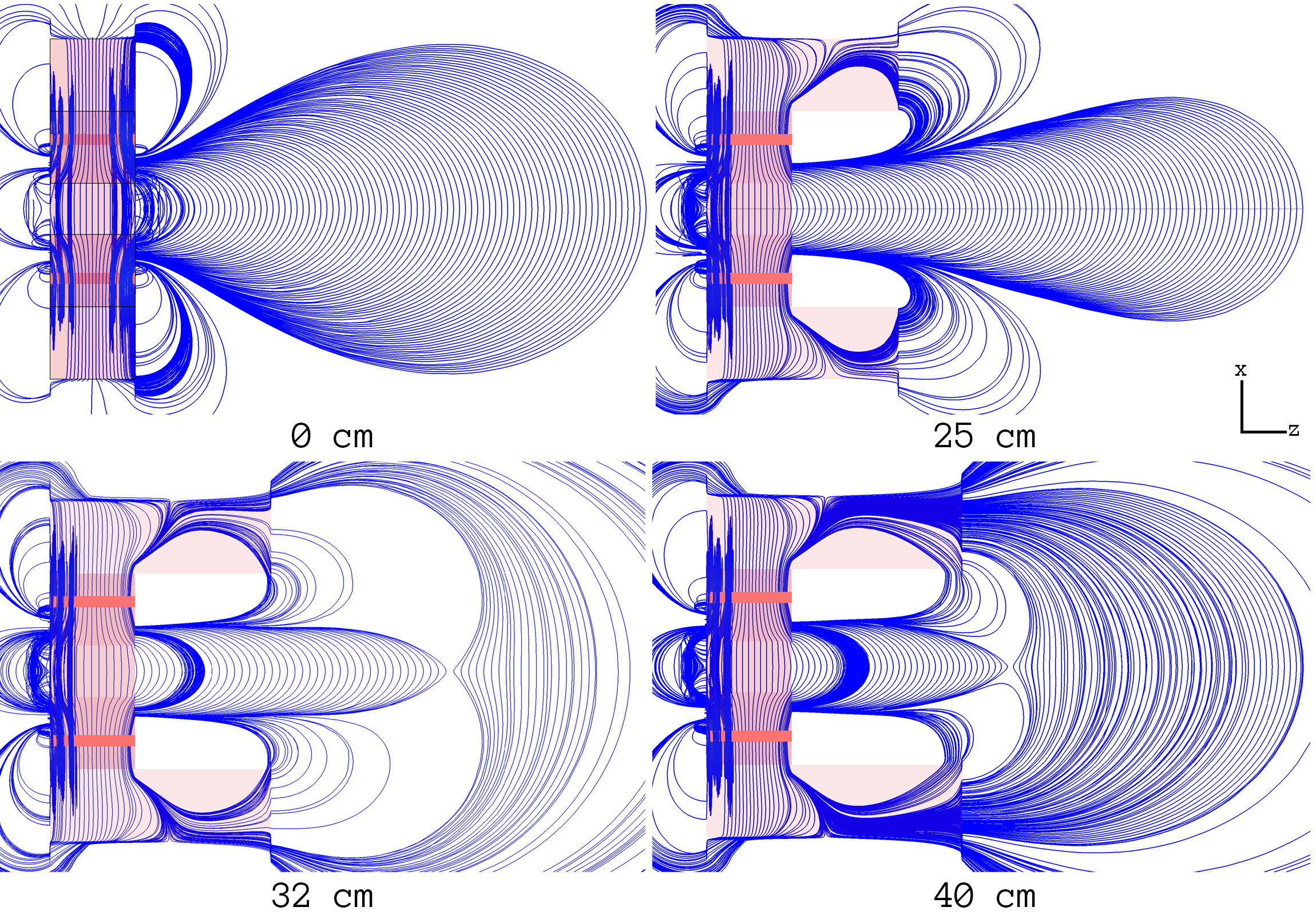}
\caption{Cross-sectional field-line views of four magnets with varying extension lengths in the plane $y=0$. The $z$ width of the standard magnet with no extensions (top left) is 20\,cm; the uniform field between the pole faces is approximately 1\,T. Extensions of length 25\,cm, 32\,cm, and 40\,cm are shown. A region of approximately constant radius of curvature is observed for an extension of length 32\,cm (bottom left).  The density of field lines is arbitrary.
}
\label{fig:magnet_fieldlines}
\end{figure}

By introducing symmetric iron extensions to the side walls of such a magnet, above and below the air gap, it is possible to turn the dipole-like field into a region of field with a constant radius of curvature. The effect of the extensions, modeled in Figure~\ref{fig:magnet_fieldlines} as rectangular bars, is to divert some of the flux away from the return yokes and back into the vacuum above and below the transverse plane to be recycled into the air gap. The resulting flux pressure constrains the expansion of the original flux radius, and with the right extension length, a channel of field lines with apparently constant radius of curvature can be created. The effect is minimal if the extensions are too short; if the extensions are too long or approach each other too closely, a flux loop between them is closed and a quadrupole point is formed in the center region where the field goes to zero and switches direction. The effect of extensions of varying length on the field profile is shown in Figure~\ref{fig:magnet_fieldlines}.

\begin{figure}[h!]
\centering
\includegraphics[width=1\textwidth]{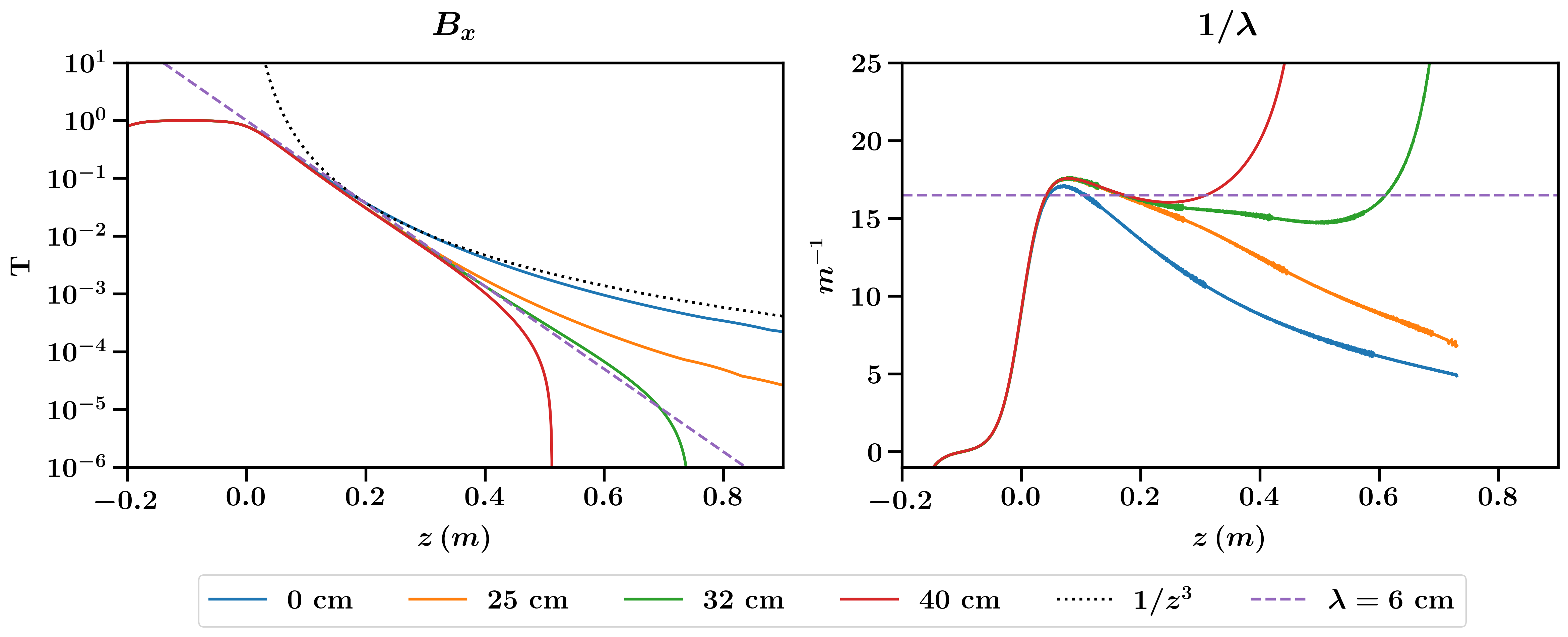}
\caption{ $B_x(z)$ and $1/\lambda (z)$ along the $z$ direction for the four magnets. Black dotted line on left chart shows the dipole-like character of the original field without extensions. The dashed line is $e^{-z/\lambda}$.
}
\label{fig:magnet_plots}
\end{figure}

The extensions need not be rectangular and fine adjustments to the location of the quadrupole point and the variance of $\lambda$ can be made by varying the shape. Such methods can be used in principle to achieve an arbitrary level of precision in $\lambda$; however, this is unnecessary as what is important is not that the field has a specific or exactly constant value of $\lambda$, but that the opposing drift components cancel out at each point in $z$ along the trajectory of the electron. This can be accomplished regardless of small deviations in $\lambda$ if the $e^{-z/\lambda}$ term of the drift components, which is used to set the voltages on the filter electrodes, is replaced by the sampled values from a precision magnetic field map of the magnet in use. Concretely, the $B_x$-component of the field is sampled along $z$ then normalized to the nearly constant $B_x$ at the air gap $z=0$ (Figure~\ref{fig:magnet_plots}):
\begin{equation}
    B_x(z)/B_x(z=0) \approx e^{-z/\lambda} \ .
\end{equation}
Nonetheless, in practice it is beneficial to avoid hard right-angles when manipulating large amounts of flux in a material like iron, as non-linear saturation at pinch points results in large temperature increases and material instability.

\begin{figure}[h!]
    \centering

    \resizebox{\textwidth}{!}{
    
    \includegraphics[height=6cm, keepaspectratio]{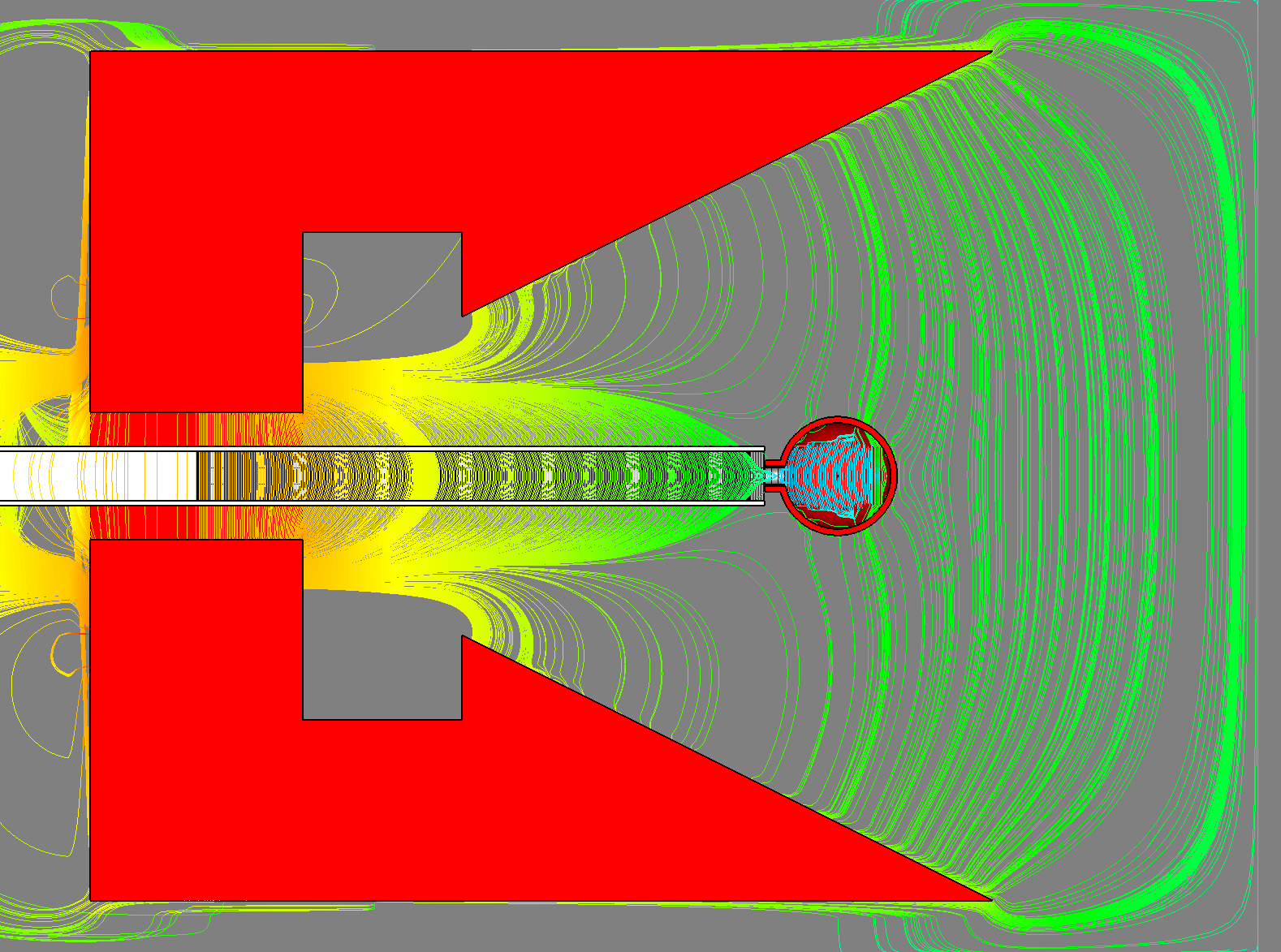}
    \hspace{1em}
    \includegraphics[height=6cm, keepaspectratio]{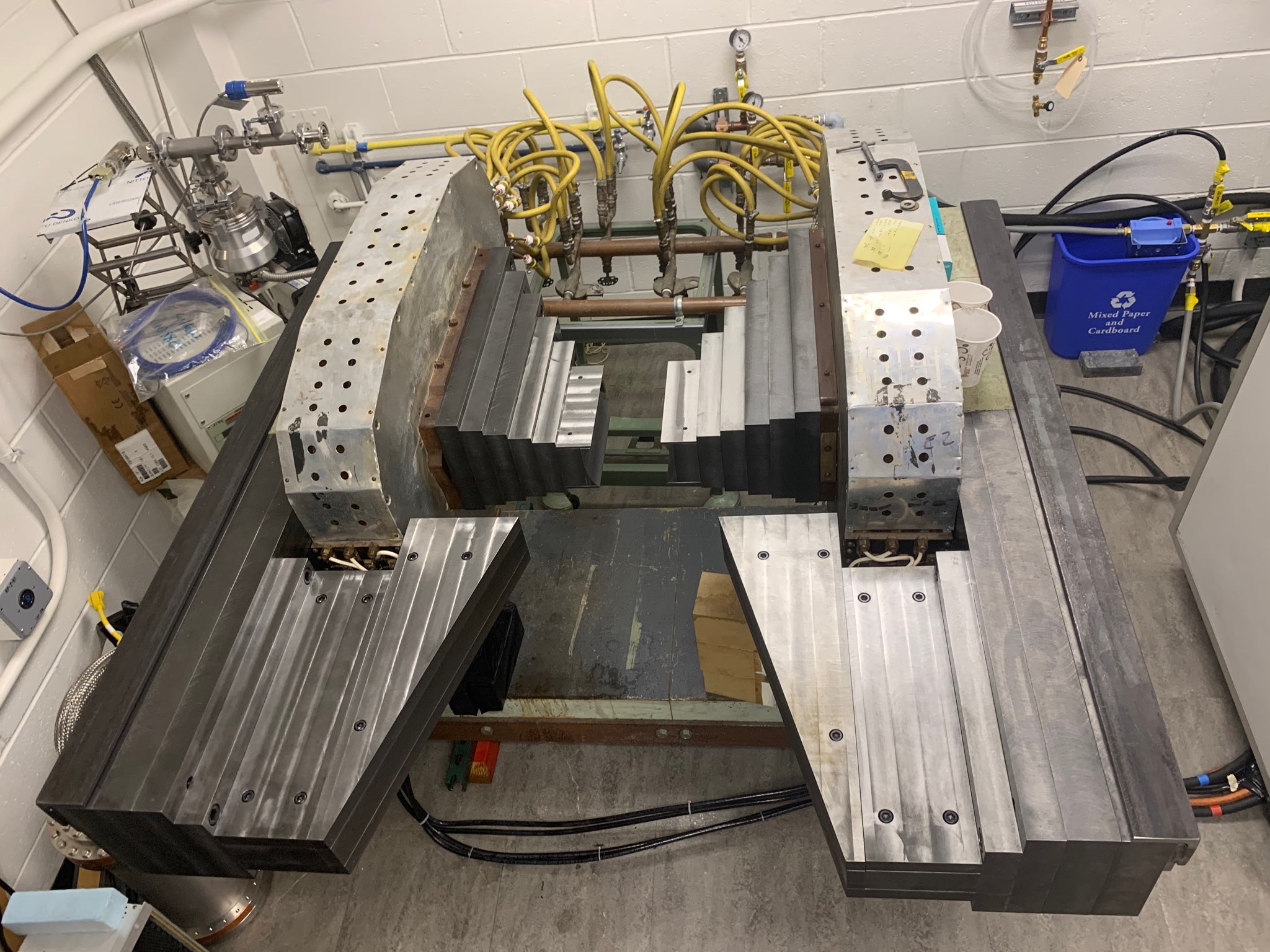}
    \hspace{1em}
    \includegraphics[height=6cm, keepaspectratio]{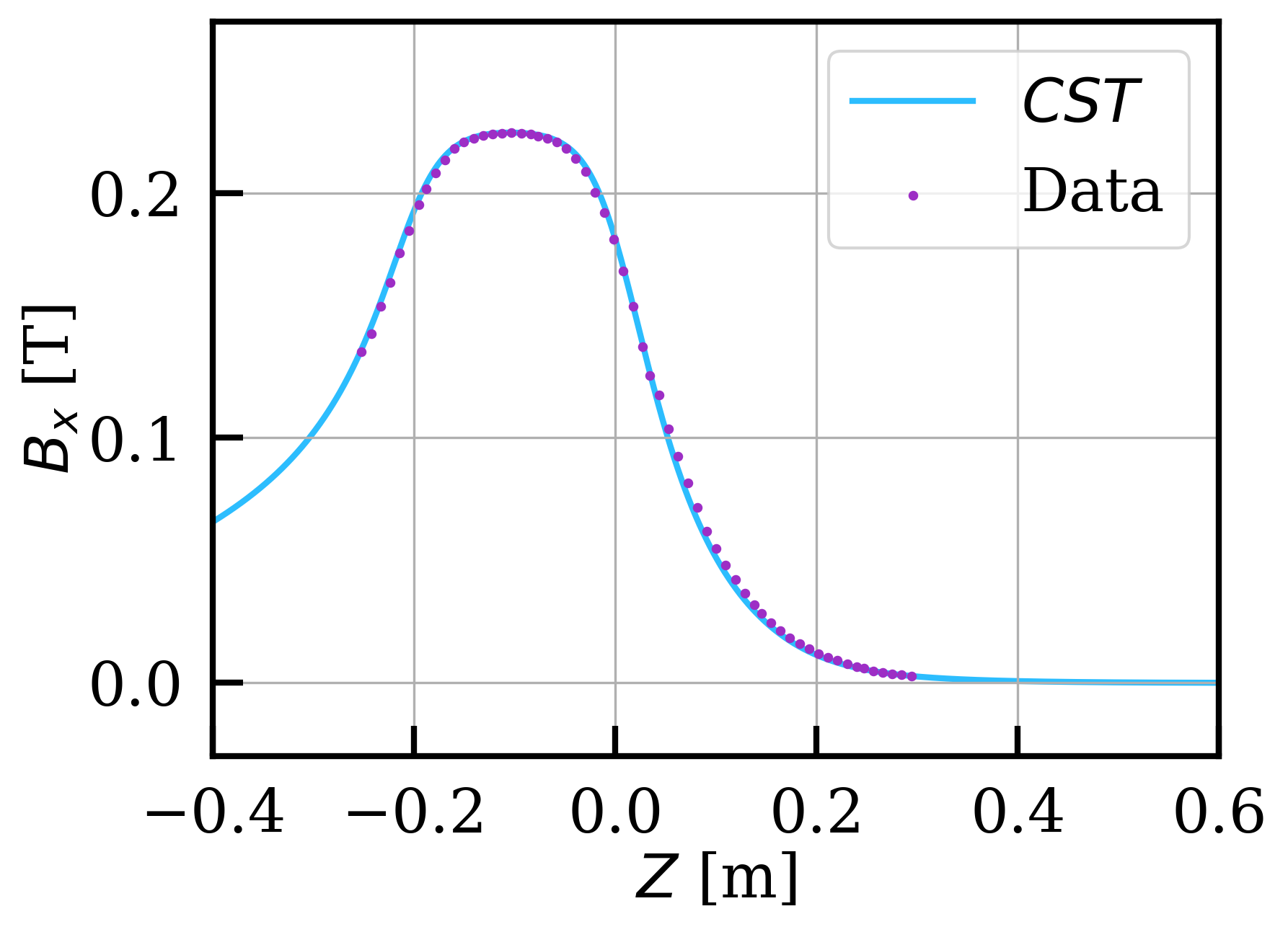}
    }

    \caption{Conceptual magnet design with field-shaping iron extensions and experimental realization of a prototype in Jadwin Hall. An old accelerator magnet was retrofit with iron extensions to produce the required field profile. }

    \label{fig:filter_mechanical}
\end{figure}

\begin{figure}[h!]
    \centering

    \resizebox{0.55\textwidth}{!}{
    
    \includegraphics[height=6cm, keepaspectratio]{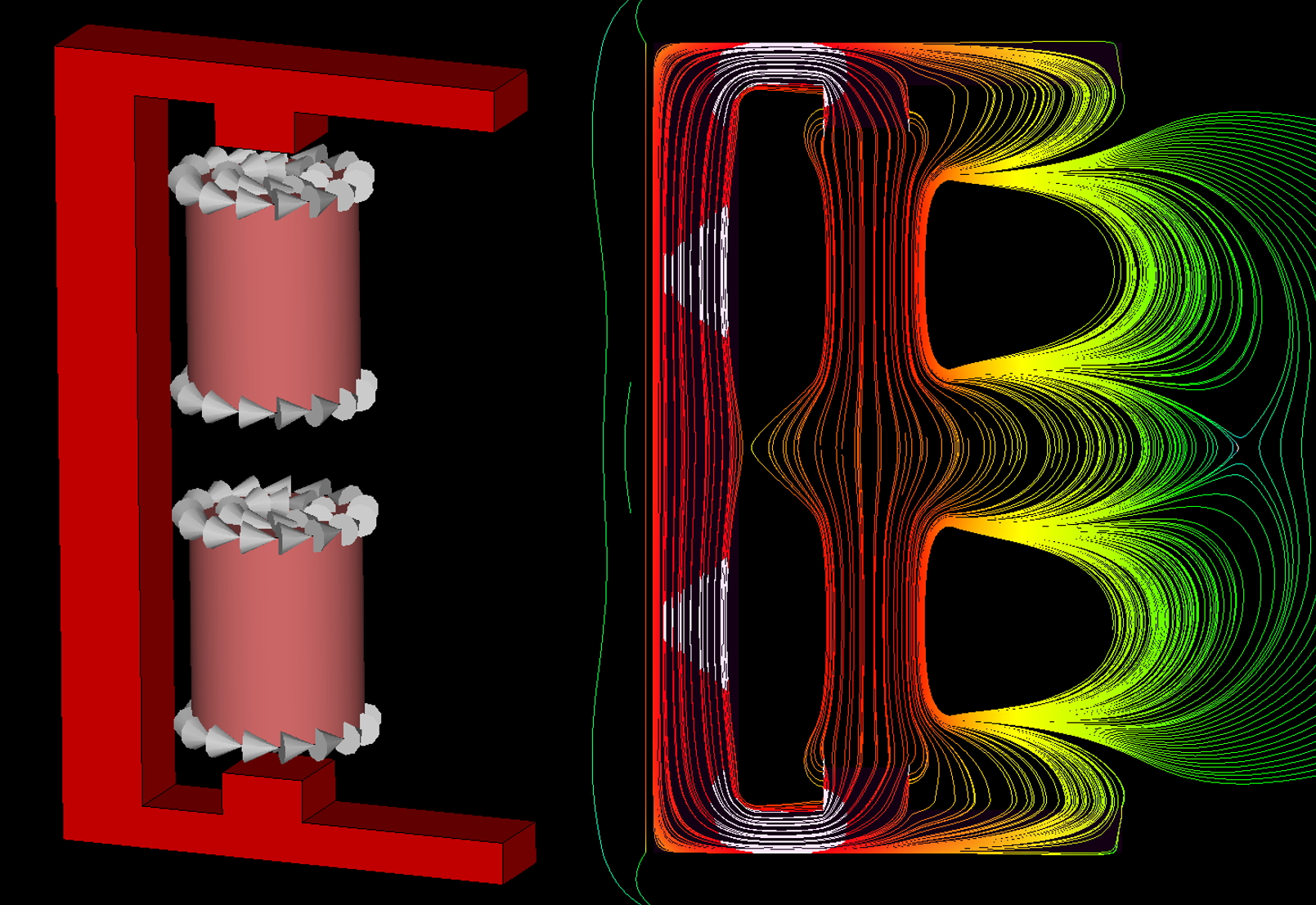}
    }

    \caption{Flux recycling concept using two superconducting solenoid magnets and an iron frame to redirect the flux. }

    \label{fig:sc_concept}
\end{figure}

A prototype magnet based on this design was constructed in Jadwin Hall, using an old accelerator magnet retrofit with iron extensions. The sampled field data showed good agreement with the simulated prediction~\ref{fig:filter_mechanical}.
The design was then expanded and scaled up for a PTOLEMY demonstrator apparatus currently being constructed at the Gran Sasso National Laboratory in Italy.
\begin{figure}[h!]
    \centering

    \resizebox{\textwidth}{!}{
    
    \includegraphics[height=6cm, keepaspectratio]{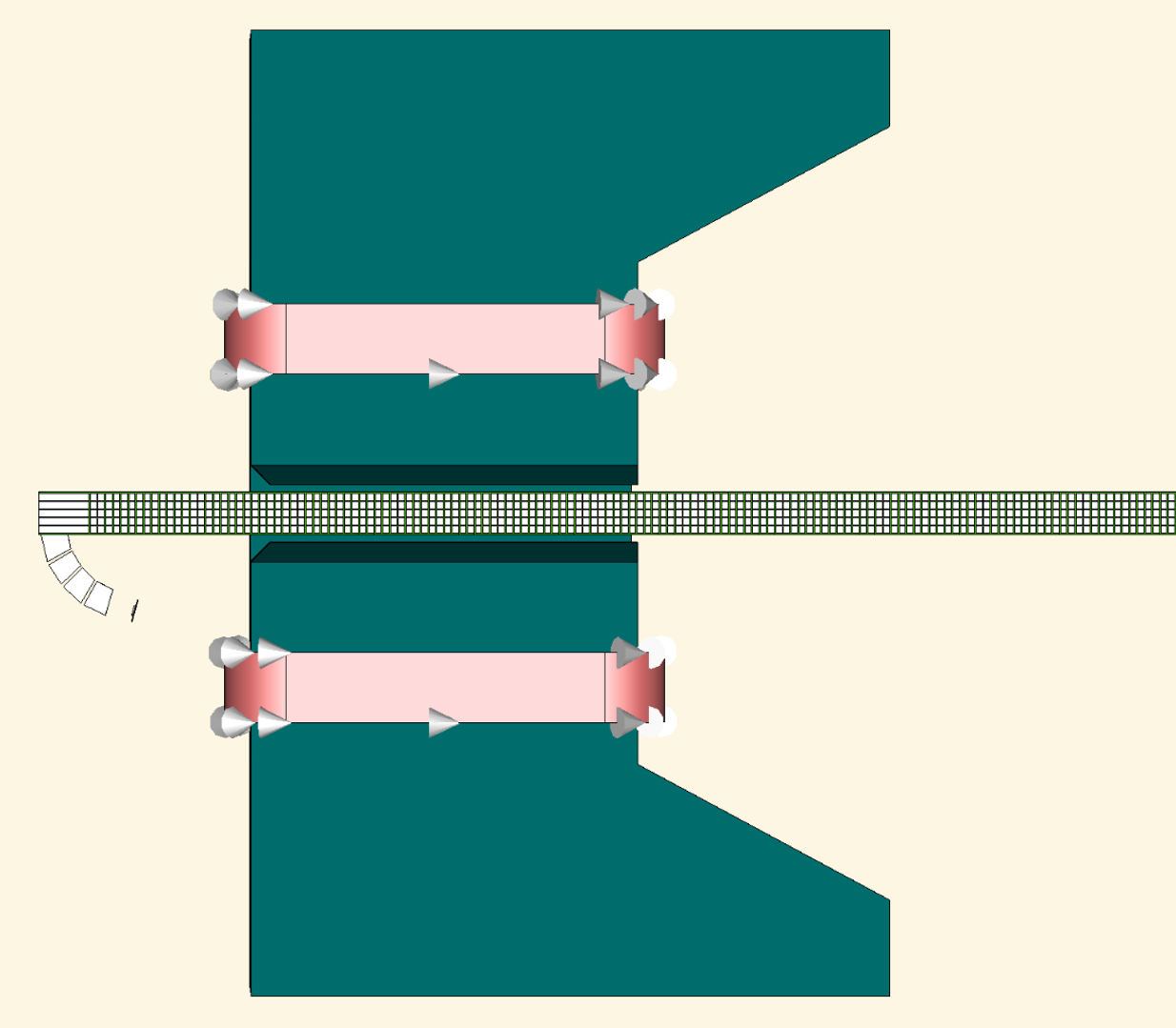}
    \hspace{1em}
    \includegraphics[height=6cm, keepaspectratio]{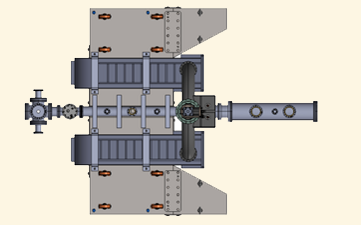}
    }

    \caption{Conceptual design for the LNGS demonstrator and mechanical rendering with coils and vacuum chamber for the filter. }

    \label{fig:lngs_magnet}
\end{figure}
The concept of recycling the flux lines to constrain the expansion of the field lines can be generally applied. A version using two superconducting solenoids and an external iron frame to recycle the flux is shown in Figure~\ref{fig:sc_concept}.

\subsection{Filter geometry}

\begin{figure}[h!]
\centering
\includegraphics[width=0.8\textwidth]{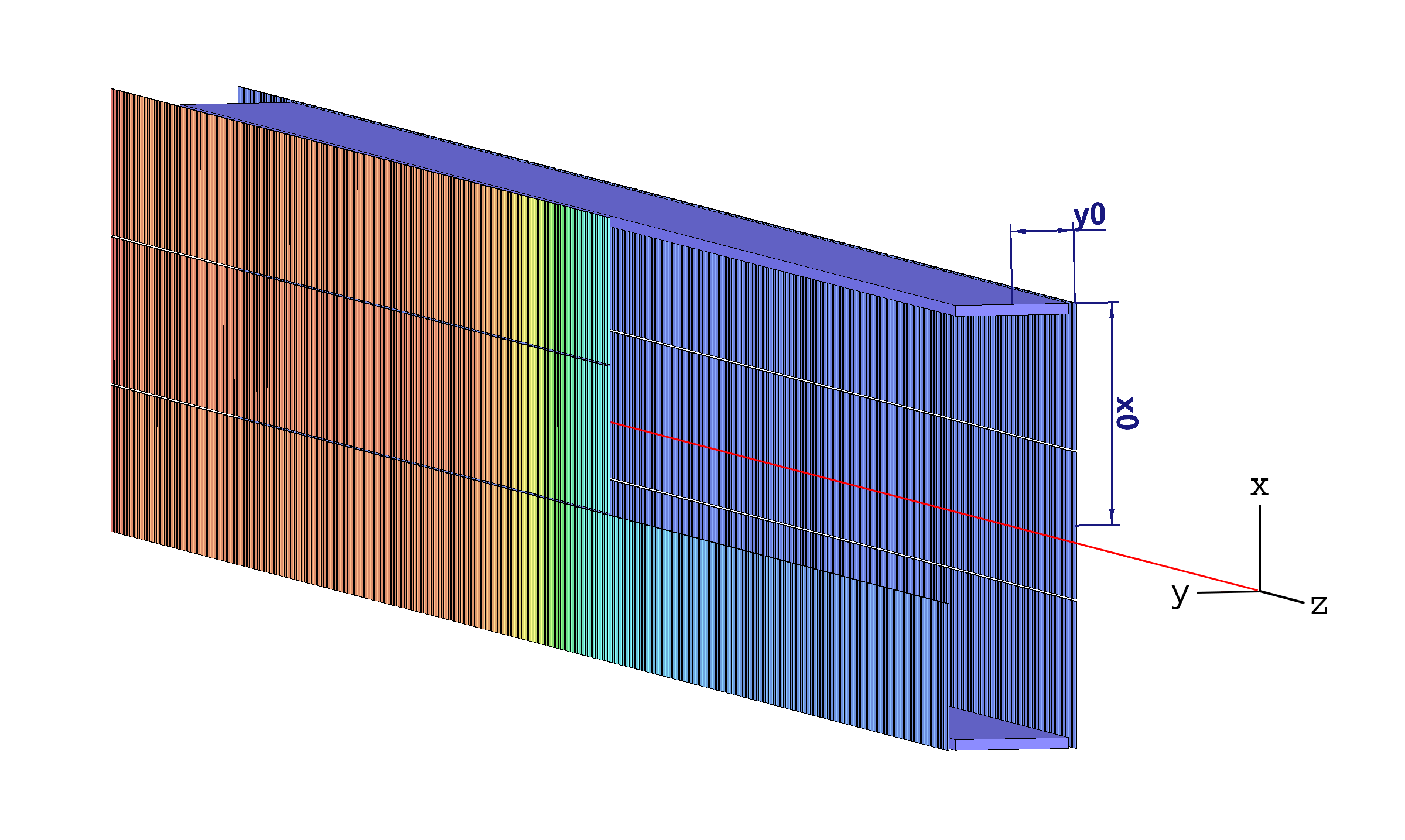}
\includegraphics[width=0.18\textwidth]{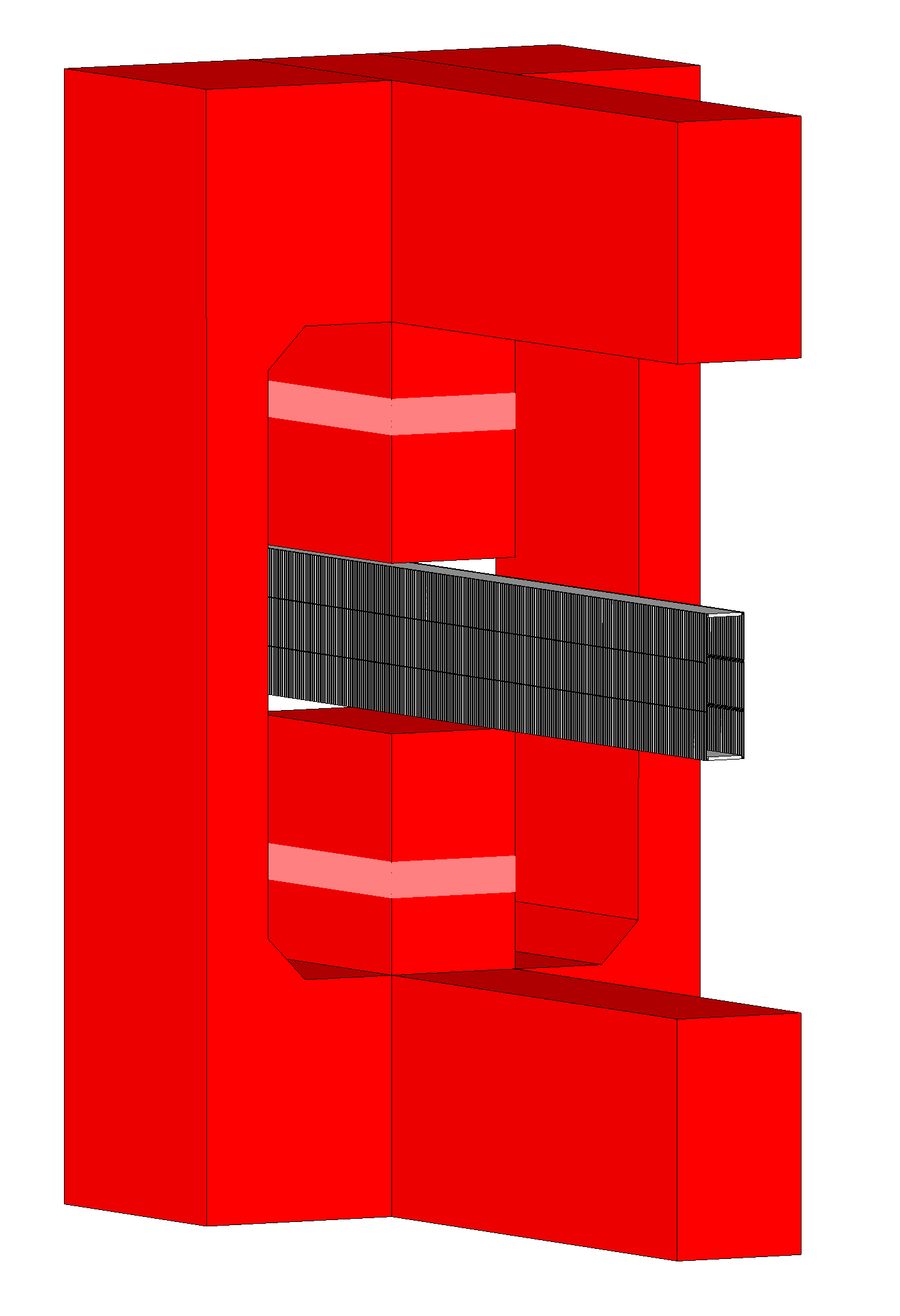}
\caption{({\it left}) Coordinate system and parameters for the filter. Part of the positive-$y$ side filter electrodes have been hidden from view. The center line of the filter, red, is the nominal GCS trajectory of the electron. Electrodes are colored by voltage. The negative-$y$ electrodes are all at constant voltage while the positive-$y$ electrodes vary. The total voltage differential along the center line is the energy drained. ({\it right insert}) The filter placed along the transverse plane of the magnet.
}
\label{fig:filter_coords}
\end{figure}

The filter geometry and coordinate system are shown in Figure~\ref{fig:filter_coords}. The magnetic field and motion of the electron parallel to it is in $x$; the transverse plane is $y-z$. The electron enters the filter through the region of uniform field in the air gap and the overall transverse drift of the trajectory is in the $+z$-direction. The electrodes responsible for the $\bm{E} \times \bm{B}$ drift are referred to as the filter electrodes and are placed a distance $\pm y_0$ from the center line ($y=0$) and span a length $2 x_0$ in $x$. Placed at $\pm x_0$ are two long electrodes that extend the full length of the filter in $z$. These are referred to as the bounce electrodes; they close off the volume enclosed by the filter electrodes and contain the electron within the filter by reflecting the parallel momentum at each end.

The parameters $x_0$ and $y_0$ are largely determined by the dimensions of the air gap of the magnet. A small aspect ratio $y_0/x_0$ is favorable to maintain field uniformity inside the filter. Nonetheless, $y_0$ should be large enough to accommodate the final cyclotron radius $\rho_c(z)$ of the electron as it grows with decreasing $B(z)$. The value of $x_0$ is best if maximized as close to the size of the air gap as possible; since the arch of the field lines is bookended by the two pole faces, it turns out $x_0 \approx \lambda$. The air gap in Figure~\ref{fig:magnet_fieldlines} is 12\,cm with $\lambda \approx 6$\,cm. In Figure~\ref{fig:filter_coords}, $x_0 = 5$\,cm.

To minimize $y_0$, we need the cyclotron radius $\rho_c$ as a function of $z$,
\begin{equation}
    \rho_c(z) = \frac{\sqrt{2m(T_\perp(z))}}{|q|B(z)} ,
\label{eq:radius}
\end{equation}
where $q$ is the charge of the electron, $T_\perp(z) = \mu B(z)$ the transverse kinetic energy of the electron in the GCS frame, and $\mu$ the orbital magnetic moment of the electron. In terms of $B_0$, the initial field magnitude in the uniform region, and the initial cyclotron radius, $\rho_0 = \rho_c(z=0)$,
\begin{align}
    \rho_c(z)  & = \rho_0 \sqrt{B_0/B(z)} \\
                & = \rho_0 \sqrt{e^{z/\lambda}} 
\end{align}
where in the last equality we use the approximation $B(z) = B_0 e^{-z/\lambda}$. For an electron with initial $T_\perp$ = 18.6\,keV and $B_0=1$\,T, the initial radius is $\rho_0 \approx$ 0.45\,mm. With $\lambda = 6$\,cm this becomes approximately 1.25\,cm at $z = 40$\,cm, corresponding to a final kinetic energy of $\approx 20$\,eV or three orders of magnitude reduction. To accommodate a final radius of 1.25\,cm we choose $y_0 = 1.5$\,cm, which yields an aspect ratio $y_0/x_0 = 0.3$. At ratios larger than this, the field uniformity inside the filter degrades rapidly.

The ability to accommodate the growth of the cyclotron radius while maintaining field-uniformity for drift balancing is the biggest challenge against further reduction in $T_\perp$. One possibility to extend filter performance is to introduce a more elaborate filter geometry in which the aspect ratio $y_0/x_0$ is kept small while the overall dimensions increase with the radius. In general however, the additional field gradients that come with an expanding or varying geometry make this procedure complex. A more efficient way to increase filter performance is to leave the geometry intact and increase the initial $B$.

The filter power increases as $B^2$ in that a factor of three increase in $B$ results in a factor of nine decrease in the final kinetic energy. The cyclotron radius goes as $1/B$ while the energy, for a fixed radius, goes as $\rho_c^2$. If the filter dimensions are unchanged and the starting field is increased to 3\,T, the same radius 1.25\,cm is achieved at $z \approx 53$\,cm, corresponding to a final kinetic energy of $\approx 2$\,eV or four orders of magnitude reduction. As shown in Figure~\ref{fig:sc_concept}, the same principle of using iron extensions to redirect flux can be implemented with superconducting coils in place of an iron core to produce the necessary field.
\subsection{Voltage setting and optimization}

The setting of the potential in an idealized, infinite-plane scenario was solved earlier and in~\cite{betti2019design}; a net $y$-drift of zero along the central line is achieved if the potential along the central line ($x$ = $y$ = 0) satisfies
\begin{align}
    \phi(z)|_{x,y=0} & = \phi_0 - \frac{\mu B_0}{|q|}\left( 1 - e^{-z/\lambda} \right) \\
    & = \phi_0 - T_\perp^0 \left( 1 - e^{-z/\lambda} \right)
\label{eq:centralphi}
\end{align}
where $\phi_0$ and $B_0$ are the initial potential and magnetic field magnitude at $z$ = 0, $T_\perp^0$ the initial transverse kinetic energy in eV, and $\mu$ is the orbital magnetic moment of the electron. To turn this into voltages for the filter electrodes, we begin with the simplifying assumption that the potential at $z$ is just the average between the two filter electrode voltages at that $z$, i.e. $\phi(z)|_{x,y=0} = [V_{y_+}(z) + V_{y_-}(z)]/2$, where $V_{y_+}(z)$, $V_{y_-}(z)$ are the voltages on the positive- and negative-$y$ side electrodes. The accuracy of this approximation increases as the aspect ratio $y_0/x_0$ decreases.

All of the negative-$y$ electrodes are set at a constant voltage which determines the total kinetic energy drained and only the positive-$y$ electrode voltages vary along $z$,
\begin{align}
    V_{y_-}(z) & = \phi_0 - T_\perp^0 \\
    V_{y_+}(z) & = \phi_0 + T_\perp^0 \left( 2e^{-z/\lambda} -1 \right)
\label{eq:filtervoltages}
\end{align}
This is a schematic equation for the filter electrode voltages; as noted in the previous section, the $e^{-z/\lambda}$ term is substituted for by the normalized sampled $B_x$-component along the center line when the filter voltages are actually set.

\begin{figure}[h!]
\centering
\includegraphics[width=1\textwidth]{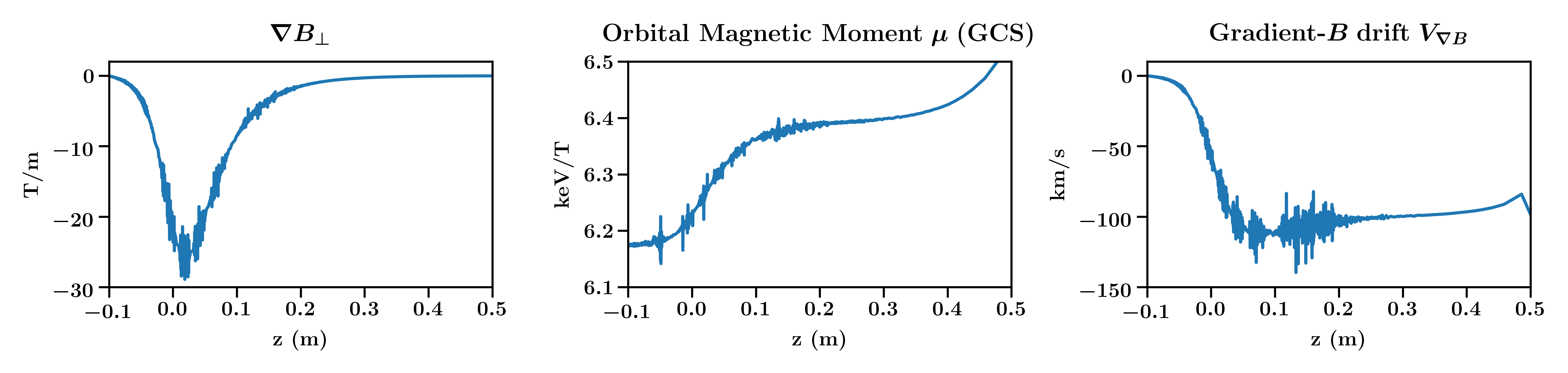}
\caption{$\nabla B_\perp$, $\mu$, and $V_{\nabla B}$ for a pitch 90$^\circ$ electron with initial transverse kinetic energy 18.6\,keV in a 3\,T initial magnetic field.
}
\label{fig:gradB_mu_VgB}
\end{figure}

\begin{figure}[h!]
\centering
\includegraphics[width=1\textwidth]{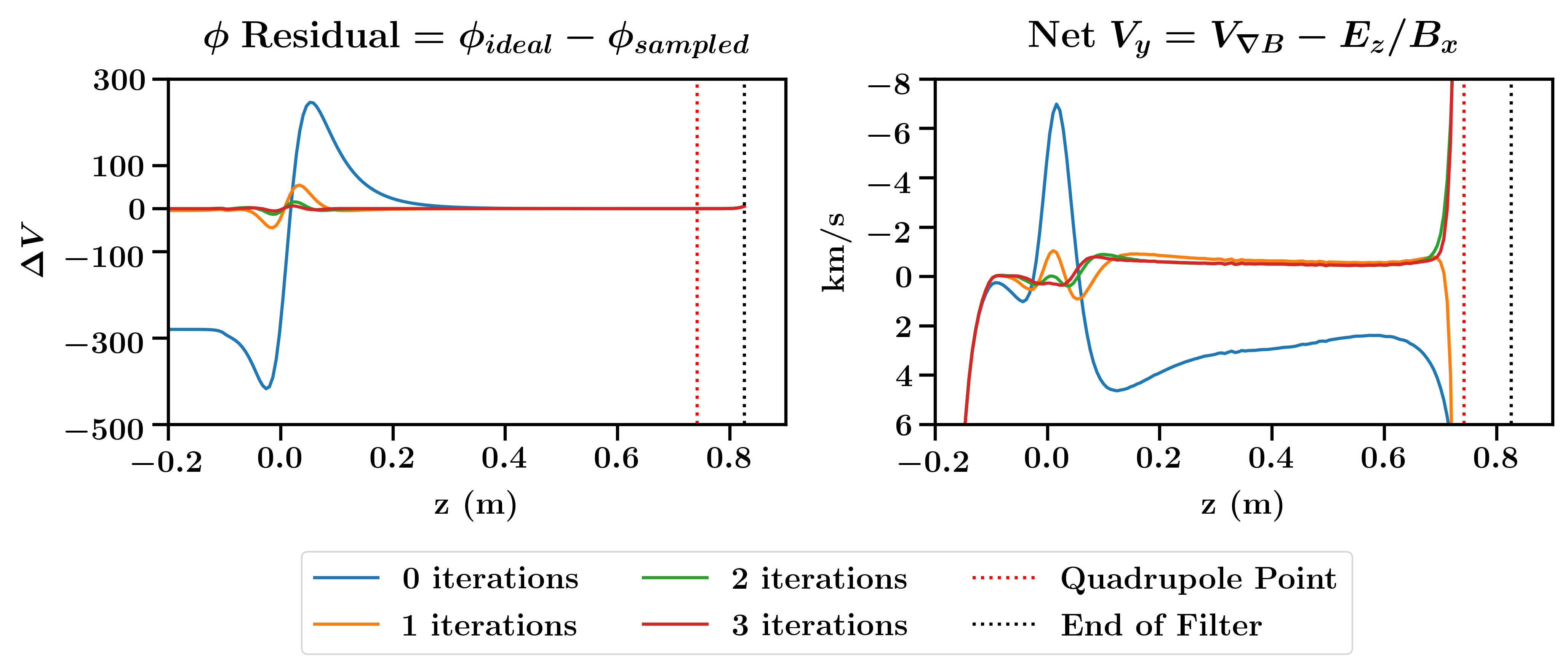}
\caption{$\Delta \phi$ and net GCS $y$-velocity along the center line for several rounds of iteration starting with a 3\,T initial field. The quadrupole point of the magnetic field and the end of the filter electrodes are indicated with dotted lines.
}
\label{fig:phidiff_netvy}
\end{figure}

Unlike the analytical conditions used in~\cite{betti2019design}, the non-zero aspect ratio of the filter and the introduction of a transition region from uniform to decaying field require corrections to the above voltages until precise drift balancing is achieved. Drift balancing can be calculated explicitly with the precision magnetic field map. The gradient-$B$ drift is nominally
\begin{equation}
    \bm{V}_{\nabla B}(z)|_{x,y=0} = - \frac{\bm{\mu} \times \bm{\nabla_\perp B(z)}}{q B(z)}
\end{equation}
where $\mu$ is taken to be adiabatically invariant in areas of low magnetic field gradient. In reality, the transition from uniform to decaying field introduces a region of high gradient at the beginning of the filter and $\mu$ is increased within the level of a few percent as shown in Figure~\ref{fig:gradB_mu_VgB}, leading to a corresponding change in gradient-$B$ drift.  Along the center line, the $B_y$ and $B_z$ components are nearly zero and therefore the magnitude $B(z) \approx B_x(z)$, and the transverse gradient $\nabla_\perp B(z) \approx d B_x/dz$, leading to
\begin{equation}
    \bm{V}_{\nabla B}(z)|_{x,y=0} = - \frac{\mu}{qB_x} \frac{d B_x}{dz} {\bm{\hat y}}
\label{eq:gradBdrift}
\end{equation}
The $y$-component of $\bm{E} \times \bm{B}$ drift that counteracts the gradient-$B$ drift is
\begin{equation}
    \bm{V}_{E \times B}^{y}(z)|_{x,y=0} = \frac{\bm{E} \times \bm{B}}{B_x^2} = \frac{E_z B_x {\bm{\hat y}}}{B_x^2} = \frac{E_z}{B_x}{\bm{\hat y}}
\label{eq:ezbxdrift}
\end{equation}
The sum of the two drifts should be zero; this is the drift balancing condition and yields an expression for $E_z$ that leads to the potential~\eqref{eq:centralphi}. Electrons can veer off the mid-plane trajectory if there is a source of vertical drift imbalance that is not accounted for in the setting of the electrode potentials, including from trajectories that begin off-center, i.e. $y \neq 0$. If the voltages~\eqref{eq:filtervoltages} are used as-is in the filter geometry presented in the previous section, the actual net-drift that results is non-zero (Figure~\ref{fig:phidiff_netvy}) owing to the unaccounted-for field transitions at the entrance and exit of the filter.

To correct for these field changes we use a simple iterative method in which the residual between the observed potential and the idealized potential~\eqref{eq:centralphi} along the center line is added as a correction term to the voltages~\eqref{eq:filtervoltages}.  The adjustment is made only to the positive-$y$ voltages. This procedure is repeated until a desired level of convergence with~\eqref{eq:centralphi} or desired level of filter performance is achieved. Explicitly, the voltages for the $i$th iteration, $V_{y_+}[i]$, are set by
\begin{equation}
    V_{y_+}[i] = V_{y_+}[i-1] + 2 \left( \phi_{\text{ideal}} - \phi[i-1]\right)
\label{eq:correctioneq}
\end{equation}
where $V_{y_+}[i-1]$ and $\phi[i-1]$ are the voltages and potential from the previous iteration, and $\phi_{\text{ideal}}$ is the solution~\eqref{eq:centralphi}.  A factor of two is attached to the residual here to reduce the number of iterations.

\begin{figure}[h!]
\centering
\includegraphics[width=1\textwidth]{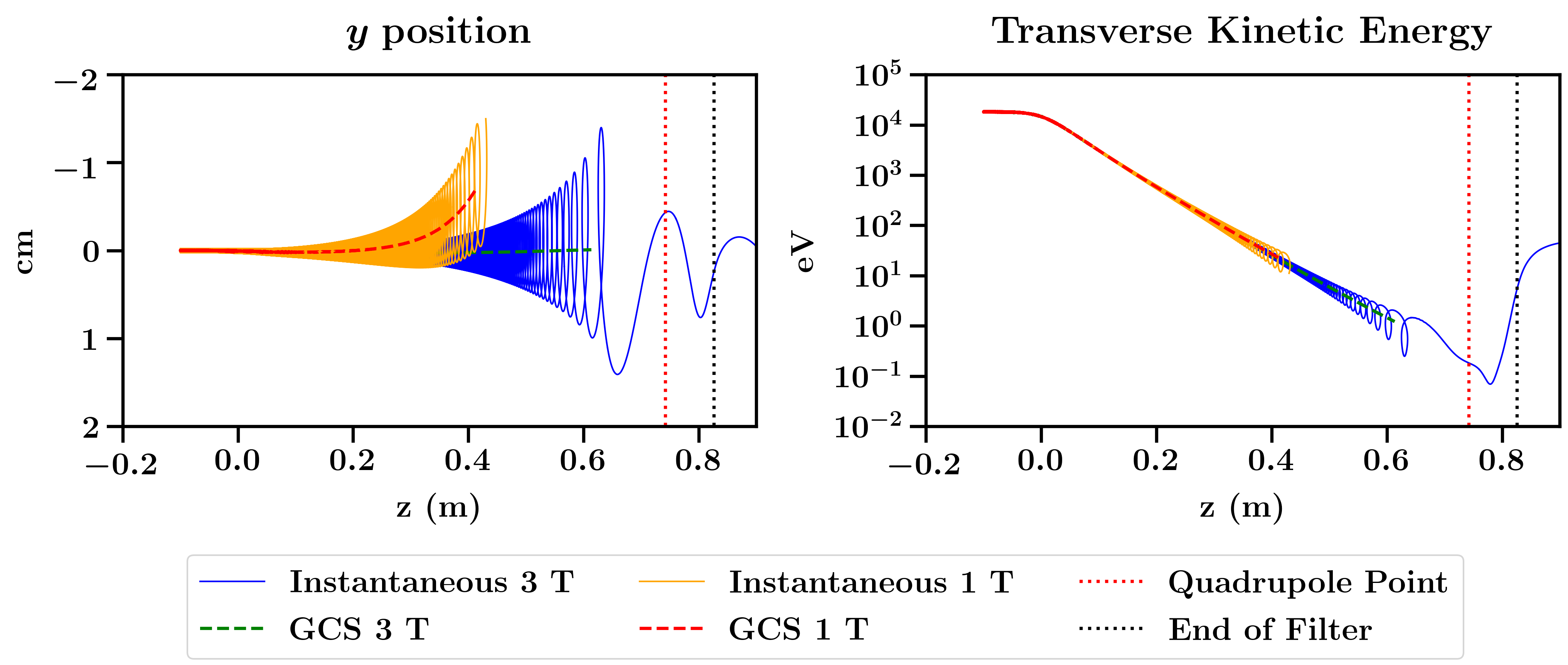}
\caption{Difference in filter performance for 1\,T vs. 3\,T starting magnetic field. Instantaneous and GCS-averaged values are shown. The initial transverse kinetic energy of the electron is 18.6\,keV. The final GCS transverse kinetic energy is 1.2\,eV for 3\,T and 9.3\,eV for 1\,T. 
The growth of the cyclotron radius of the electron as $B$ decreases puts a ceiling on filter performance for a given $y_0$. The GCS trajectories are calculated from the instantaneous trajectories by averaging values over one cyclotron orbit. The beginning and end of a single cyclotron orbit is defined by intervals in which the instantaneous $y$ and $z$ velocities of the electron change sign twice in an alternating fashion, indicating circular motion. 
}
\label{fig:y_tke}
\end{figure}

For dimensions $y_0/x_0 =1.5/5$\,cm and a starting field of 3\,T, the difference in $\phi$, i.e. the residual, compared to the ideal and the net $y$-velocity along the center line for several rounds of iteration are shown in Figure~\ref{fig:phidiff_netvy}. It is favorable to minimize the potential difference in the transition region $z = 0$ to keep the electron from falling off the center line early on; the net $y$-drift is not in practice exactly zero but is nearly constant for the majority of the filter. Residual net $y$-displacements accumulated over the length of the filter from a non-zero vertical drift imbalance can be offset by adjusting the nominal starting $y$-position of the electron entering the filter, as long as the magnitude of the $y$-displacement off the center-line is small compared to the filter dimensions. Figure~\ref{fig:y_tke} shows the transverse kinetic energy drain of the electron for starting magnetic field values of 1\,T and 3\,T, with final GCS kinetic energies of \textless 10\,eV and \textless 1\,eV respectively.

\subsection{Transverse filter speed}

The speed at which the total kinetic energy of the electron is drained by the transverse drift filter depends on the $z$-component of the $\bm{E} \times \bm{B}$ drift, which is in the positive $z$-direction,
\begin{equation}
\label{eq:vzexb}
\left. V^z_{E \times B}\right|_{x,y=0} = -(1/B) \left. \bm{\hat{y} \cdot \nabla} V(y, z) \right|_{x,y=0} = \frac{E_y}{B_x} \ \bm{\hat{z}} .
\end{equation}
It is important that the magnitude of the $z$ drift, which diverges as $1/B$, not exceed the instantaneous transverse velocity of the electron, $|v^{*}_\perp|$, until reaching the end of the filter. The GCS approximation assumes that the transverse drift is a fraction of the cyclotron velocity, giving prolate-shaped cycloid motion in the plane of the cyclotron motion. The transition to curtate-shaped cycloid motion occurs when the transverse drift velocity overtakes $|v^{*}_\perp|$, corresponding to an unraveling of the cyclotron motion in the filter frame of reference.

\begin{figure}[h!]
    \centering
    \includegraphics[width=0.6\textwidth]{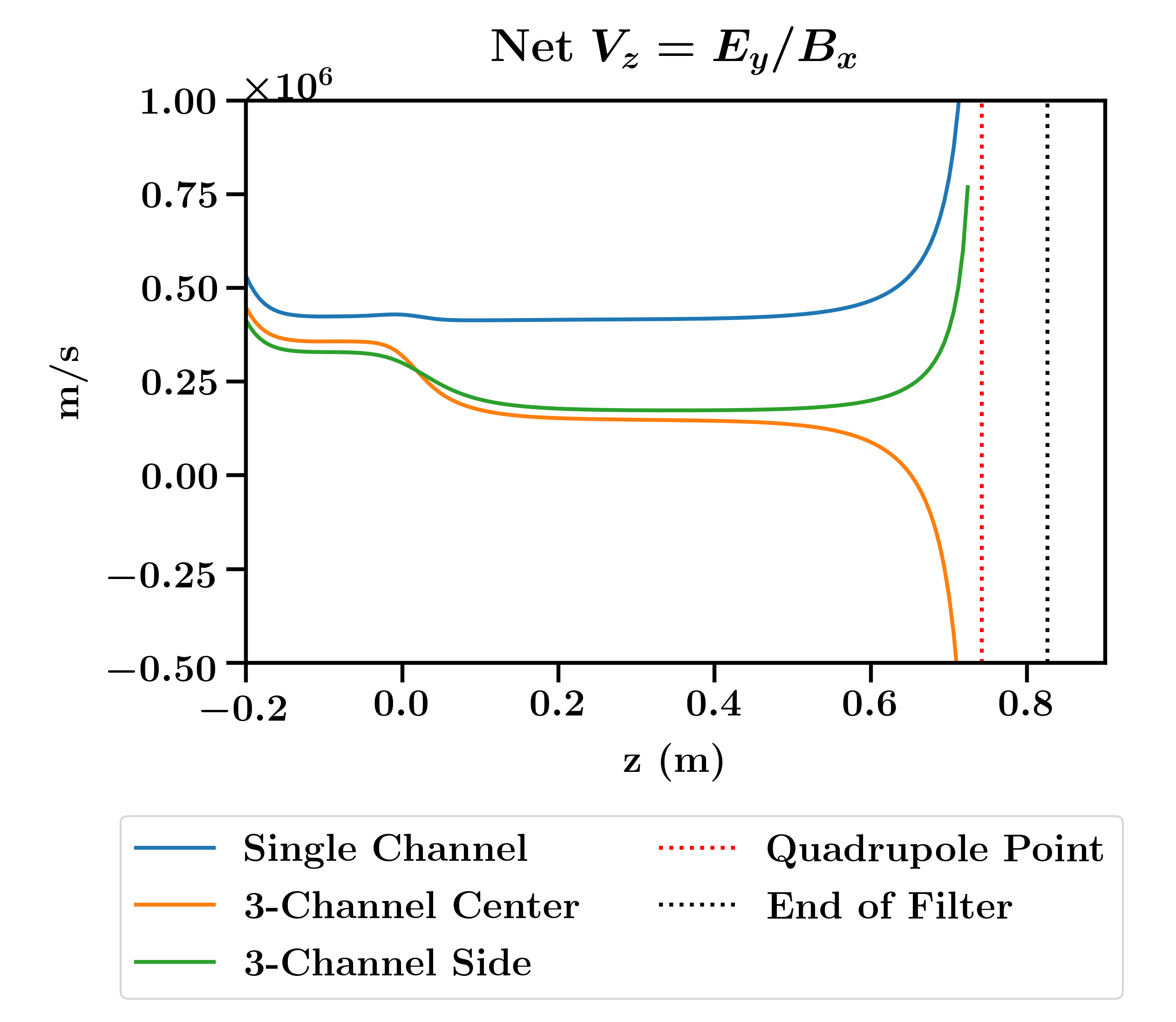}
    \caption{Net $z$-velocity along the center line in the GCS frame. The asymptotic behavior is not in general physically realized as the GCS frame breaks down before this point. Nonetheless, the increase in $z$-velocity as $B$ approaches zero can be delayed by reducing the $E_y$ component of the field at the end of the filter.
    }
    \label{fig:netvz}
    \end{figure}

\begin{figure}[h!]
\centering
\includegraphics[width=1\textwidth]{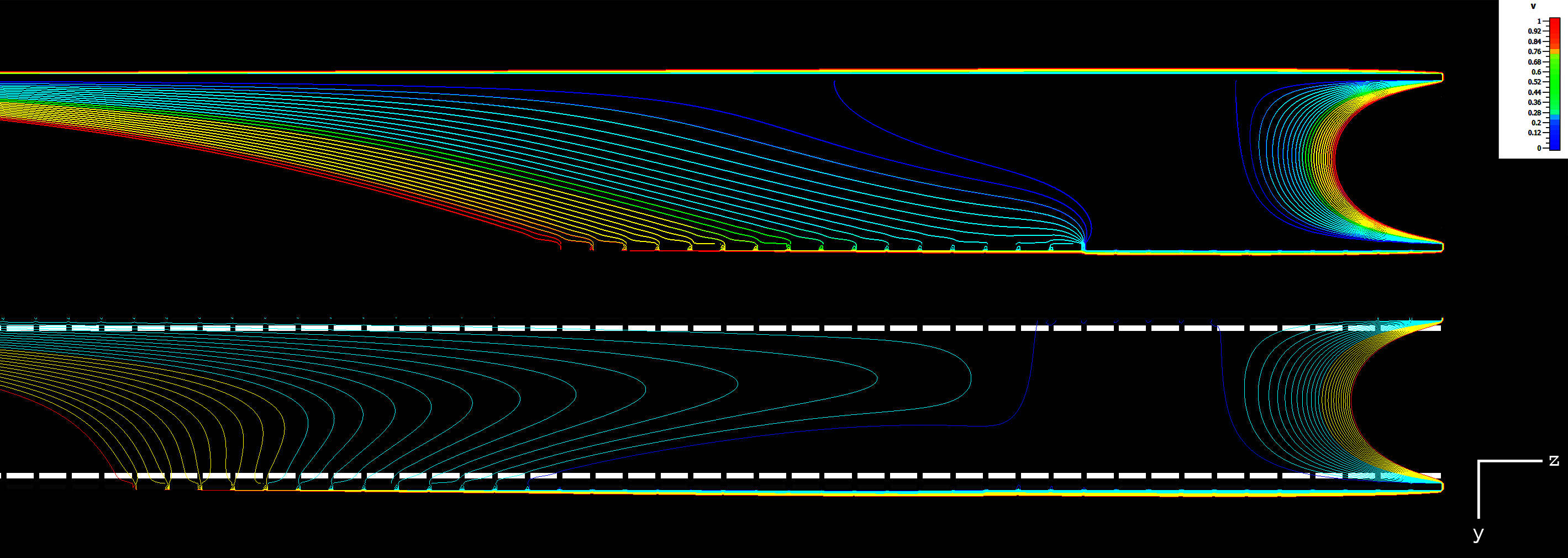}
\caption{Equipotential lines at the end of the filter in the plane $x=0$ for the single-channel design (top) and the three-channel design (bottom). In the three-channel design, the voltage settings of the side potential wells, used to drain the parallel kinetic energy, naturally form a saddle point in the center channel with a local minimum along $z$, driving the $E_y$ to zero.
}
\label{fig:eof_isolines}
\end{figure}

The divergence in $z$-velocity as $B$ decreases to zero is mitigated if the $E_y$ component of the field goes to zero faster than $B_x$ does; one way to achieve this is to form a saddle point in the potential just before the quadrupole point, with the local maximum along $x$ and the local minimum along $z$. The filter electrode voltages in this region can easily be manipulated to produce the saddle point; it also arises naturally in a three-channel filter geometry used to drain the parallel kinetic energy of the electron alongside the transverse, as shown in Figures~\ref{fig:netvz} and~\ref{fig:eof_isolines} and described in the proceeding sections.
   
\subsection{Kinetic energy parallel to the magnetic field}

The transverse drift filter was invented with the primary goal of draining the transverse kinetic energy of a charged particle using gradient-$B$ drift, and in the previous section the filter electrode voltages were calibrated under the assumption that all of the kinetic energy is transverse to the magnetic field, i.e. pitch 90$^\circ$. In practice an electron will also have a parallel component. In this section we describe a scheme in which significant amounts of longitudinal momentum can be reduced with the transverse in lockstep using a 3-channel filter design, with the voltages set according to an estimate of the initial pitch angle at the entrance of the filter, assumed here to come from the RF module. In reality it is more practical to drain as much parallel momentum as possible before the electron enters the filter, then drain the residual parallel momentum within the filter using the original 1-channel scheme, which, due to the fringe field effects from the bounce potentials, naturally exhibits the parallel-momentum draining scheme presented here.

An electron with non-zero parallel kinetic energy inside the filter will undergo periodic `bouncing' motion in $x$; the bounce electrodes reflect electrons back to the center of the filter with a projective $\bm{E} \cdot \bm{B}$ term, which in~\cite{betti2019design} was implemented as a harmonic potential.

In general the motion of an electron inside the filter is continuous cyclotron motion with forward drift in $z$ accompanied by bouncing motion in $x$. If the parallel component becomes the dominant part of the total kinetic energy, two important effects begin to manifest. The first is the non-adiabatic nature of the trajectory where less than a single cyclotron orbit is completed before the electron has completed a single bounce, i.e. traversed the full width of the filter. The second is the transverse drift known as curvature drift.

Curvature drift originates from the centripetal forces that result as a tendency of the cyclotron motion of charged particles to follow magnetic field lines. For an exponentially falling $B$ field, the characteristic length $\lambda$ and the radius of curvature are equal, $R_c = \lambda$, and
\begin{equation}
\bm{\nabla_\perp} B =  - \frac{B}{R_c} \bm{\hat{n}}
\end{equation}
where $\bm{\hat{n}}$ is the unit vector normal to the magnetic field line curvature. In vacuum, the combined gradient-$B$ and curvature drifts are given by
\begin{equation}
\label{eq:gradbcurv}
\bm{V}_{\nabla B-C} =  \frac{1}{2}m(v_\perp^2 + 2 v_\parallel^2)  \frac{\bm{B} \times \bm{\nabla_\perp}B} {qB^3} 
 = (T_\perp + 2 T_\parallel) \frac{\bm{B} \times \bm{\nabla_\perp}B}{qB^3} 
\end{equation}
in the non-relativistic approximation~\cite{roederer2014particle}. For an equal amount of kinetic energy, the curvature drift is apparently a factor of two greater than the gradient-$B$ drift. Unlike the gradient-$B$ drift however, which is constant along the filter owing to the first adiabatic invariant $\mu$, the curvature drift depends on the instantaneous value of $v_\parallel^2$, which is therefore rapidly averaged over successive bounces in the filter to have an effective bounced-averaged value $\langle v_\parallel^2 \rangle$. For a filter geometry symmetric in $x$ the maximum value $|v_\parallel^{max}|$ is attained at $x=0$, and for a harmonic bounce potential along the $B$ field direction, the average value of $\langle v_\parallel^2 \rangle = (v_\parallel^{max})^2/2$. Plugging this back into~\eqref{eq:gradbcurv}, the factor of two relative to the gradient-$B$ drift disappears. 

Therefore, the total gradient-$B$ and curvature transverse drift is proportional to the total kinetic energy $T(z)$ of the electron at $x=0$ in the filter and points in the negative $y$-direction,
\begin{equation}
\label{eq:kedrain}
\bm{V}_{\nabla B-C} \sim \left. -\frac{T(z)}{B(z) R_c} \right|_{x=0}
\bm{\hat{y}} \ .
\end{equation}
Additionally, as described in~\cite{roederer2014particle} in the discussion following Eq.~(2.13) on particle inertia, the component of transverse drift along the normal of the $B$ field, i.e. along $z$, introduces an additional contribution to the curvature drift that scales as $|v_\parallel V^z_{E \times B}|$ rather than $v_\parallel^2$.  Therefore, at the end of the filter, the effects of curvature drift can be dramatic if $V^z_{E \times B}$ is a substantial fraction of $|v^{*}_\perp|$. The $V^z_{E \times B}$ drift is mitigated by use of a saddle point, nonetheless it is advantageous to also drain the parallel energy faster relative to the transverse to avoid this runaway drift.

\subsection{Bounce electrodes, field wires and side-well potentials}

The parallel kinetic energy of an electron inside the filter can be drained by modifying the design into a three-channel geometry as follows. Consider an electron undergoing bounce motion in the filter. At the turning points of the motion, the parallel kinetic energy goes to zero while the transverse kinetic energy is largely unchanged. When the electron is reflected back towards the center of the filter by the bounce electrodes, the parallel kinetic energy returns to its maximum value at $x=0$. However, because the electron is also moving forward in $z$ as the bounce occurs, the potential at $x=0$ upon return to the center is not the same as it was when it left; it has increased. In the GCS frame, since this increase in potential is along the direction of the bounce motion, and not in the transverse, the parallel kinetic energy is the one that is drained, in an amount nominally equal to the potential difference at $x=0$ before and after the bounce.

\begin{figure}[h!]
\centering
\includegraphics[width=0.9\textwidth]{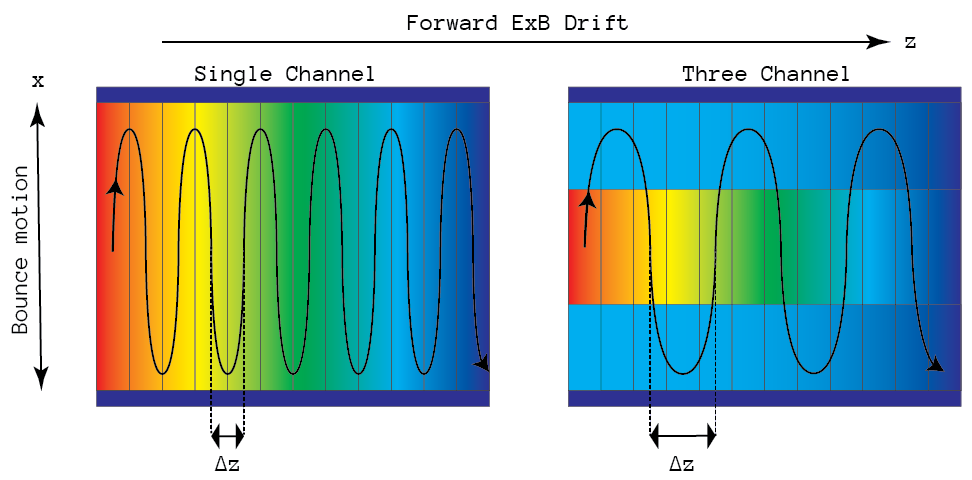}
\caption{Schematic drawing of parallel momentum draining in the single channel vs. three channel filter designs. Color indicates potential, from red (lowest) to blue (highest). Dark blue strips at the top and bottom are bounce electrodes, with filter electrodes drawn as vertical bars along the trajectory. In the three channel design, the side channels are set at a higher voltage than the center, approximately equal to the parallel kinetic energy along the trajectory. The electron lingers in the side well before returning to the center, thus re-entering at a higher potential and draining more parallel energy than it would have in the single channel design.
}
\label{fig:parallel_drain_schematic}
\end{figure}

This effect can be observed in the original single-channel filter, but the parallel drain is small since the forward displacement in $z$ between bounces is small and the potential gradient is not in general oriented in the bounce direction. The amount of parallel drain can be increased if the electron is made to linger in the side region of the filter before returning to the center, increasing the forward displacement in $z$. A schematic drawing of this mechanism is shown in Figure~\ref{fig:parallel_drain_schematic}.

Since the total kinetic energy of the electron is assumed to be at the tritium endpoint, the pitch angle gives an estimate of the initial transverse/parallel energy split necessary to configure the three-channel design. The interior of the filter is split into three discrete potential wells; we call them the center- and side-wells. To the center potential, previously configured only for the transverse kinetic energy drain, is now added the parallel draining term. The center potential at the end of the filter remains the total energy drained. The side potential at $z$ is increased (decreased, as it were, for an electron) by an amount equal to the parallel kinetic energy of the electron at $z$. Since the potential step between side-center-side is in the direction of parallel motion, when the electron enters the side-well from the center its parallel kinetic energy is reduced to nearly zero for an extended period before bouncing back, compared to the single-well design where the bounce is nearly instantaneous, thus causing it to linger there while undergoing transverse drain forward in $z$, before eventually bouncing back to the center as before, achieving the desired mechanism.
\begin{figure}[h!]
\centering
\includegraphics[width=0.9\textwidth]{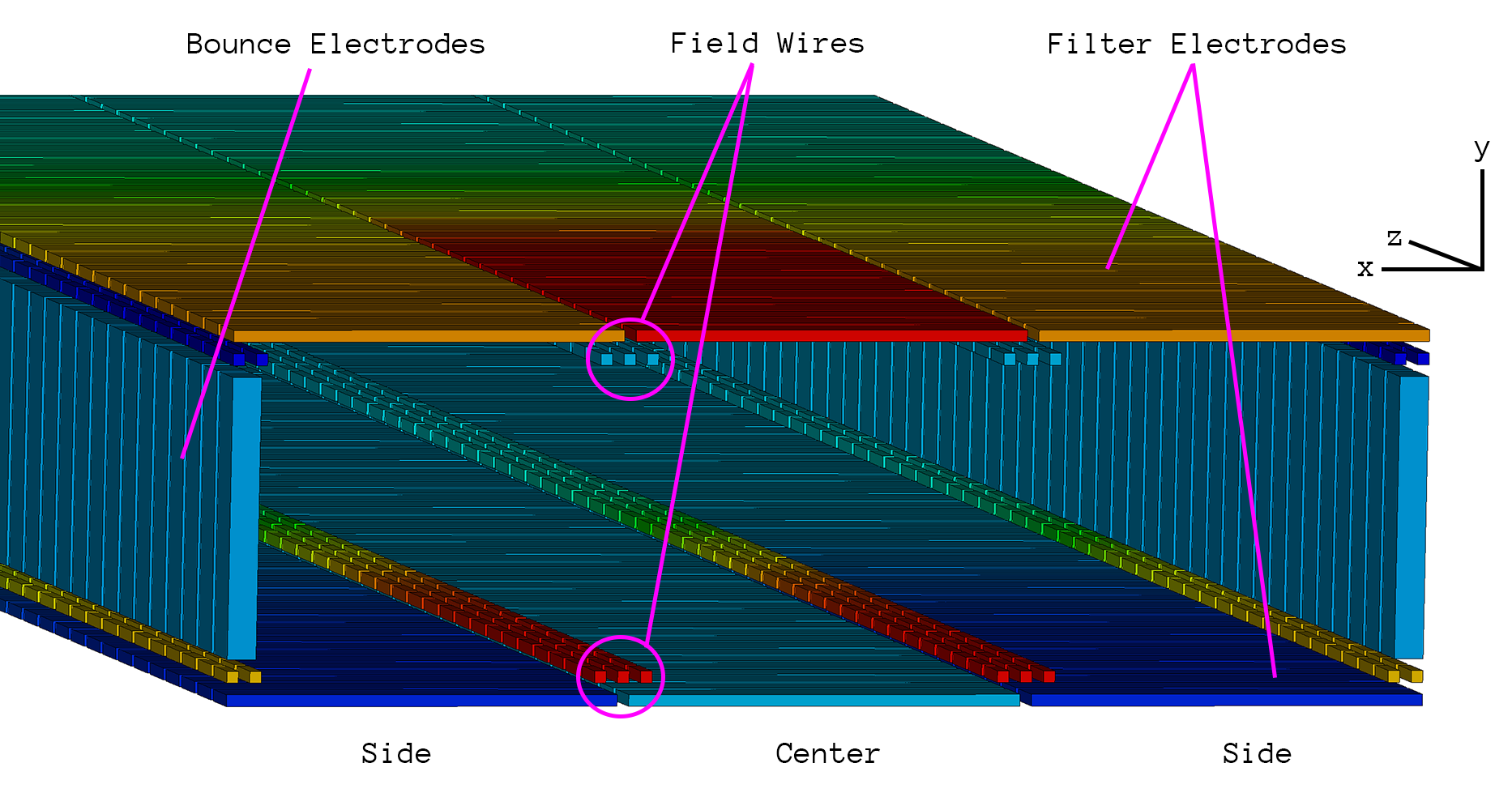}
\caption{Transverse perspective view of the three-channel filter geometry.
}
\label{fig:transview_pers}
\end{figure}
Explicitly, with $T_\perp^0$ and $T_\parallel^0$ the initial transverse and parallel kinetic energies, respectively, the idealized potentials are now, setting the reference offset $\phi_0 = T_{total}$,
\begin{align}
\label{eq:side_potentials}
    \phi_{cent}(z)|_{y=0}   & =  T_\perp^0\,e^{-z/\lambda_{cent}} + T_\parallel^0\,e^{-z/\lambda_{side}}  \\
    \phi_{side}(z)|_{y=0}   & =  \phi_{cent}(z) -T_\parallel^0\,e^{-z/\lambda_{side}} \nonumber \\
                            & =  T_\perp^0\,e^{-z/\lambda_{cent}} \ ,
\end{align}
where $\lambda_{side}$, which sets the rate of parallel energy drain, is not required to be the same as $\lambda_{cent}$. If $\lambda_{side} = \lambda_{cent}$,
\begin{align}
\label{eq:side_potentials_equal}
    \phi_{cent}(z)|_{y=0}  & =  T_{total}\,e^{-z/\lambda} \\
    \phi_{side}(z)|_{y=0}  & =  T_\perp^0\,e^{-z/\lambda} \ ,
\end{align}
which makes clear that the center potential is the total energy drained and the side potential is just the potential for an electron with the same transverse kinetic energy but zero parallel component, i.e. transverse-only draining. The corresponding electrode voltages are
\begin{align}
\label{eq:3ch_filtervoltages1}
    V_{cent,\,y_-}(z)   & = 0 \\
    V_{cent,\,y_+}(z)   & = 2\,T_{total}\,e^{-z/\lambda} \nonumber \\
                        & = 2\,T_\perp^0\,e^{-z/\lambda_{cent}} + 2\,T_\parallel^0\,e^{-z/\lambda_{side}} \\
    V_{side,\,y_-}(z)   & = 0 -2\,T_\parallel^0\,e^{-z/\lambda_{side}} \\
    V_{side,\,y_+}(z)   & = 2\,T_\perp^0\,e^{-z/\lambda_{cent}} + 2\,T_\parallel^0\,e^{-z/\lambda_{side}} \ .
\label{eq:3ch_filtervoltages2}
\end{align}

In contrast to the center, the addition of the parallel energy term to the side voltages is split across the top and bottom electrodes rather than applied only to the top electrodes. The total potential difference across the top and bottom in the side well is also increased relative to the center in order to increase the $E_y$ magnitude and therefore the forward drift in $z$ to account for the slight backwards motion in $z$ due to the curvature of the field lines.

\begin{figure}[h!]
\centering
\includegraphics[width=0.9\textwidth]{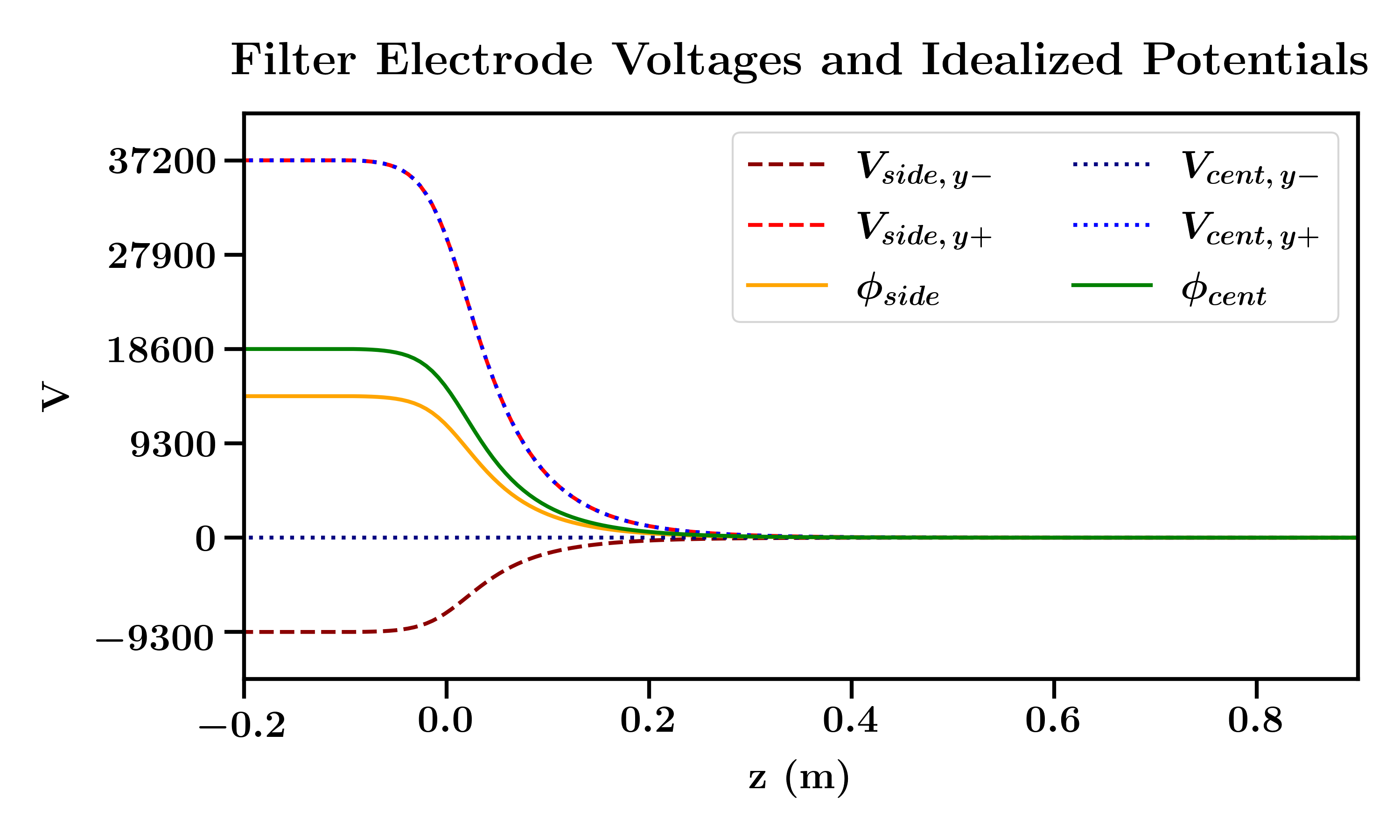}
\caption{Filter electrode voltages and idealized potentials along the center line of the respective wells. Voltages are set for a pitch 60$^\circ$ electron with total kinetic energy 18.6\,keV, with $\lambda_{side} = \lambda_{cent}$.
}
\label{fig:3ch_voltages_idealized_potentials}
\end{figure}

Given~\eqref{eq:3ch_filtervoltages1}-\eqref{eq:3ch_filtervoltages2}, the voltages are iterated as before, with the correction term also split across top and bottom for the side voltages.
\begin{align}
    V_{cent,\,y_-}[i] & = 0 \\
    V_{cent,\,y_+}[i] & = V_{cent,\,y_+}[i-1] + 2 \left( \phi_{ideal,\,cent} - \phi_{cent}[i-1]\right) \\
    V_{side,\,y_\pm}[i] & = V_{side,\,y_\pm}[i-1] + \left( \phi_{ideal,\,side} - \phi_{side}[i-1]\right)
\label{eq:3ch_correction}
\end{align}

\begin{figure}[h!]
\centering
\includegraphics[width=0.9\textwidth]{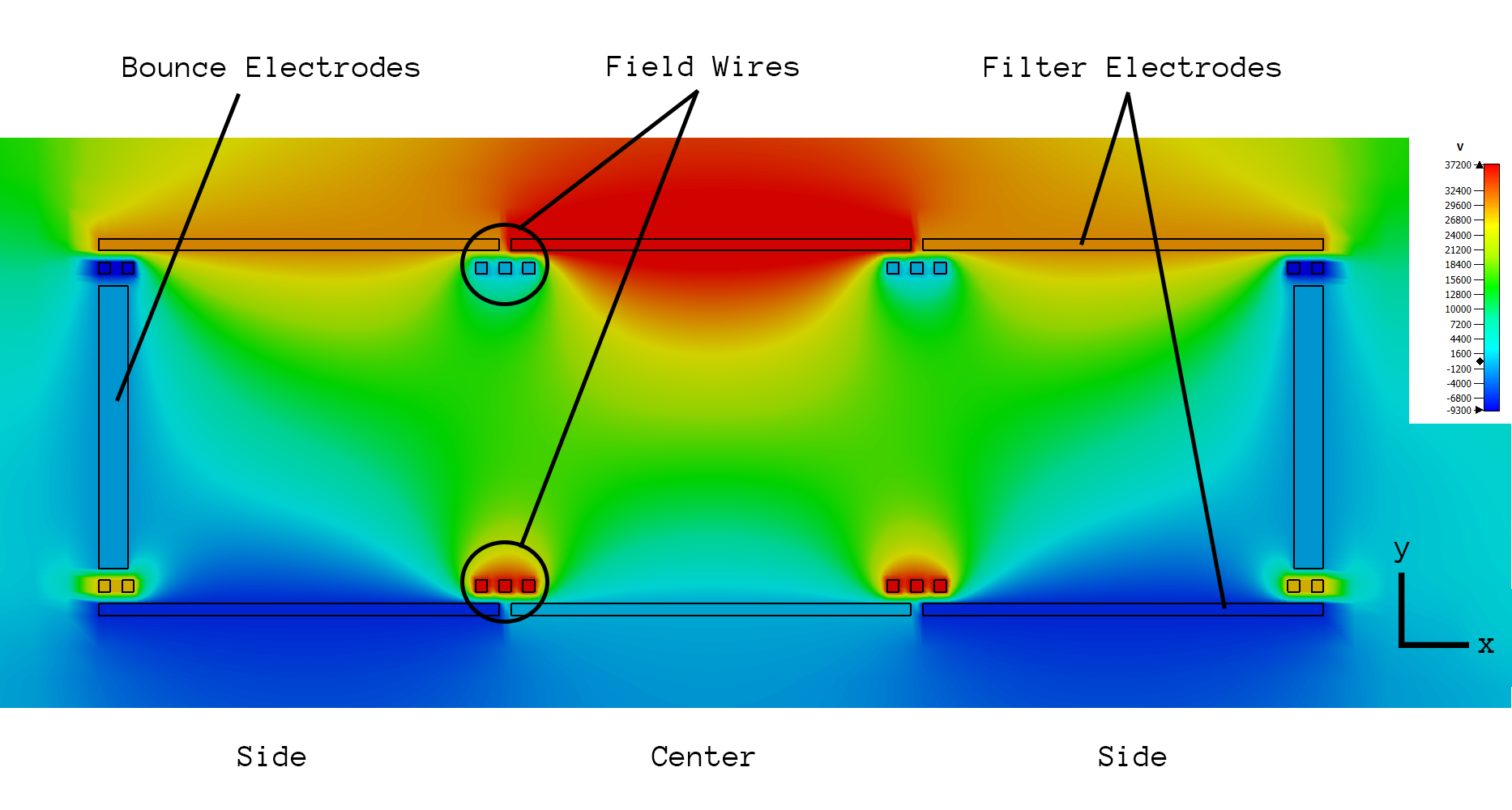}
\includegraphics[width=0.9\textwidth]{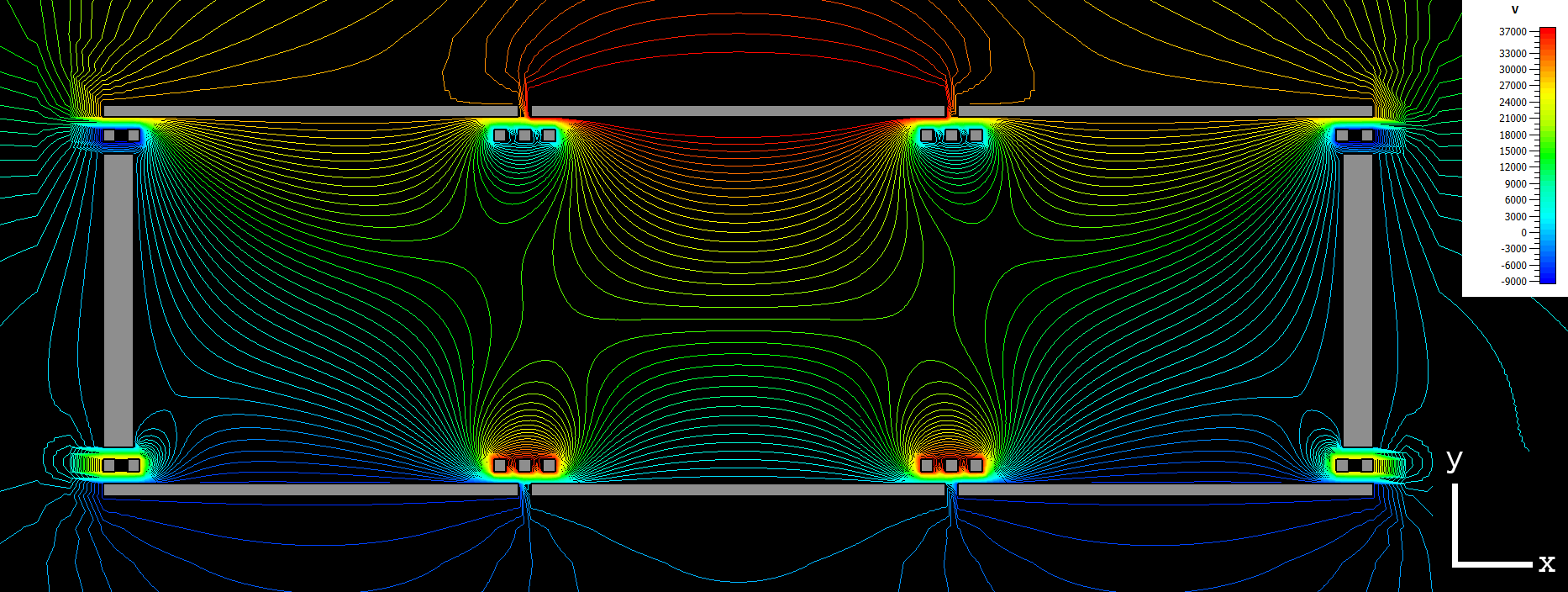}
\caption{Transverse cross-section of the three-channel filter geometry in the $xy$-plane at the start of the filter, with potential field overlay. Annotations (top) show the placement of the electrodes and field wires. The field wire voltages mirror the opposite side filter electrode voltages. Equipotential lines (bottom) show the sharpness of the transition between the side and center wells.
}
\label{fig:transview}
\end{figure}

\begin{figure}[h!]
\centering
\includegraphics[width=0.9\textwidth]{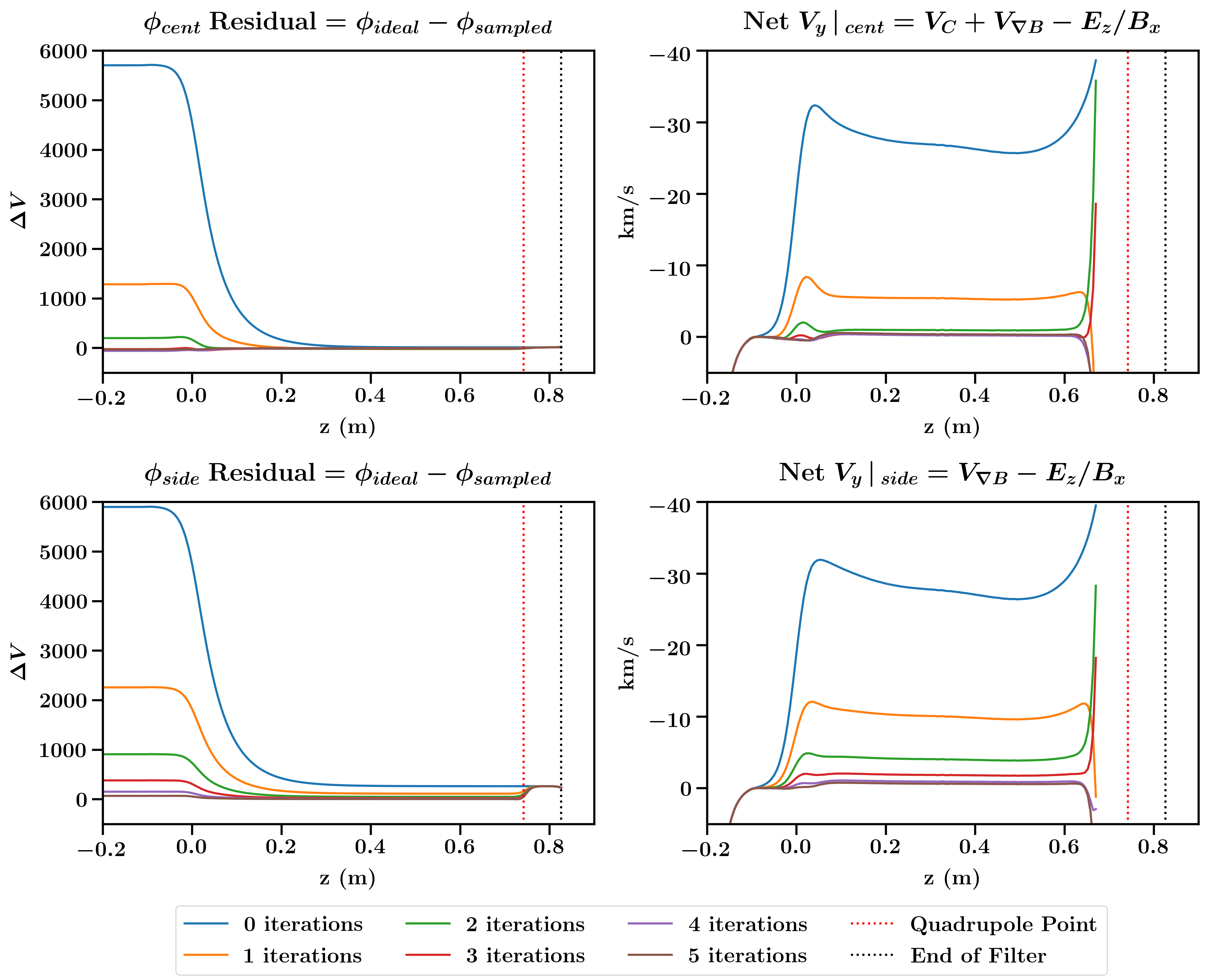}
\caption{Voltage iteration results for the three-channel filter geometry. Top row plots are for the center channel, bottom row plots for the side channels. The curvature drift $V_C$ for the center channel is calculated by assuming the parallel kinetic energy decreases as $T_\parallel^0 \, e^{-z/\lambda_{side}}$ in Eq.~\eqref{eq:gradbcurv}.
}
\label{fig:parallel_drain}
\end{figure}

\begin{figure}[h!]
\centering
\includegraphics[width=0.9\textwidth]{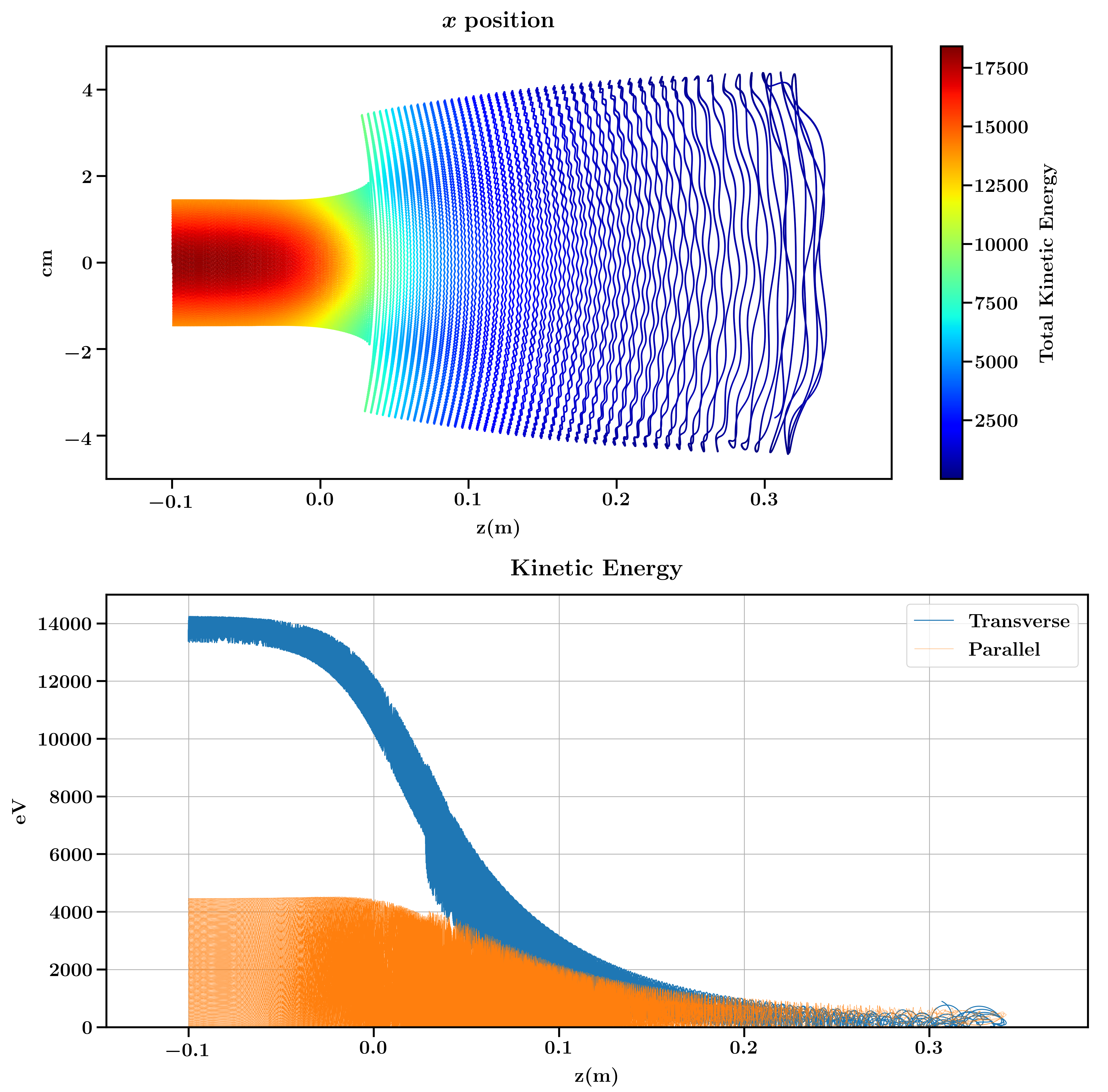}
\caption{$x$-position and kinetic energy drain for a pitch 60$^\circ$ electron in the first 35\,cm of the three-channel filter showing the simultaneous draining of parallel and transverse kinetic energies.
}
\label{fig:p60_x_ke}
\end{figure}

Finally, to increase the sharpness of the transition between the center and side wells, additional field wires are placed along the splits in the filter electrodes. To maintain maximum relative uniformity of the side and center potentials down the entire filter length $z$, both the field wires and bounce electrodes can be segmented in $z$ in the same fashion as the filter electrodes. In practice this level of discrete granularity in the field wires is not necessary along the entire filter length if the aspect ratio is small enough. The field wire voltages mirror the opposite side filter electrode voltages and the bounce electrodes, to ensure continuous bouncing but not so much so that the bounce potential contaminates the side and center potentials, are set with a fixed offset plus a decay term proportional to the initial parallel kinetic energy,
\begin{align}
    V_{wires,\,y_\pm}(z)    & = V_{cent/side,\,y_\mp}(z) \\
    V_{bounce}(z)   & = \phi_0 - V_{fixed} - \alpha T_\parallel^0\,e^{-z/\lambda} \ .
\label{eq:bounce_voltages}
\end{align}
where $\alpha < 1$. A transverse cross-section of the three-channel filter geometry is shown in Figure~\ref{fig:transview}. Figure~\ref{fig:transview_pers} shows a perspective view. For $\lambda_{side} = \lambda_{cent}$, results of the voltage iteration are shown in Figure~\ref{fig:parallel_drain} for a pitch 60$^\circ$ electron. Figure~\ref{fig:p60_x_ke} shows the x-position parallel to the field lines and the simultaneous draining of transverse and parallel kinetic energies with the three-well scheme.

\subsection{Interface to microcalorimeter}

At the end of the filter, the electron kinetic energy has been drained down to a few eV and the magnetic field is around one milliTesla.  To extract the electron for final readout, the cyclotron motion must be unraveled into a linear trajectory leading to the calorimeter. A first scheme to do that is presented here.

To unravel the cyclotron motion, a cylindrical $\mu$-metal housing is interfaced to the end of the filter, roughly centered around the quadrupole point of the magnetic field, inside of which the magnetic component of the Lorentz force law is zero. To direct the resulting trajectory to the calorimeter, inside the $\mu$-metal housing is an Einzel lens, a series of three conducting rings with the two outer rings set at a lower potential than the middle ring. The lens acts as a pure electrostatic focusing device analogous to a convex optical lens, focusing charged particle trajectories instead of light.

The inner radius of the Einzel lens should be at least as wide as the filter dimension $y_0$ to accommodate the full incoming radius of the cyclotron motion, and a short buffer length should be left between the end of the filter electrodes and the start of the the $\mu$-metal housing, meaning that the first Einzel lens is partly outside of the $\mu$-metal housing. This arrangement maintains a small magnetic field inside the first ring where electrons continue to undergo cyclotron motion, and due to the superposition of the filter potential and the Einzel lens potential, a pseudo-stable well is formed where the electric force is oriented back towards the positive-$y$ side filter electrodes, the magnetic gradient and curvature drifts still push the electron toward negative-$y$, and as the $E_y$ component reduces to zero, the forward-$z$ $\bm{E} \times \bm{B}$ drift goes to zero. If the Einzel lens voltages are set correctly to exploit this drift circus, an additional amount of kinetic energy, on order 10-50~eV, depending on the Einzel lens voltage, can be drained with the gradient-$B$ drift grinding the electron against the negative-$y$ side of the first Einzel ring before it passes through the second and third rings.

\begin{figure}[h!]
\centering

\includegraphics[width=1\textwidth, keepaspectratio]{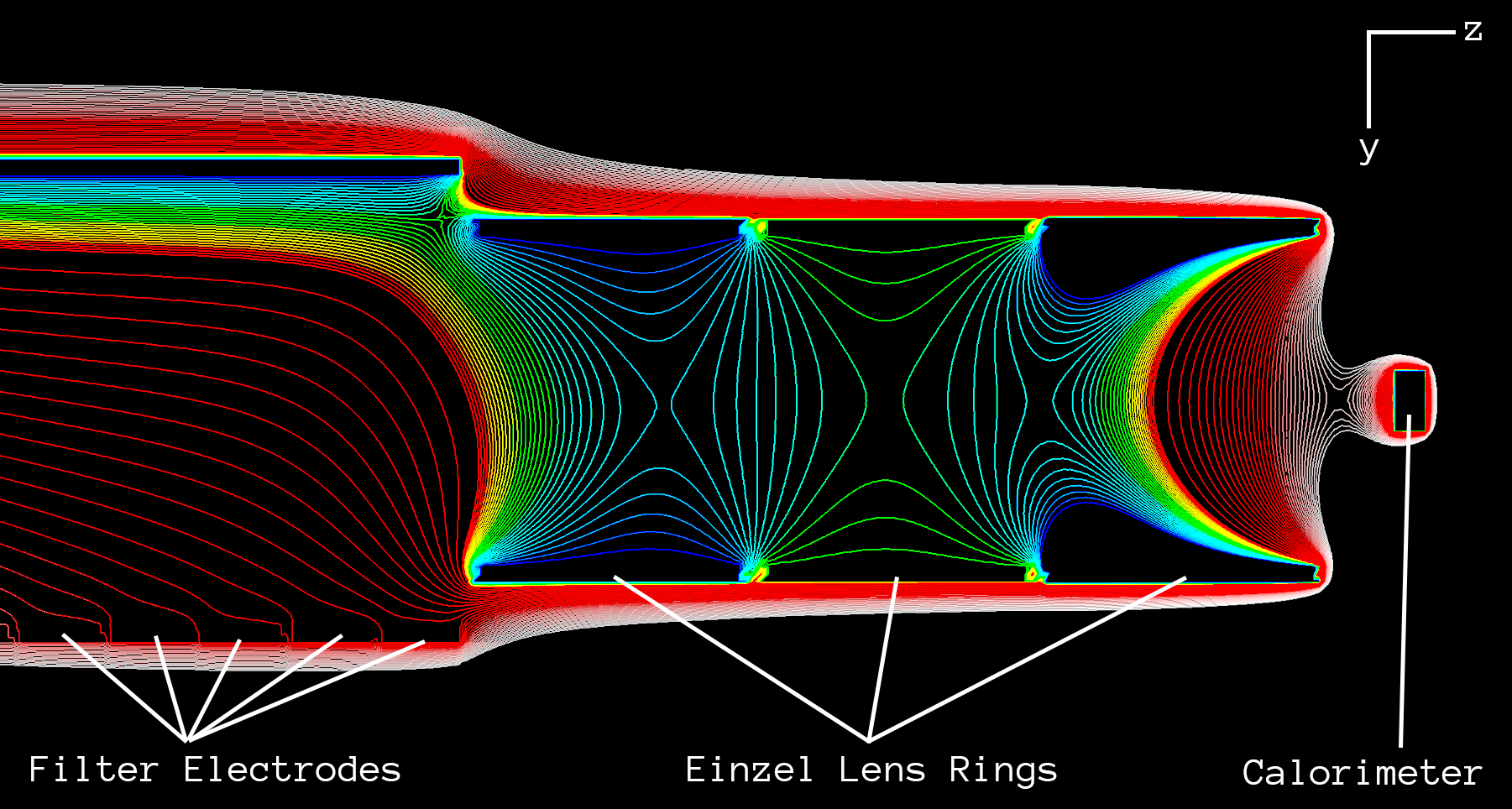}
\par\bigskip

\resizebox{\textwidth}{!}{

\includegraphics[height=6cm, keepaspectratio]{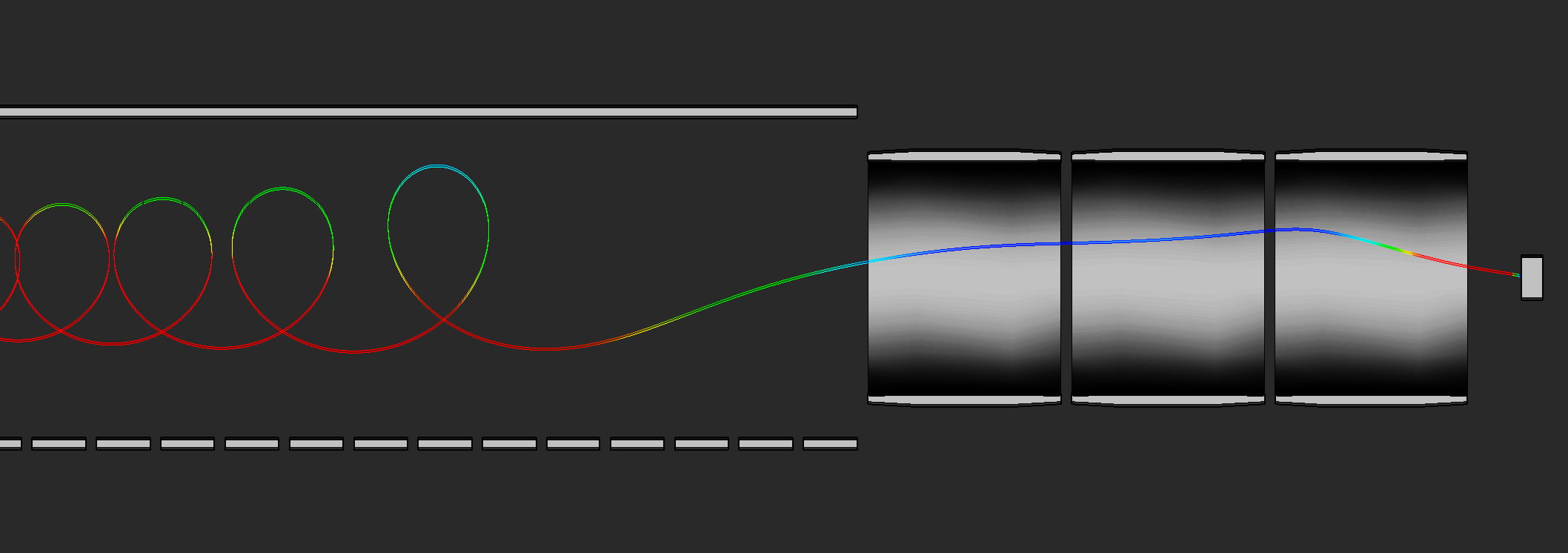}
\hspace{1em}
\includegraphics[height=6cm, keepaspectratio]{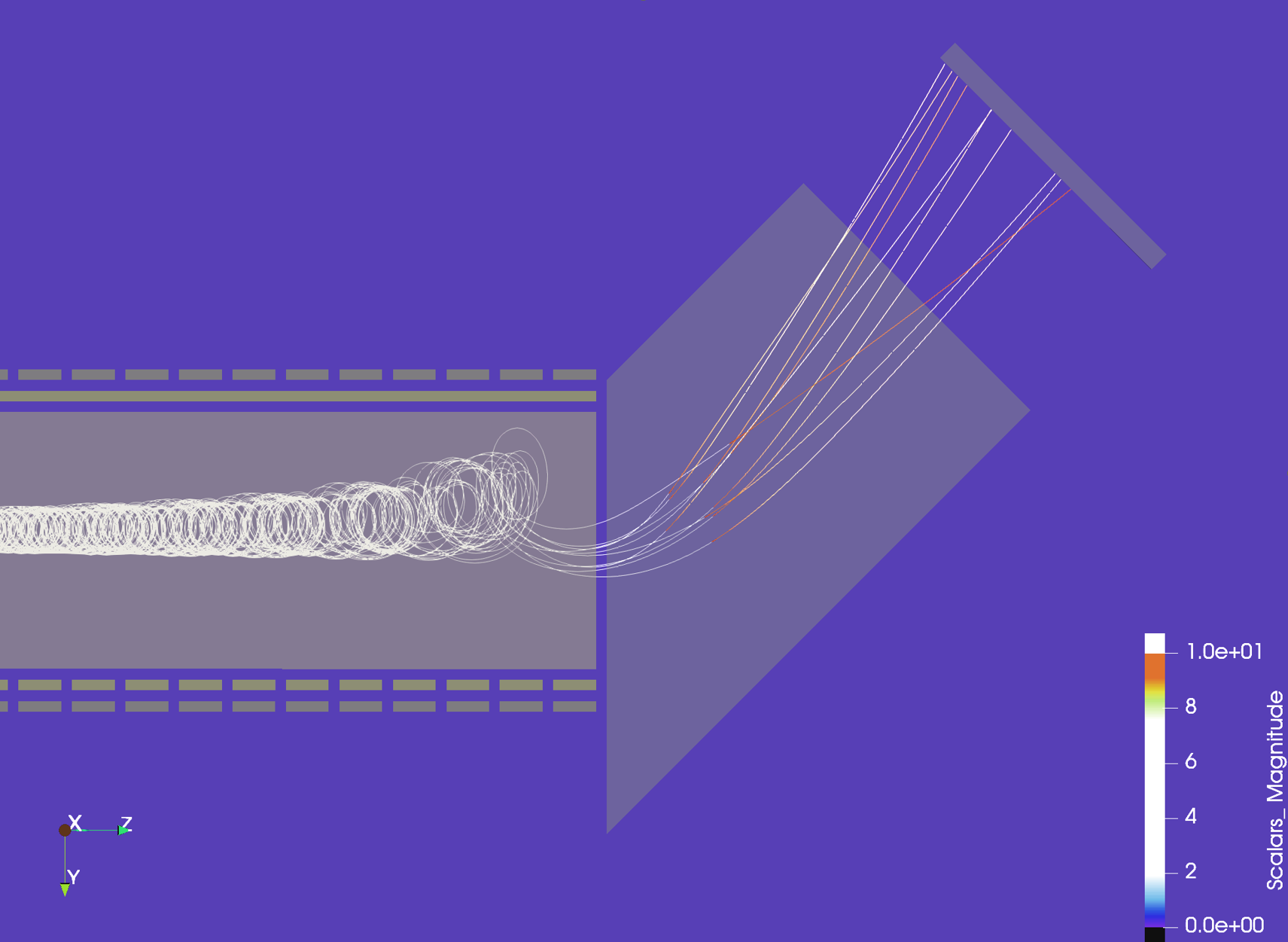}

}
\caption{ (\textit{top}) Equipotential lines in the plane $x=0$ at the end of the filter showing the Einzel lens and microcalorimeter interface. (\textit{bottom left}) Sample trajectory of an electron at the end of the filter. The cyclotron motion is unraveled and focused through the Einzel lens to the calorimeter. The $\mu$-metal housing, not pictured, is concentric with the Einzel lens. (\textit{bottom right}) A tilted single lens configuration showing the softened impact of the incident cyclotron phase angle.
}
\label{fig:einzel_lens}
\end{figure}

The calorimeter is located in the focal plane of the lens on the exit side of the filter. In practice it is favorable to tilt the Einzel lens and calorimeter both off-axis from the center line so as to reduce the line-of-sight IR heating on the calorimeter. As the electron trajectory orbits counterclockwise about the magnetic field, such a tilt would also increase acceptance as the axis of the Einzel lens approaches the tangent of the orbit. The final electron kinetic energy is uncertain at the level of 1-10~eV depending on the filter starting field and RF momentum measurements, and there is a spread in final electron trajectories distributed over the acceptance of the calorimeter. Equipotential lines and a sample trajectory demonstrating the maneuver are shown in Figure~\ref{fig:einzel_lens}.

\section{Magnetic Adiabatic Drift Collimation and Transmission}

As mentioned in the previous section, it is more practical to drain as much momentum longitudinal to the magnetic field as possible before the electron enters the filter. A robust mechanism to achieve this and end-to-end electron transport from the target into the filter and through to the microcalorimeter is described here.

\begin{figure}[h!]
    \centering
    \resizebox{\textwidth}{!}{
    \includegraphics[height=6cm, keepaspectratio]{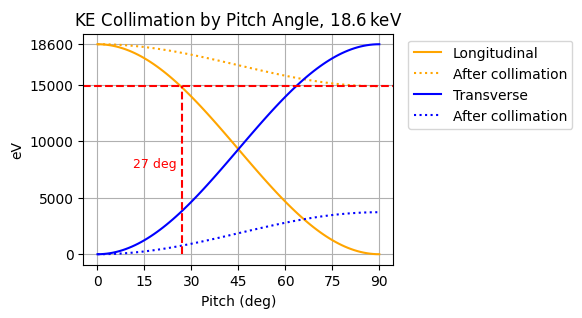}
    \includegraphics[height=6cm, keepaspectratio]{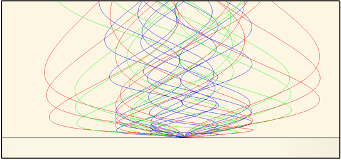}
    }
    \caption{Collimation of transverse to longitudinal kinetic energy with $B_\text{perm} \approx 0.5$ and $B_\text{ext} \approx 0.1$. The effective pitch angle range is compressed by the factor $B_\text{perm}/B_\text{ext}$. (right) Close-up view of simulated endpoint electrons released from a permanent magnet surface exhibiting collimation. }
    \label{fig:pitch_compression}
\end{figure}

The idea is to place the tritium target on the surface of a high-remnance permanent magnet in a low-field fringe region off-axis from the central filter channel. The transition from the high- to low- field region acts as a miniature, one-sided MAC-E filter just after emission of the electron. The transverse momentum of the electron is collimated into longitudinal momentum via adiabatic collimation, and electrodes are instrumented along the path of the electrons, which follow the magnetic field lines, to drain the parallel compoment of the momentum. Once the electron enters the central channel region of the filter, an accelerating drift is applied to clear the target vicinity. The bounce electrodes begin here to contain the electron. Voltages are matched in reverse to the sampled magnetic field to accelerate the electrons into the uniform field region while maintaining a quasi-linear GCS trajectory. This scheme effectively compresses the initial pitch angle range and drains most of the parallel momentum, maximizing static acceptance of the filter for a much larger range of pitch angles than previously described.

\begin{figure}[h!]
    \centering
    \resizebox{\textwidth}{!}{
    \includegraphics[height=6cm, keepaspectratio]{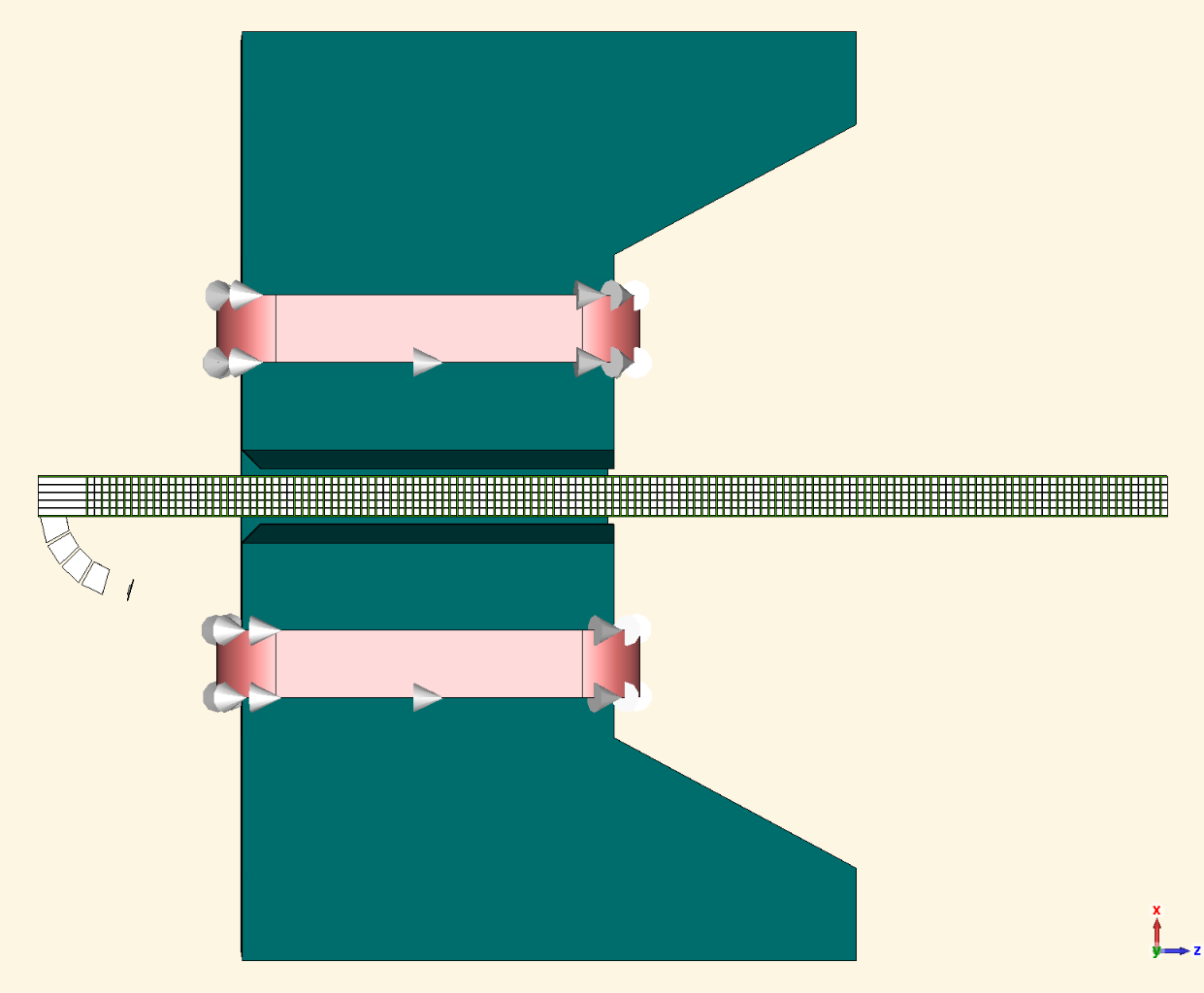}
    \includegraphics[height=6cm, keepaspectratio]{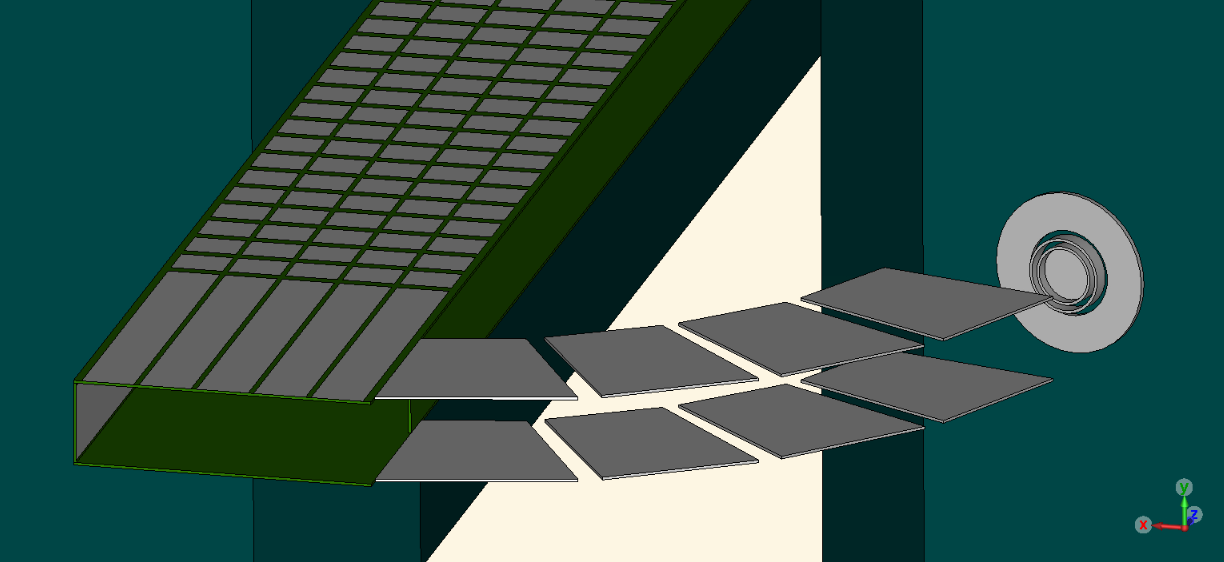}
    }
    \caption{(left) Full conceptual view of the LNGS demonstrator apparatus showing the magnet, coils, main filter channel, and target injection scheme. (right) Perspective view of the target injection region showing the permanent magnet disk, "can" electrodes, and arc-shaped electrodes following the field lines.}
    \label{fig:injection_scheme}
\end{figure}

\subsection{Permanent magnet target}

A small ~1.5 T permanent magnet disk of radius 1\,cm is placed in a low-field region of the demonstrator magnet, significantly off-axis from the central transport channel. The usable target surface area of the disk in this concept study is approximately 0.5\,cm in radius, corresponding to $\approx$ 25\,ng of tritium. The target injection design described here readily allows this mass to be scaled up to the microgram scale.

\begin{figure}[h!]
    \centering
    \includegraphics[scale=0.6]{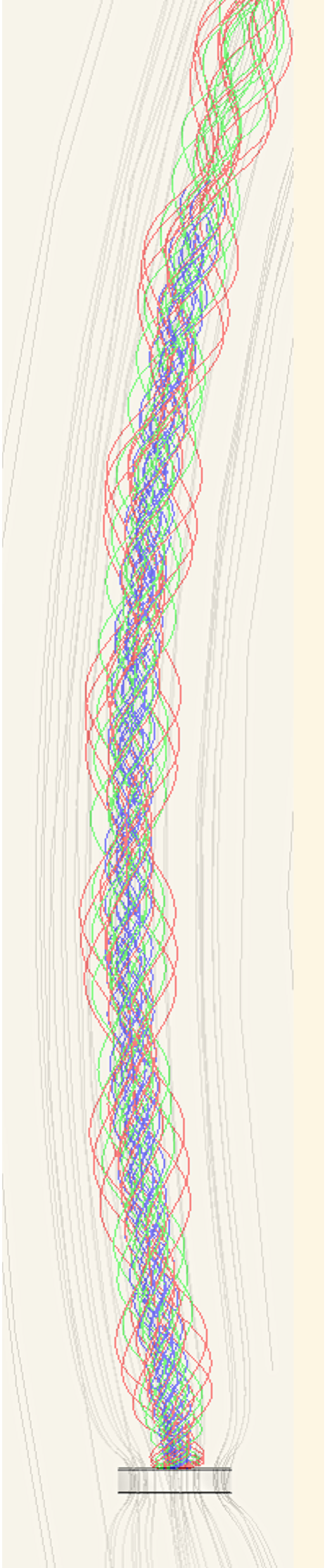}
    \includegraphics[scale=0.6]{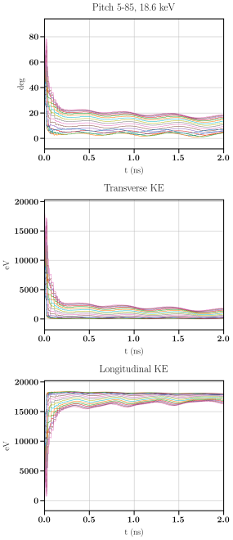}
    \caption{(left) Electron trajectories follow the magnetic field lines after emission from a permanent magnet-mounted substrate. (right) Transverse kinetic energy is collimated into longitudinal, effectively compressing the emitted pitch range. }
    \label{fig:collimation_traj}
\end{figure}

The target substrate is mounted on the surface of the permanent magnet, with the surface of the magnet oriented normal to the magnetic field lines. The principle of magnetic collimation, as used in the MAC-E filter, is used to collimate the transverse momentum of outgoing electrons from the target into longitudinal momentum as they transition from the high-field of the permanent magnet into the low-field region. This collimation compresses the effective pitch angle range of emitted electrons by a factor equal in principle to the ratio of the permanent magnet remnance to the external field strength, i.e. since $\mu = \frac{T_\perp}{B}$, it follows that $\theta_\text{coll} = \theta_\text{orig} \cdot \frac{B_\text{ext}}{B_\text{perm}}$. Figure~\ref{fig:pitch_compression} shows this relation. Figure~\ref{fig:collimation_traj} shows simulated trajectories and transverse-to-longitudinal energy collimation for 18.6\,keV electrons.

\subsection{Potential well around target}

\begin{figure}[h!]
    \centering
    \resizebox{\textwidth}{!}{
    \includegraphics[height=6cm, keepaspectratio]{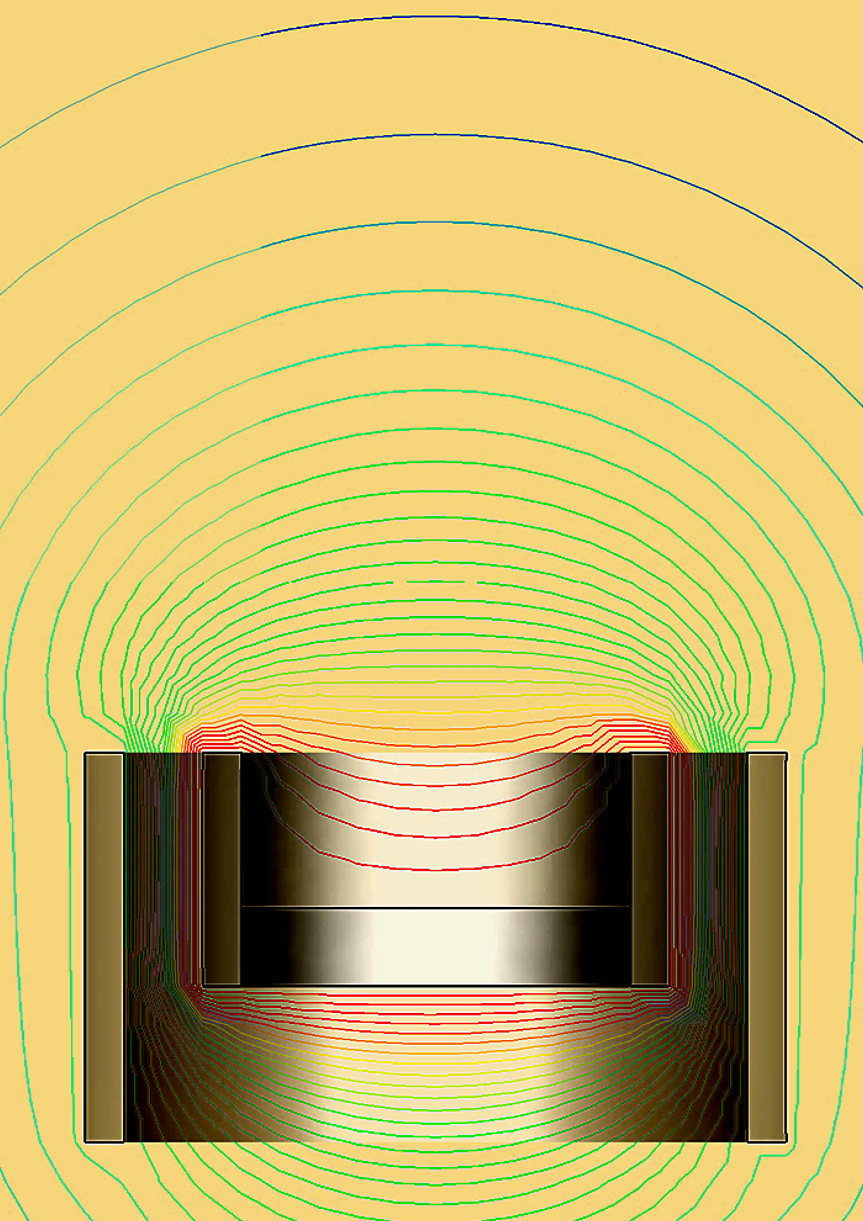}
    \hspace{1em}
    \includegraphics[height=6cm, keepaspectratio]{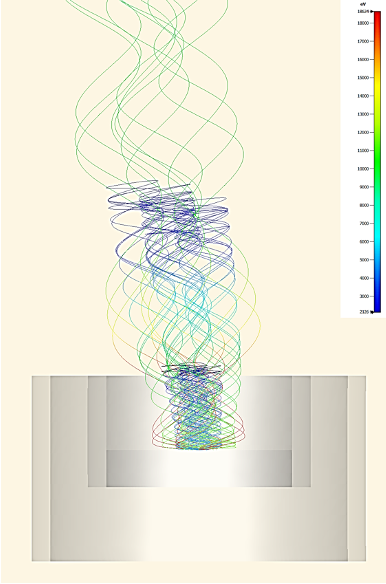}
    \hspace{1em}
    \includegraphics[height=6cm, keepaspectratio]{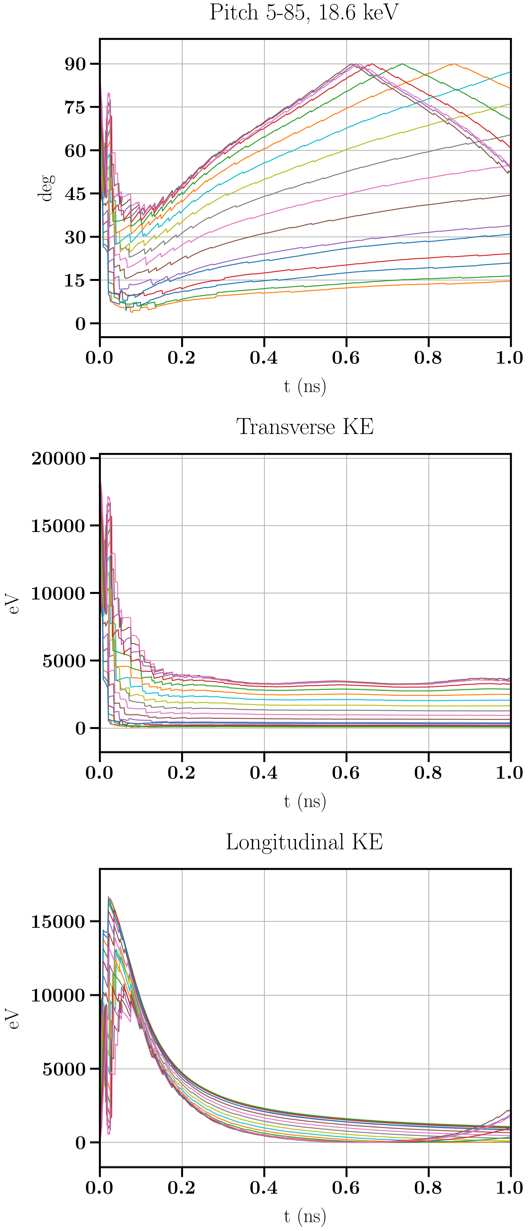}
    }
    \caption{Cylindrical electrodes set at 18.6\,kV act as a first energy-cut for emitted electrons. Low-energy electrons (11.6\,keV, 4.6\,keV) are unable to escape the potential well and are reflected back to the target. 18.6\,keV electrons (green) are able to escape.}
    \label{fig:cans_energy_cut}
\end{figure}

Cylindrical electrodes placed around the target act as an initial longitudinal energy cut on emitted electrons, which must escape the potential well of these electrodes to proceed further into the apparatus (Figure~\ref{fig:cans_energy_cut}). Pairs of equal voltage, arc-shaped electrodes are then placed along the path of the field lines emanating from the target, i.e. along the trajectories of the electrons, to create a series of potential step planes oriented orthogonal to the direction of motion of the electrons (Figure~\ref{fig:injection_schematic}). Such an orientation has the effect of draining only the parallel kinetic energy of the electrons as they make their way into the central channel of the apparatus. The shape of the fieldlines can be readily approximated by the equation of an ellipse to calculate the placements of the electrodes.

\subsection{Drain, drift, contain, and accelerate}

\begin{figure}[h!]
    \centering
    \resizebox{\textwidth}{!}{
    \includegraphics[height=6cm, keepaspectratio]{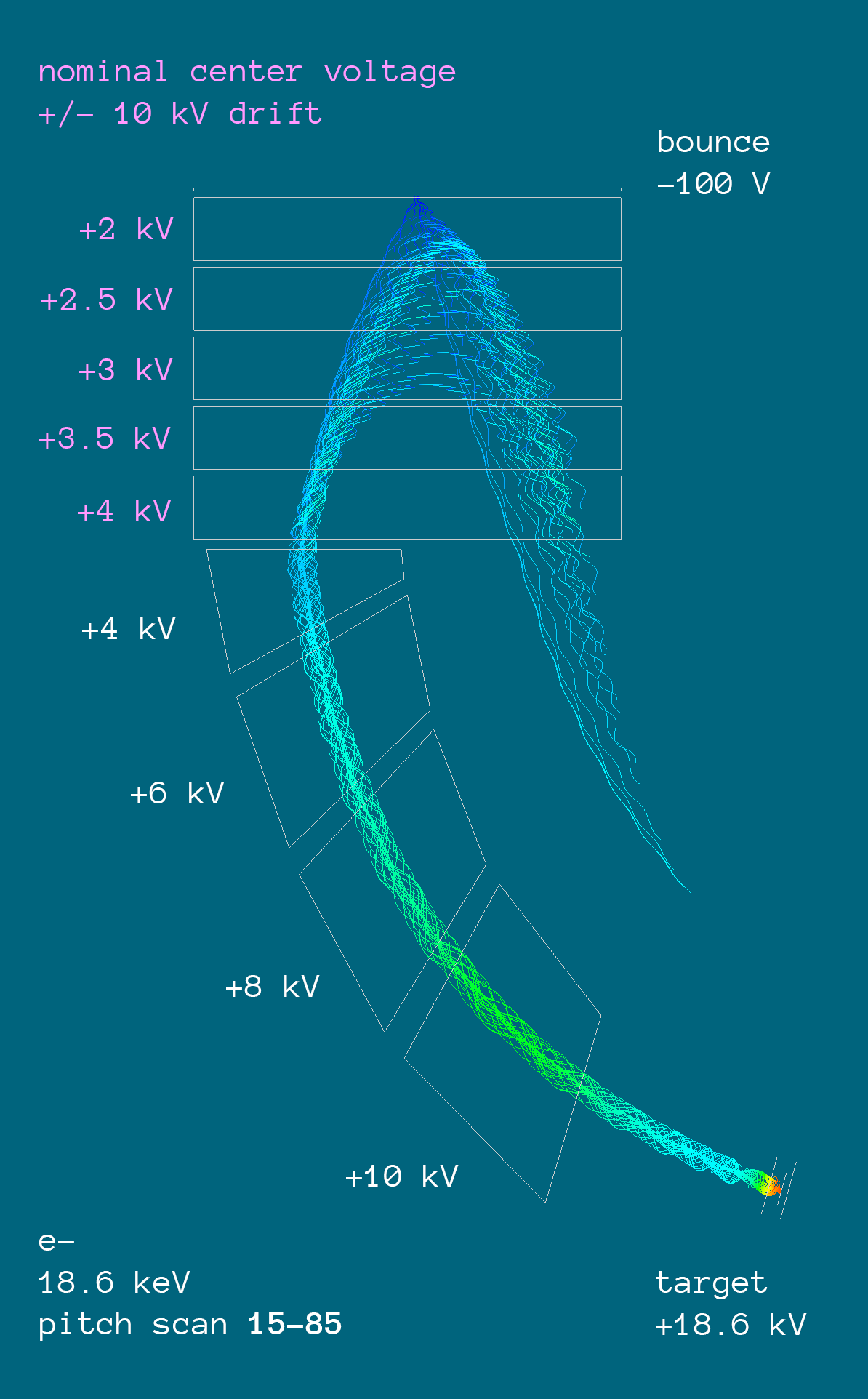}
    \includegraphics[height=6cm, keepaspectratio]{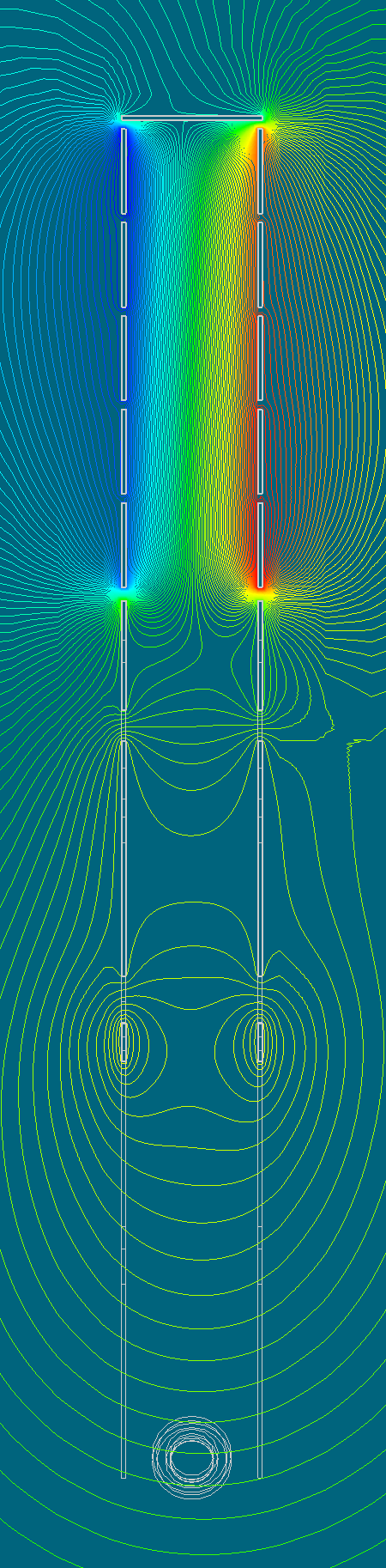}
    \includegraphics[height=6cm, keepaspectratio]{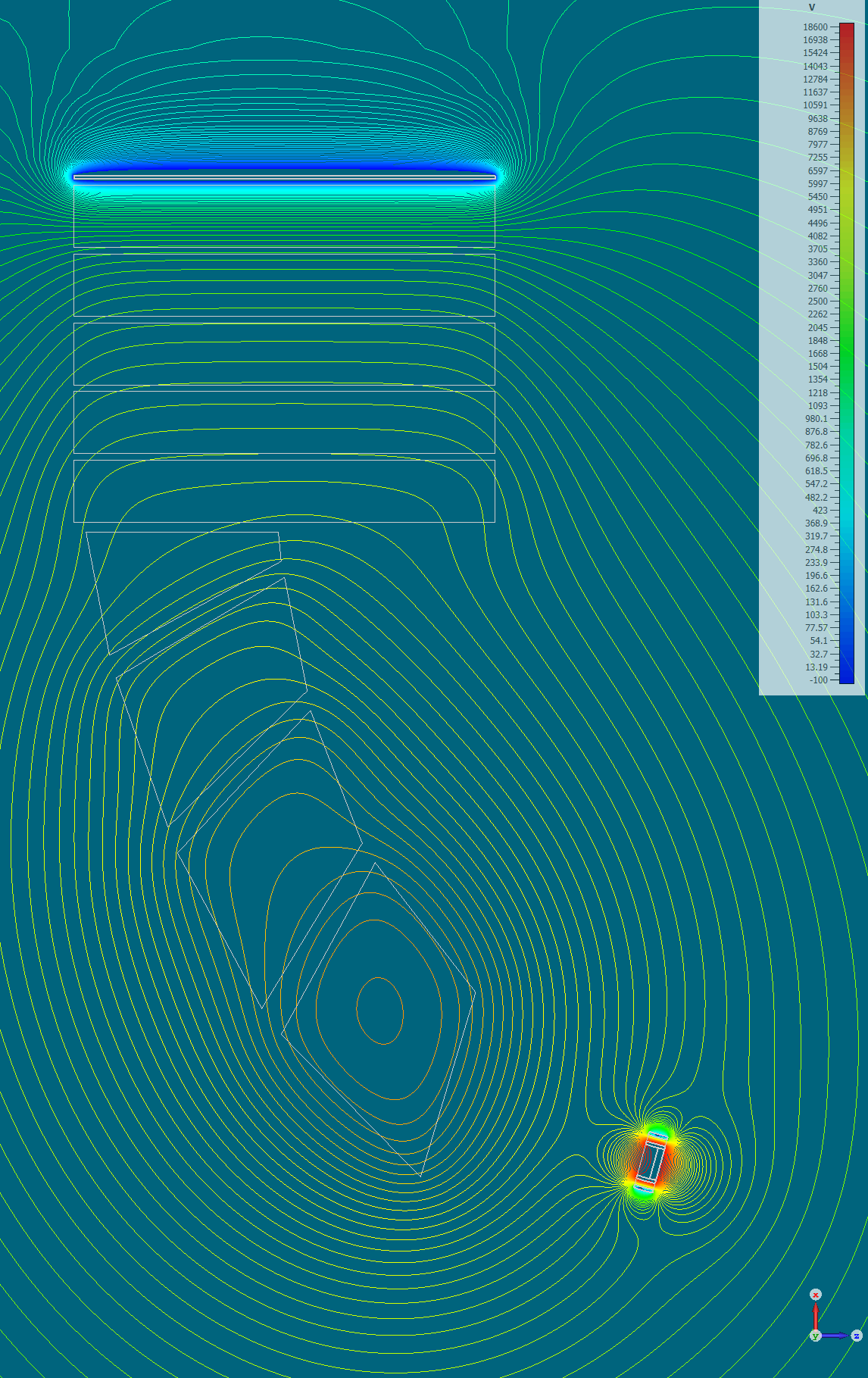}
    }
    \caption{(left) Target injection scheme with voltage labels. Arc-shaped electrodes along the trajectories drain only the parallel energy after collimation (white labels). A drift of $\pm$10\,kV is applied once the electrons enter the central channel (purple labels). Arbitrarily large pitch ranges may be transported depending on the voltage settings and collimation factor. Pitch 15-85 shown in this image. (center, right) Equipotential lines for the injection scheme showing the voltage contours in the $z$ (center) and $y$ (right) planes. The $y$ plane shows the potential contours of the arc-shaped electrodes oriented normal to the parallel momentum of the electrons. The $z$ plane shows the drift voltages when the electrons enter the central channel.}
    \label{fig:injection_schematic}
\end{figure}

When the electrons enter the central channel, an accelerating drift differential is applied to the voltages to accelerate the electrons into the high-field uniform region of the magnet (Figure~\ref{fig:injection_schematic}). The drift voltages are configured to maintain smooth transitions between the no-drift and drift regions to avoid abrupt steps in potential along the electron trajectories. The initial drift voltages at the injection-acceleration transition are matched to the arc-electrodes to ensure a smooth transition, then subsequently scaled along z to the sampled $B_x(z)$ along the central channel to achieve the same drift balancing condition between the gradient-$B$ drift and $E\times B$ drift as in the filter.

In this configuration, a high degree of collimation allows for the transport of the entire range of pitch angles 0-90$^\circ$ from the target into the central channel.
While previously the filter design implied a narrow acceptance of a small pitch angle range within which the drift balancing and subsequent trajectory were valid, with the initial collimation and large pre-drain of parallel momentum, the effective pitch angle acceptance is increased by the factor of collimation.
If the factor of collimation is high enough, it is possible to achieve a static, total-energy filter -- low-pitch electrons have their transverse energy drained early in the filter but stay on trajectory as their residual parallel energies continue to be drained by the natural rocking motion introduced by the fringe field effects of the bounce potentials, while high-pitch electrons continue to have their transverse energy drained.

\subsection{Drift collimation across a wide pitch angle range}

When the electrons enter the filter region, the phase space of momenta are spread over some distance in $y$. In the original formulation of the filter, a specific transverse energy at a specific position is selected for which the voltages are configured to enable a straight GCS trajectory, and electrons without the requisite energy or not at the required position drift off of the center. Based on the geometrical aperture calculations done in earlier in this thesis, we describe a method for which it is possible to collimate the trajectories of a wide range of pitches and starting positions onto a distribution in $z$.

\begin{figure}[h!]
    \centering
    \includegraphics[width=0.5\textwidth]{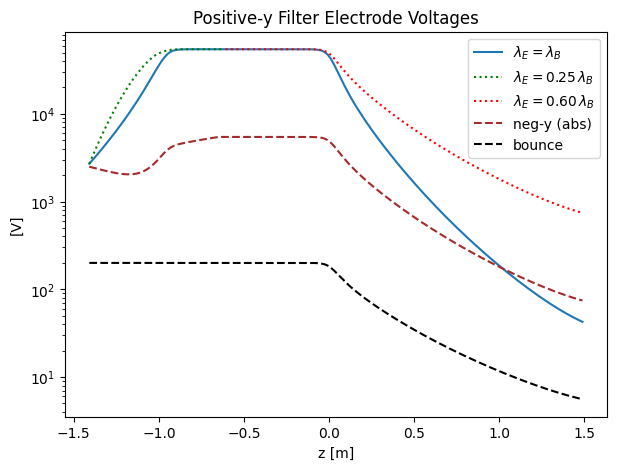}
    \includegraphics[width=\textwidth]{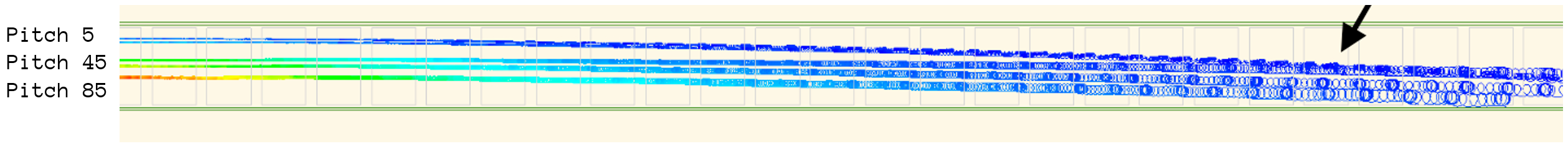}
    \caption{(top) Voltage settings for the collimation injection and filter. (bottom) Convergence of trajectories for different pitches and starting positions due to the imbalance in $\lambda_E$ to $\lambda_B$.}
    \label{fig:transmission_voltages}
\end{figure}

The crux of the idea is that while the $\lambda_B$ of the magnetic field is set by the magnet, the $\lambda_E$ of the electric field is still variable depending on how the filter voltages are set. We intentionally introduce imbalances in the $\lambda_E$, scaled as a factor of $\lambda_B$ for simplicity, to surgically introduce imbalances in the $y$-drift terms at injection and filter entry. The calculations done earlier state that the trajectories of any electron can be calculated and predicted in $y$ and $z$ as a function of the field voltage $V_0$ and the transverse kinetic energy $T_{\perp 0}$. We use this fact to empirically dial the scaling of $\lambda_E$ to force the trajectories for different pitch angles and starting $y_0$ positions to converge at some point in $z$. (Figure~\ref{fig:transmission_voltages})

Relaxing the $\lambda_E$ has the desired effect of enabling a wider range of pitch angle acceptance, but the tradeoff is that since the voltages no longer converge at the end of the filter, the $E\times B$ drift tends to run high. Separately, the cyclotron radius explodes in growth.

\begin{figure}[h!]
    \centering
    \includegraphics[width=0.75\textwidth]{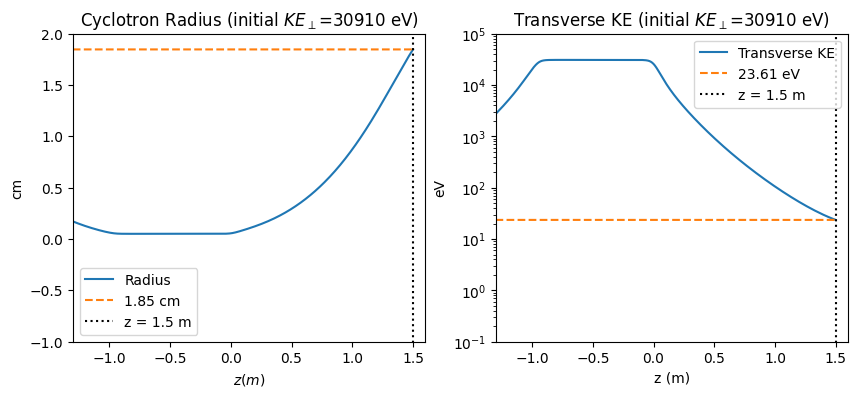}
    \caption{Cyclotron radius and kinetic energy of the electron through the apparatus.}
    \label{fig:transmission_radius}
\end{figure}

\begin{figure}[h!]
    \centering
    \resizebox{\textwidth}{!}{
    \includegraphics[height=6cm, keepaspectratio]{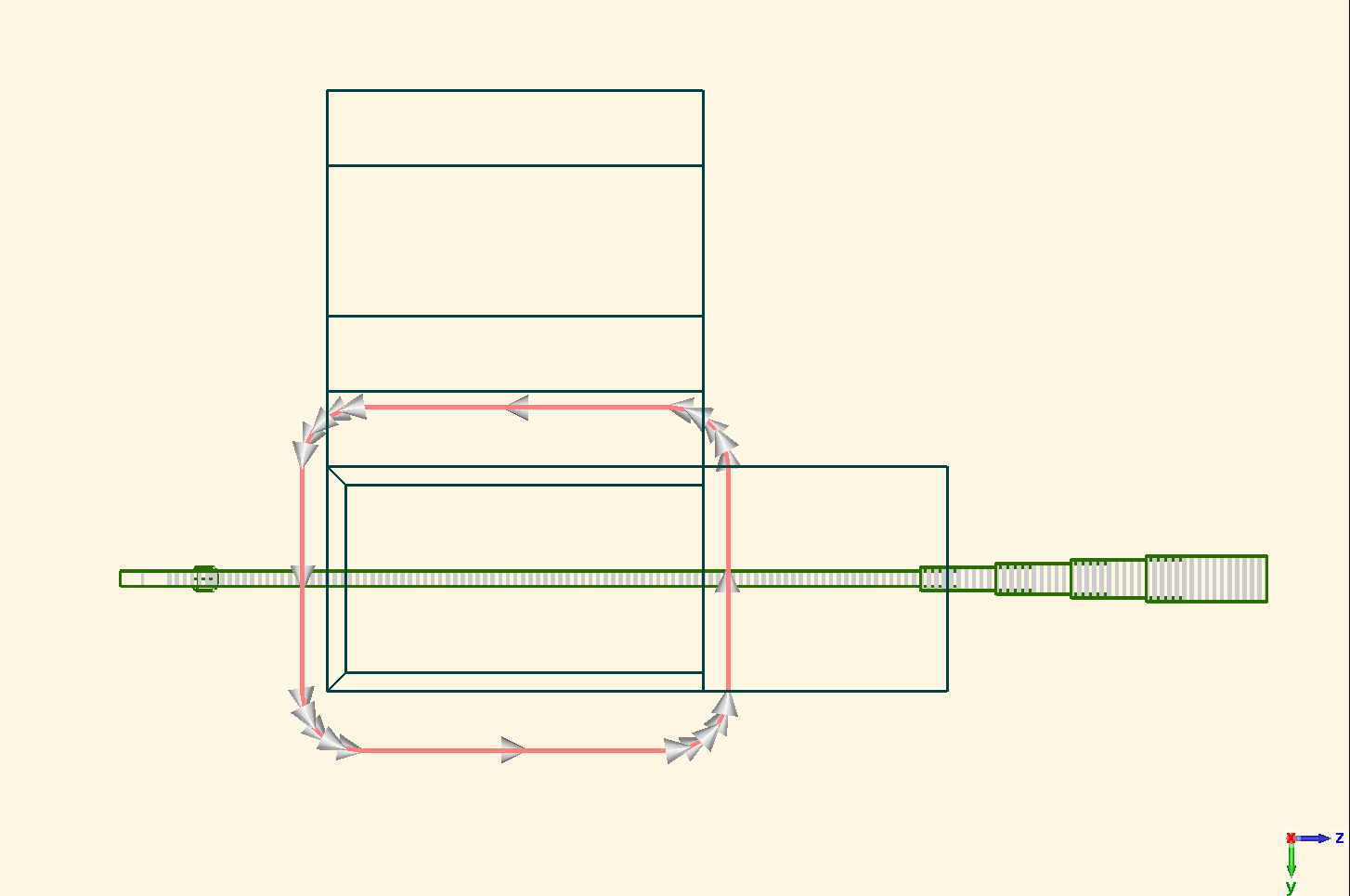}
    \includegraphics[height=6cm, keepaspectratio]{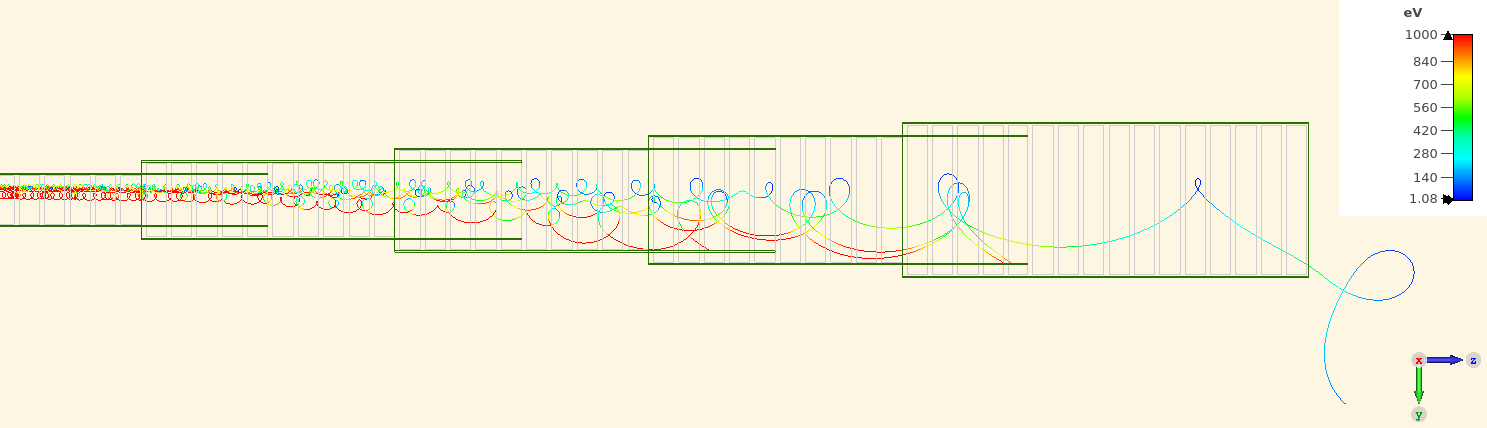}
    }
    \caption{$y-z$ plane views of the 5-stage telescoping filter design.}
\label{fig:5stage}
\end{figure}

Both of these issues can be mitigated with the introduction of a 5-stage telscoping design for the filter dimensions (Figure~\ref{fig:5stage}). The wider separation in $y$ contains the cyclotron radius and decreases the $E\times B$ drift owing to the larger spacing. The decrease in aspect ratio proves to be advantageous as the fringe field effects at the end of the filter contribute more readily to the draining of the residual parallel momentum as shown in Figure~\ref{fig:pitch_analysis}

\subsection{Design considerations and scaling up mass capacity}

The most important parameters that set the final acceptance for such a design are the initial factor of collimation and the initial drift voltages at the injection-acceleration transition. It is advantageous to compress the effective pitch range into as small a window as possible to minimize the transverse energy spread later in the filter. Similarly, the initial drift voltages set the voltage scale for the rest of the filter electrodes in the apparatus, so it is advantageous to have lower drifts to minimize the total potential spread at the entrance of the filter.

The target mass can be scaled up in several ways. First, instead of placing the target off-axis in $z$, it can be placed off-axis in $y$ instead and brought to the center in $z$ (i.e. $x$=0). The target substrate can then be affixed to both sides of the desk, and the natural curvature drift of the magnetic field lines can be used in a long initial magnetic bottle bounce to clear the emitted electrons away from the target. On this side of the magnet, the gradient-$B$ drift would act in the positive-$y$ direction for electrons. This method can be shown to scale up the mass capacity for the LNGS demonstrator to 1 $\mu$g using a double-sided disk of approximately 6\,cm.

Another method is to duplicate the injection scheme along the magnetic field lines by a rotation about $z$. The entire setup can then be duplicated by a reflection about $x$. While such a scale-up necessitates fine-tuned optimizations of the geometry for maximum acceptance, the available physical volume of the low-field region is in principle compatible with on order 1 micrograms of tritium.

\subsection{Transmission Function}

\begin{figure}[h!]
    \centering
    \includegraphics[width=\textwidth]{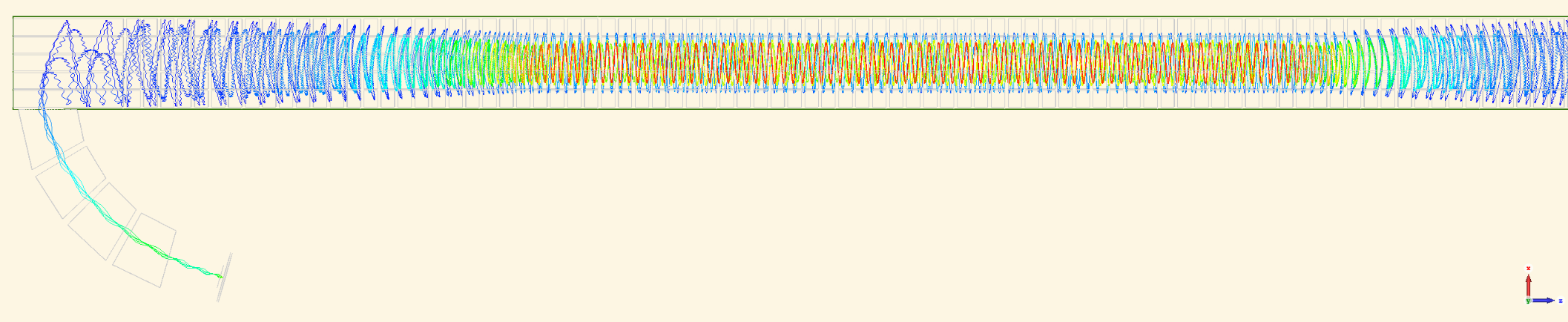}
    \caption{$y$-plane view of 18.6\,keV electron trajectories showing injection from the target into the central channel, acceleration into the uniform-field region, and transport into the filter. Pitch 85, 65, 45, 25, 5 shown, illustrating the effective transport of a large pitch angle range owing to the initial collimation.}
    \label{fig:5ch}
\end{figure}

\begin{figure}[h!]
    \centering
    \includegraphics[width=\textwidth]{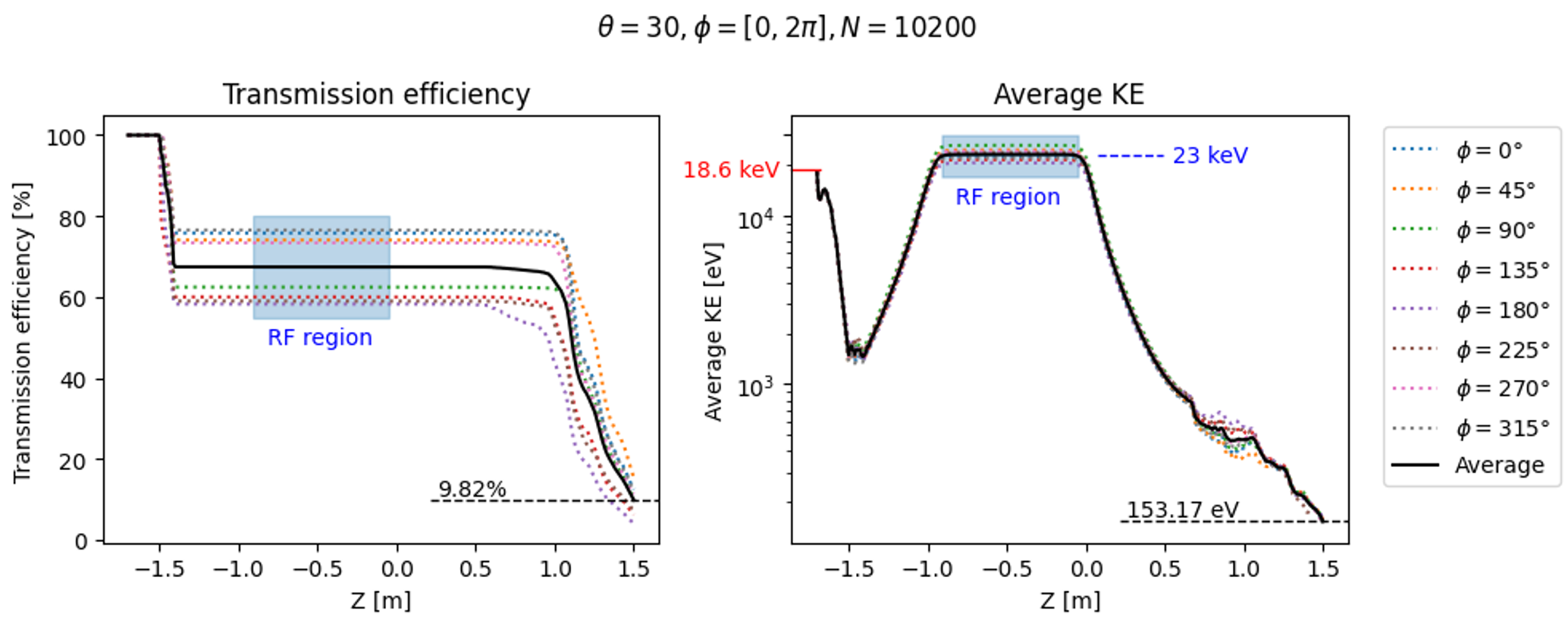}
    \includegraphics[width=\textwidth]{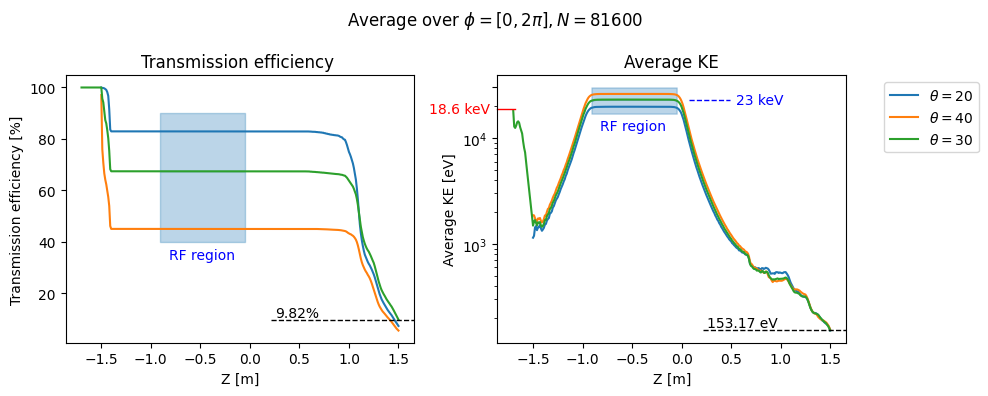}
    \caption{Static transmission curves from the target to the end of the filter.}
    \label{fig:static_transmission}
\end{figure}

\begin{figure}[h!]
    \centering
    \resizebox{\textwidth}{!}{
    \includegraphics[height=6cm, keepaspectratio]{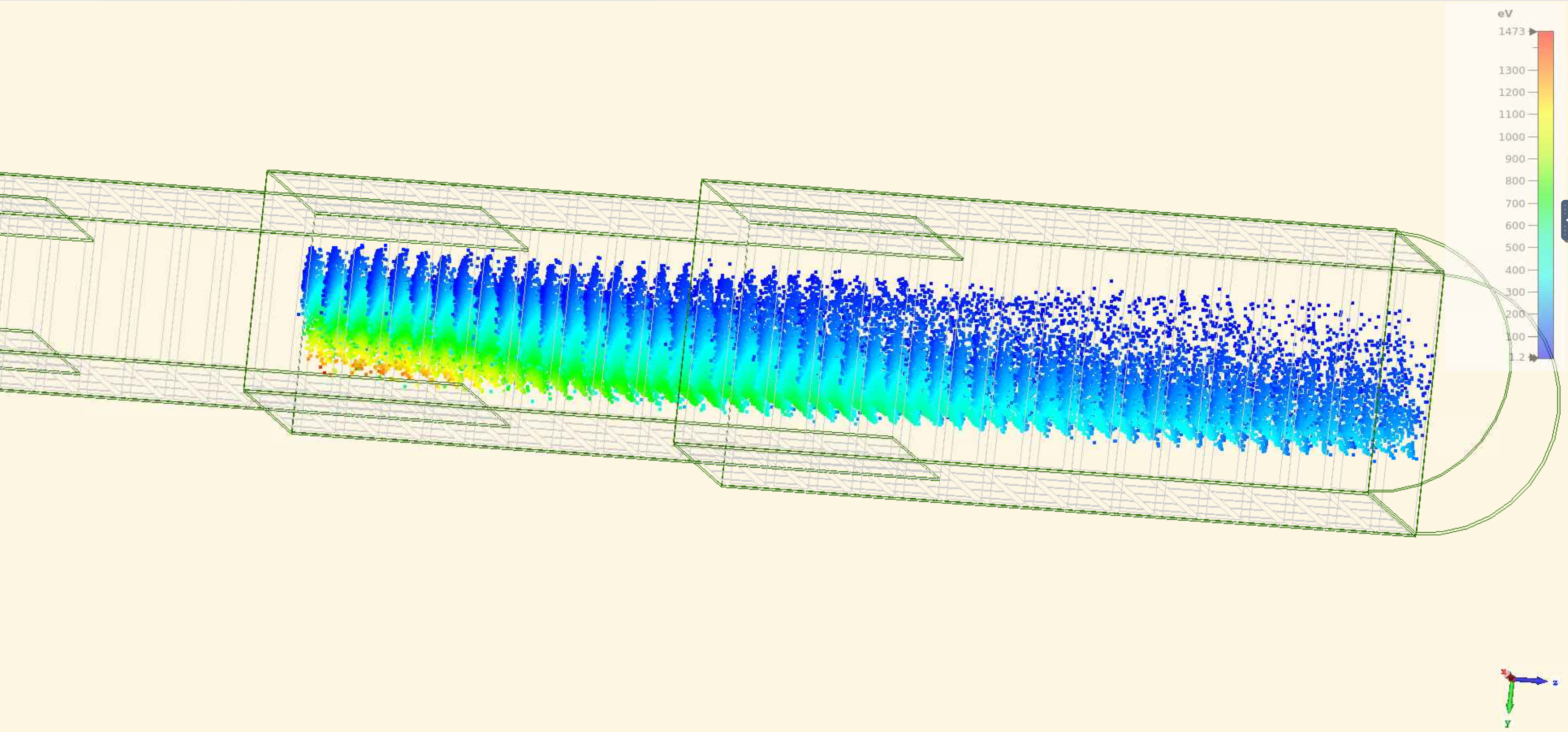}
    \includegraphics[height=6cm, keepaspectratio]{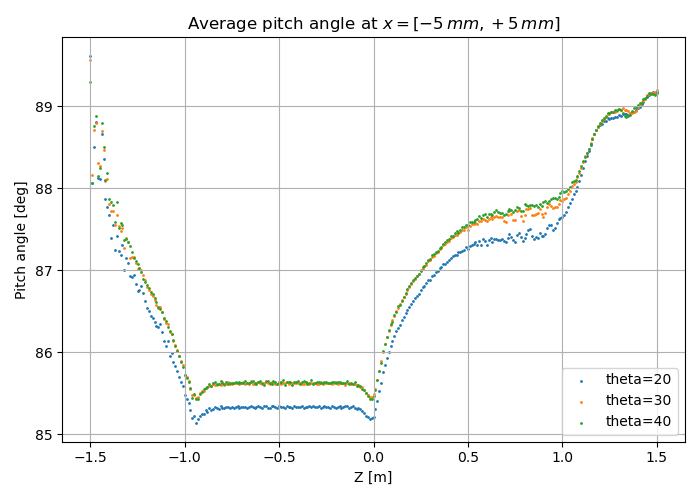}
    }
    \caption{(left) Sampled trajectory points at the end of the filter showing the transmission efficiency. (right) Effective pitch along the filter, showing the reduction of the residual parallel momentum due to the natural rocking motion caused by the fringe fields of the bounce potentials. The pitch approaches 90$^\circ$ as desired at the end of the filter.)}
    \label{fig:pitch_analysis}
\end{figure}

A static transmission function for this scheme is characterized. Endpoint electrons are emitted from the surface of a 1\,mm radius target in an isotropic, radial distribution, corresponding to about 25\,ng of tritium at 50\% loading on monolayer graphene. A uniform distribution of pitch angles is used in the range 5-85$^\circ$. The cyclotron phase angle $\phi$ at emission is a uniform distribution in the range [0,2$\pi$]. The total sample size is N=81,600 electrons per 10-degree pitch window, for N=652,800 endpoint electrons total. Using the voltage values presented above, representing an entirely un-optimized, intuitive proof-of-concept, we achieve approximately 10\% transmission efficiency from target to end of filter for the pitch range 15-45$^\circ$, or roughly 5\% for the entire pitch range 5-85$^\circ$ (Figure~\ref{fig:static_transmission}). The instantaneous pitch angles show the desired reduction of the residual parallel momentum due to the rocking motion caused by the fringe field effects of the bounce potentials inside the filter (Figure~\ref{fig:pitch_analysis}).

\subsection{Sensitivity to the neutrino mass}

Based on the static transmission scheme presented here, the PTOLEMY experiment expects an idealized sensitivity, without taking into account systematics, to a neutrino mass of approximately 150\,meV at 90\% confidence level for 1\,$\mu$g of tritium over a 3-year run.

\begin{figure}[h!]
    \centering
    \resizebox{\textwidth}{!}{

    \includegraphics[height=6cm, keepaspectratio]{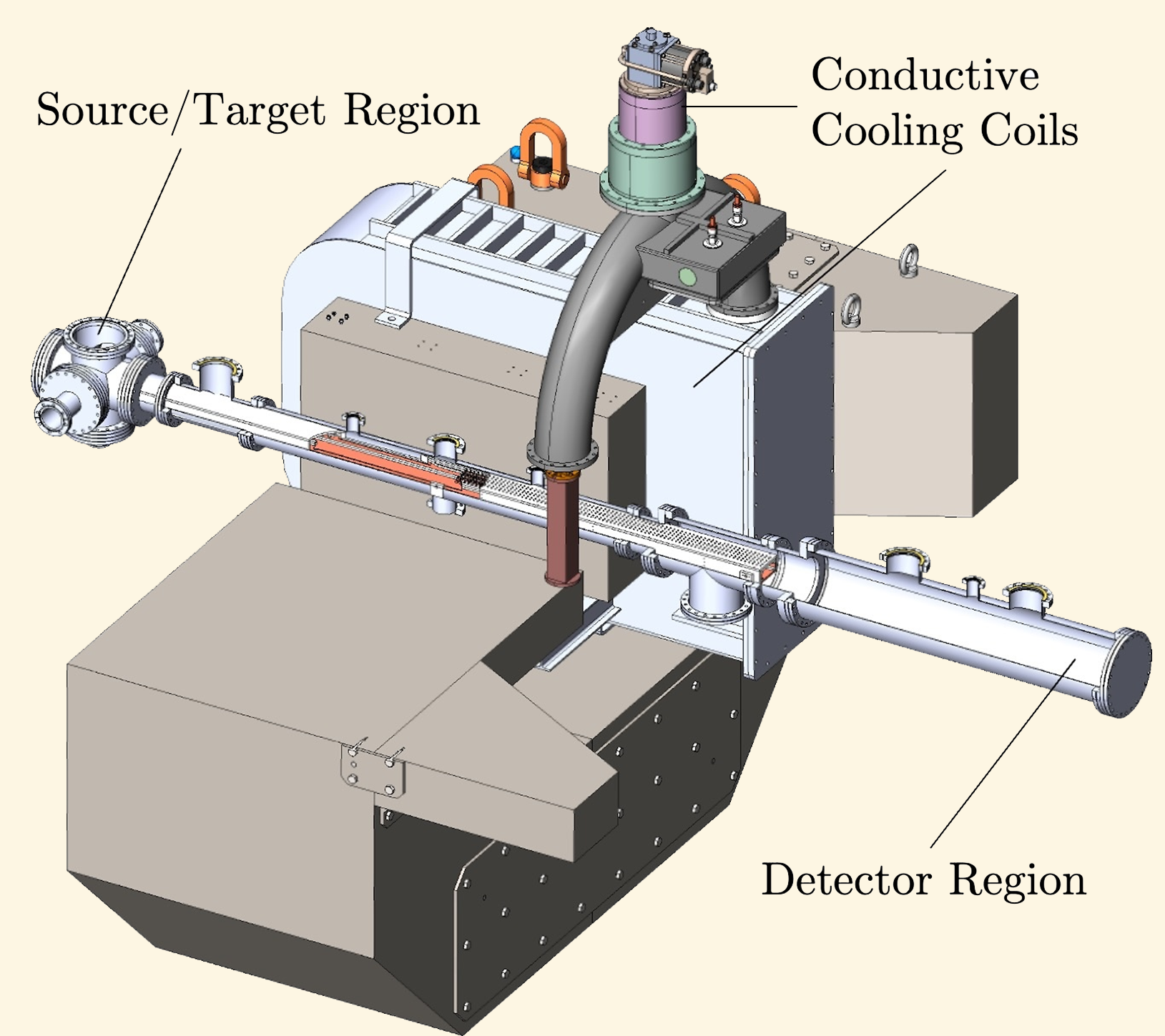}
    \includegraphics[height=6cm, keepaspectratio]{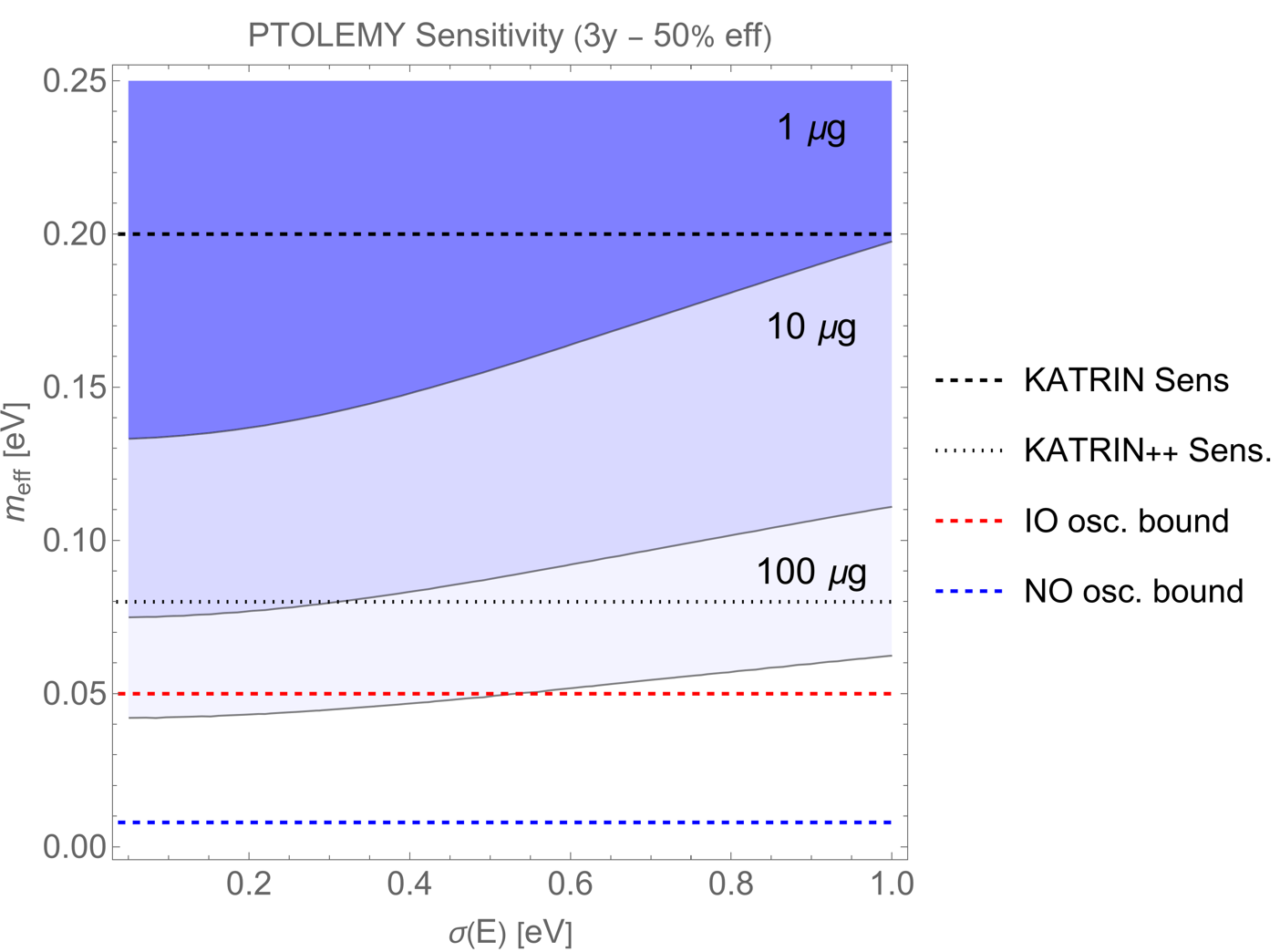}

    }
    \caption{(left) LNGS demonstrator apparatus. (right) Idealized sensitivity to the neutrino mass, no systematics present. Sensitivity is likely optimistic by a factor of 2x.}
    \label{fig:sensitivity}
\end{figure}
\section{A new type of particle accelerator}

When the filter is run in reverse, accelerating electrons into a high field region as is done in the target injection scheme of the previous section, it acts as a new kind of charged particle accelerator. Direct injection of charged particles into magnetic fields is a useful capability in many applications such as fusion plasma heating, space propulsion, and lithography. The application to fusion plasma heating as an alternative to neutral beam injection is detailed in~\cite{fusion_bookchapter}

A reversed filter starts with a charged particle at low kinetic energy and has a reversed direction of gradient-$B$ with mirrored filter voltages.  With the same drift balancing conditions, this accelerates the charged particle's transverse kinetic energy along the center line rather than drain it, hence a reversed filter is a new type of particle accelerator.  There is no linear acceleration in this process.  The electron's magnetic orbital angular momentum is accelerated during a process of constant drift into high magnetic field.

\begin{figure}[h!]
\centering
\includegraphics[width=1\textwidth]{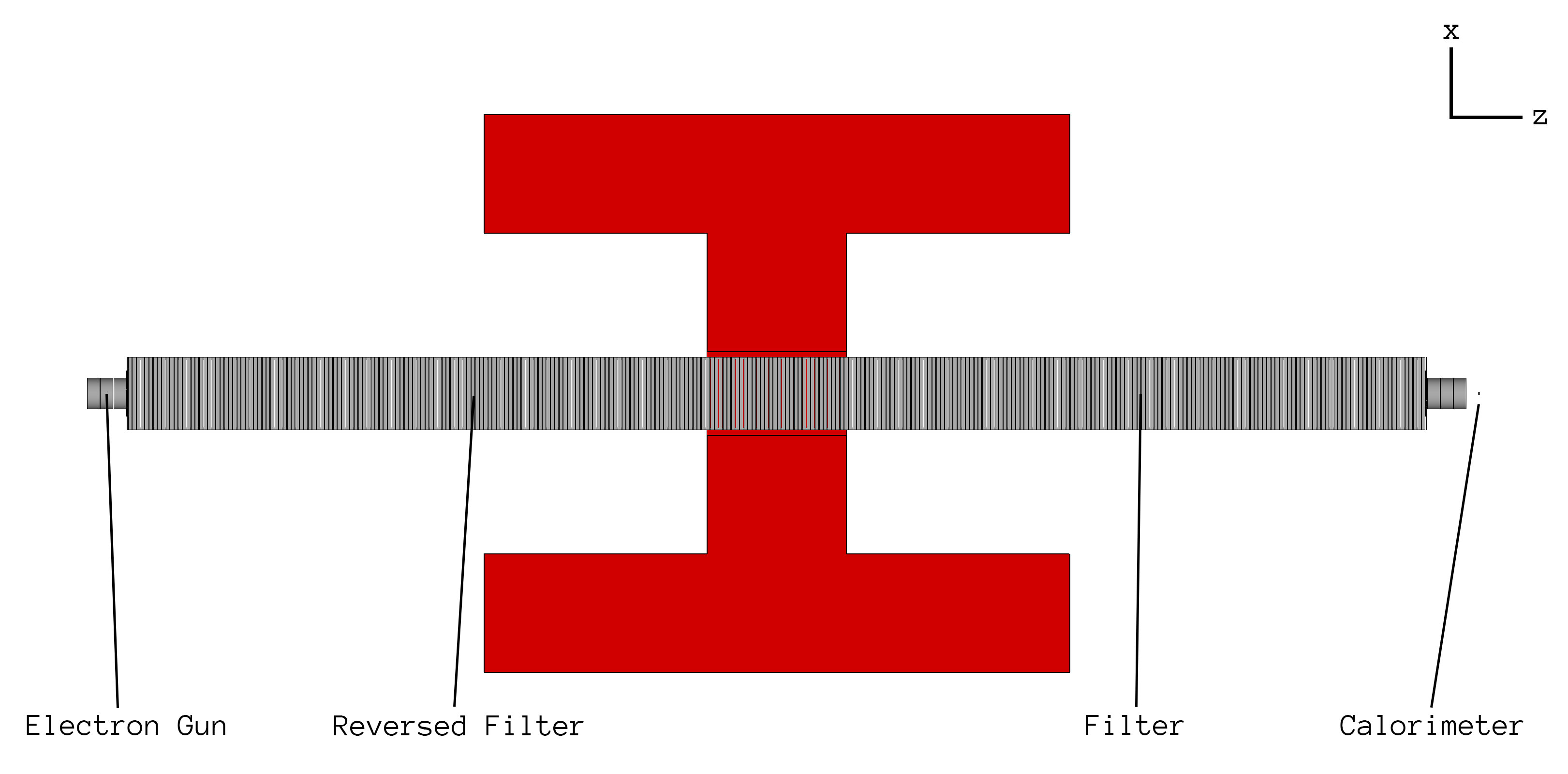}
\includegraphics[width=1\textwidth]{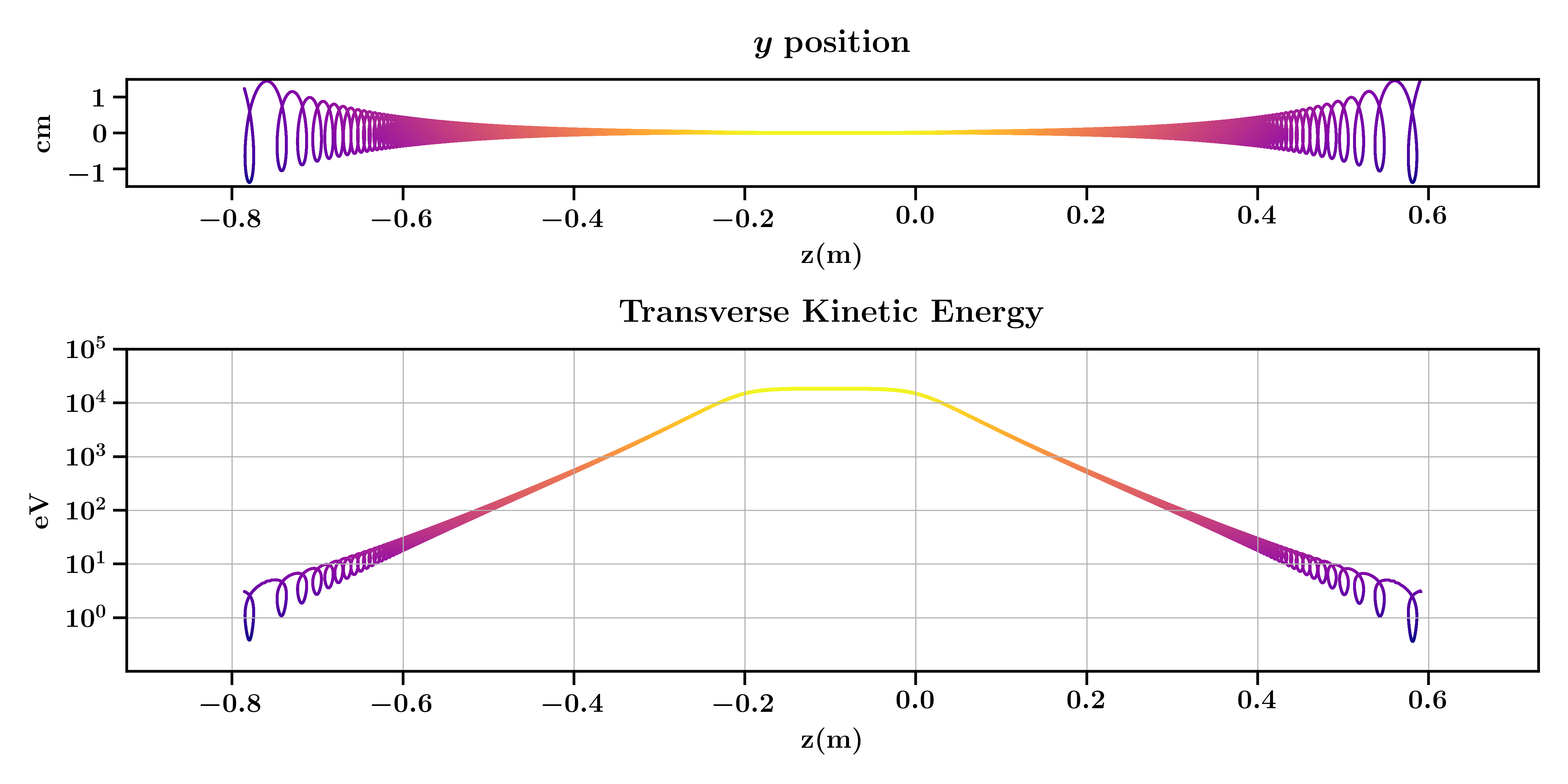}
\caption{(\textit{top}) Schematic of a reverse filter setup. Both magnet and filter are mirrored about the center of the magnet. Injected electrons on the left are accelerated by gradient-$B$ drift using the same drift conditions as the draining process. (\textit{bottom}) Trajectory and transverse kinetic energy for a pitch 90$^\circ$ electron in the reverse filter setup. The electron starts at $\approx$ 1\,eV, is accelerated by gradient-$B$ drift to $\approx$ 18.6\,keV, then drained back down to $\approx$ 1\,eV. Highlighted overlays are GCS values.
}
\label{fig:reversed_filter}
\end{figure}

A reversed filter can be realized with the same magnet design principles described in this thesis by mirroring the magnet extensions and filter electrodes about the center of the magnet (Figure~\ref{fig:reversed_filter}). In this setup an electron is emitted from an electron gun at around 1\,eV, accelerated into the uniform $B$ region to 18.6\,keV, then filtered back through the filter down to 1\,eV again.

\chapter[Dual-Readout Calorimetry at a Future Lepton Collider]{Dual-Readout Calorimetry at \\ a Future Lepton Collider \\ \large SCEPCAL}
\chapteropeningtext

\noindent
The next collider is likely to be an $e^+e^-$ or $\mu^+\mu^-$ machine. It is possible that accelerator technology makes a great leap in the next decade. Nonetheless, the design for a general-purpose detector can remain largely accelerator-agnostic up to the machine-detector interface. 

Designing a detector to be built 15 years from now using today's technologies can be successfully done if the methodology is flexible enough to admit late-stage changes when needed. Today that means modularized and differential simulation. Physics requirements can be tuned along with material properties and geometry configurations. We take advantage of early-adopting the new key4hep~\cite{key4hep} software framework for future colliders in this respect.

A differentiable full detector simulation has been implemented for a dual-readout, segmented crystal calorimeter. The SCEPCal provides excellent intrinsic EM resolution and is designed to be paired with the IDEA dual-readout fiber HCAL~\cite{ideadetectorpaper} and has recently been included as part of the IDEA baseline design. The simulation features a fully automated and configurable geometry enabling differentiation of all detector dimensions, including crystal dimensions and tower configurations. The software architecture, development environment, and necessary components to implement a new detector concept from scratch are described.

The simulation we have developed also allows for the customization of the detector response function, which is rarely, if ever, done. We introduce a new, original concept of bridging the domain gap in collider reconstruction with synthetic data from such a customized detector response function. We perform qualitative assessments and a first implementation of this idea, demonstrating its interpretability in physics-based machine learning methodologies.

\section{The Segmented Crystal Precision EM Calorimeter}
Dual-readout calorimetry has a recent but rich history~\cite{25years} and offers a genuinely new type of calorimetry for use in future collider detectors. First pioneered by the DREAM collaboration and RD52 experiment, it has since been proposed for a future lepton collider by the IDEA collaboration~\cite{idea}, the benefits of improving EM resolution by adding a longitudinally-segmented homogeneous crystal electromagnetic calorimeter in place of the preshower in the IDEA detector has been studied in several recent works~\cite{Lucchini_2020,calorimetryatfccee}.

\begin{figure}[h!]
  \centering
  \includegraphics[width=0.8\textwidth]{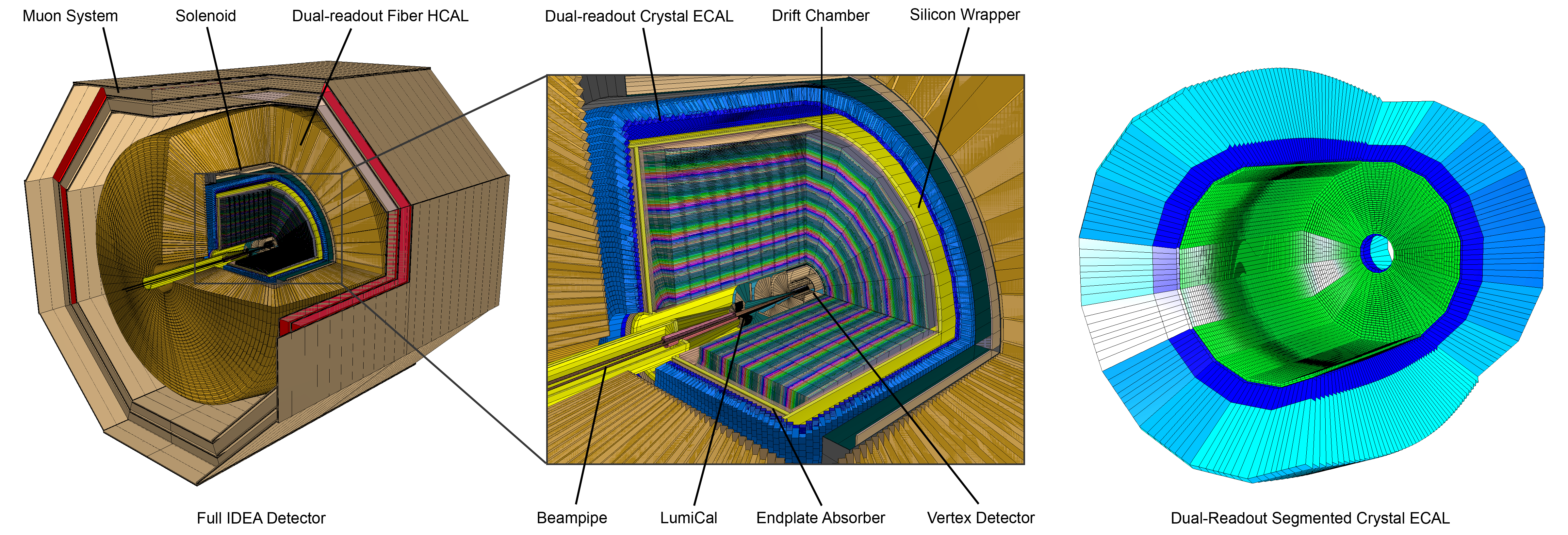}
  \caption{The Segmented Crystal EM Precision Calorimeter (SCEPCal). Crystal dimensions exaggerated for visibility. Dark/light blue: main layer front/rear quasi-projective crystals. Green: projective timing layer crystals.
  }
\label{idea_full}
\end{figure}

The SCEPCal~\cite{Lucchini_2020} is a quasi-projective geometry of homogeneous crystals much like the CMS or L3 detectors, but with each crystal longitudinally segmented into a front and rear compartment, with the lengths in a ratio of 6:16 radiation lengths, totaling 22\,$X_0$ for full EM shower containment. SiPMs are instrumented to the front and rear faces of the respective segments to avoid the introduction of dead material between the crystals. It is likely only the rear crystals require both scintillation and Cerenkov SiPMs  since hadrons tend to shower late in the crystal.

\begin{figure}[h!]
  \centering
  \includegraphics[width=0.8\textwidth]{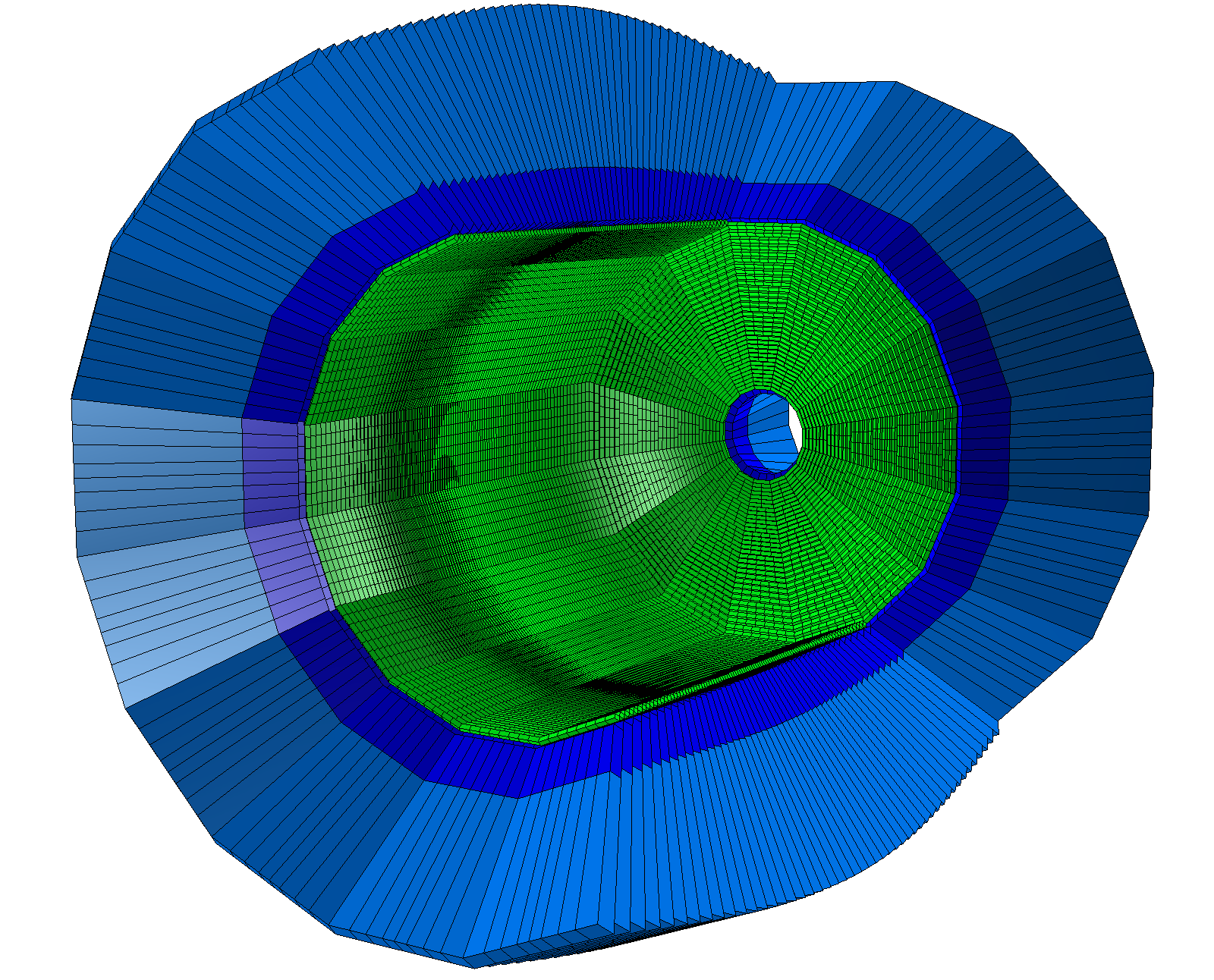}
  \caption{The Segmented Crystal EM Precision Calorimeter (SCEPCal). Crystal dimensions exaggerated for visibility. Dark/light blue: main layer front/rear quasi-projective crystals. Green: projective timing layer crystals.
  }
\label{scepcal_3d}
\end{figure}

A pure-projective precision crystal timing layer made of a fast scintillator is instrumented in front of the main segmented layer with SiPMs instrumented at each end of the crystals similar to the CMS BTL~\cite{cmsmtdtdr}. The layers are shown in the insert of Figure~\ref{phisegment}, which shows a single phi segment of the detector.

Lead tungstate (PbWO$_\text{4}$) is used for the projective layer crystals, and LYSO for the timing layer. LYSO is used as a baseline because it is used in the CMS BTL, but is likely to change as its performance in a lepton collider environment is studied. Material properties have a significant impact on simulation results and remain an active area of research, with interesting possibilities such as lattice-oriented crystals and chromatic calorimetry.

Angular offsets from the interaction point are used to mitigate projective cracks. The work in this chapter has been published in~\cite{calor2024}, the FCC feasibility study report~\cite{fcc-fsr}, and the IDEA detector paper~\cite{ideadetectorpaper}.

\subsection{Differentiable full detector simulation}
So-called differentiable simulation is a recent and active area of research in which a simulated process is made differentiable with respect to its inputs or initial conditions. In collider physics, early work of this nature has typically focused on event generation and fast simulations of detector response in a bid to optimize the performance of existing methods. For a future collider detector, the most evident use of differentiable simulation is to dynamically re-configure geometry parameters such as crystal dimensions and sensor positions to optimize fine-grained elements of the detector design with respect to reconstruction parameters and hyperparameters in a process called bilevel optimization. We present such a system in this work.

Another aspect of full detector simulation which is often overlooked is the detector response function, which fundamentally dictates the topology of the generated signals. The typical method is to consider the energy deposited in a material volume at each step of a simulated particle track and to save or discard hits based on a given threshold, typically 1\,keV. We show that using a response function motivated by detector-specific capabilities instead, i.e. wavelength-based tracking of optical photons for dual-readout calorimeters, produces new forms of synthetic data that are understood to be non-physical representations of a given shower process that may nonetheless be used as training data for neural networks.

The following sections specifically address the technical implementation of such a detector in the latest key4hep software infrastructure~\cite{key4hep}, which unifies the software dependencies and simulation programs for all future collider experiments. The use of so-called compact XML files in the detector description toolkit for high-energy physics (dd4hep~\cite{dd4hep}) to define the high-level placements of subdetectors already allows for a high level of configurability in re-arranging and interchanging subdetectors. Certain parameters, such as the number of layers in a tile-based sampling calorimeter, may also be specified in the compact XML file, allowing for dynamic re-configuration, provided the detector construction routine has implemented the detector elements in question in a re-configurable manner. Material properties may also be changed in the compact XML.

In calorimetry, dd4hep currently provides cylindrical geometries in the nomenclature of staves and layers, used in certain tracker designs and tile-based sampling calorimeters. To implement a new detector concept based on instead a projective geometry, which is not currently implemented in dd4hep, it is necessary to write the projective geometry calculations from scratch to construct the individual detector element volumes and rotate and place them in the correct positions with exact hermiticity. We take the opportunity to write a fully automated and dynamically configurable projective geometry with longitudinal segmentation capabilities, wherein all detector dimensions are dynamically calculated from only a few input specifications.

To simulate the detector response, a sensitive action must additionally be written to process the particle steps that take place inside of the subdetector volumes marked as sensitive, which, for a homogeneous crystal calorimeter, is all of the crystal volumes. For dual-readout calorimetry, we record not only the energy deposits but also the counts of scintillation and Cerenkov photons generated in each volume and their average time of generation with respect to the beginning of the event. Propagation and generation of optical photons can be fully simulated at the expense of time, or a scaling factor and Poisson smearing to the photon counts can be applied offline instead without running the full optical physics; we generally opt for the latter.

The sensitive action processes and saves hit information at each particle step. A sensitive action filter, which is defined separately, determines which particle steps are passed to the sensitive action at all - this is typically done as a function of the deposited energy as mentioned earlier, typically 1\,keV. For this work we additionally define a new sensitive action filter for optical photons specifically, which passes judgment not on the energy deposited in a single step, but on the wavelength of the primary photon track. Photon steps belonging to a track with a wavelength above a given threshold, e.g. 300\,nm, are passed to the sensitive action and recorded, even if the energy deposited is very small or zero.

\subsection{Detector geometry}
A idealized projective geometry can be fully determined and parameterized with the parameters listed in Table~\ref{scepcal_inputs}, which are then used to calculate the secondary parameters listed in Table~\ref{scepcal_secondary}. The inner radius, Z-extent of the barrel, and number of total segments in phi set the overall scale of the detector. A single global phi segment is further segmented in theta and phi into blocks of crystal towers, within which different granularities of front and rear crystals are possible. The nominal values of the square crystal tower face-widths and number of front and rear divisions are used to calculate the final dimensions. The side lengths and orientations for each asymmetric trapezoidal volume in the projective geometry are then calculated automatically and placed in their respective positions. Final crystal counts and other outputs are shown in Table~\ref{scepcal_final_parameters}.

\begin{figure}[h]
    \centering
    \includegraphics[width=\textwidth]{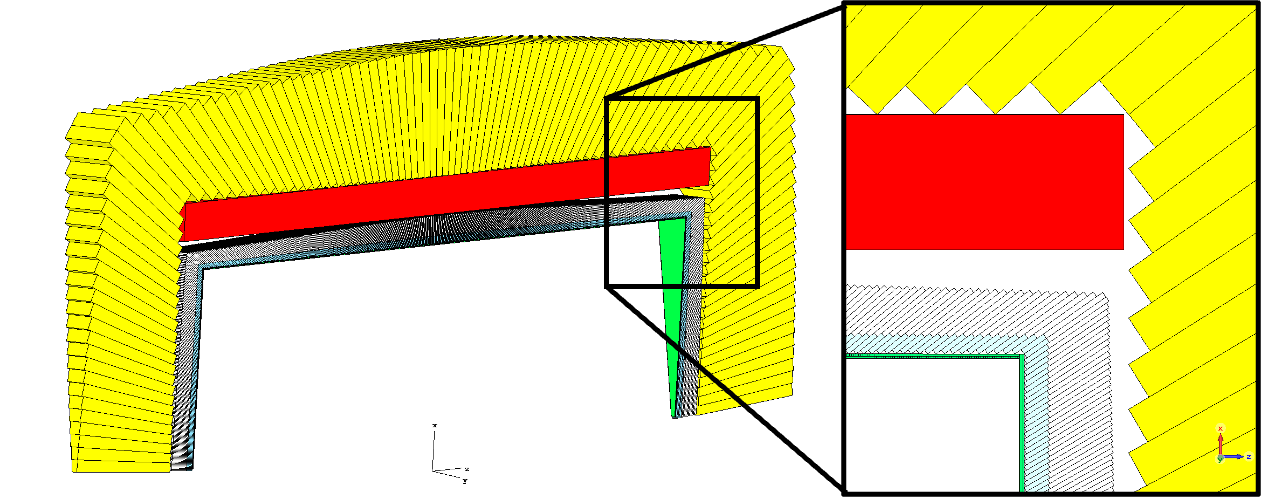}
    \caption{
    A single phi segment of a combined calorimeter concept with a precision timing layer (green) and segmented crystal layer (front crystals: blue, rear crystals: white) comprising the ECAL placed in front of the solenoid (red) of an IDEA-like dual-readout fiber HCAL (yellow towers; fibers not shown). (Insert) Cross-sectional view of the barrel/endcap junction showing the ECAL, solenoid, and HCAL towers.
    }
    \label{phisegment}
\end{figure}

\begin{table}[h!]
\centering
\caption{Input parameters for the SCEPCal fully parameterized geometry construction routine. Table from~\cite{calor2024}.}
\label{scepcal_inputs}
  \begin{tabularx}{\textwidth}{Xll}
  \toprule
  Description                                         & Variable           & Value  \\\hline
  \midrule
  Half Z-extent of the barrel                         & $Z_\text{B}$      & 2.40\,m \\
  Inner radius of the barrel                          & $R_\text{inner}$  & 2.25\,m \\
  Global number of phi segments                       & $N_\Phi$          & 128 \\
  Nominal square face width of a crystal tower        & $C_\text{fw}$     & 10\,mm \\
  Number of front crystal divisions per tower         & $N_\text{F}$      & 1 \\
  Number of rear crystal divisions per tower          & $N_\text{R}$      & 1 \\
  Front crystal length                                & $F_\text{dz}$     & 50\,mm \\
  Rear crystal length                                 & $R_\text{dz}$     & 150\,mm \\
  Pointing offset from IP                             & $P_\text{offset}$     & 100\,mm \\
  \bottomrule
  \end{tabularx}
\end{table}
  
\begin{table}[h!]
\centering
\caption{Secondary parameters for the SCEPCal calculated from input parameters. Table from~\cite{calor2024}.}
\label{scepcal_secondary}
  \begin{tabularx}{\textwidth}{Xll}
  \toprule
  Description                                                   & Variable              & Formula  \\\hline
  \midrule
  Angular size of a single phi segment                            & $d\Phi$        & 2$\pi$/ $N_\Phi$ \\
  Angular size of barrel region                                   & $\Theta_B$     & atan($Z_\text{B}/R_\text{inner}$) \\
  Angular size of endcap region                                   & $\Theta_E$     & atan($R_\text{inner}/Z_\text{B}$) \\
  Number of barrel towers in $\theta$                             & $N\theta_B$    & floor($2~Z_\text{B}/C_\text{fw}$) \\
  Number of endcap towers in $\theta$                             & $N\theta_E$    & floor($R_\text{inner} / C_\text{fw}$) \\
  Angular size of a single barrel tower in $\theta$               & $d\theta_B$    & $(\pi-2~\Theta_E) / N\theta_B$ \\
  Angular size of a single endcap tower in $\theta$               & $d\theta_E$    & $\Theta_E/N\theta_E$ \\
  Number of barrel segments in $\phi$ in a single phi segment     & $N\phi_B$      & floor($2\pi R_\text{inner}/(N_\Phi C_\text{fw})$) \\
  Number of endcap segments in $\phi$ in a single phi segment$^*$ & $N\phi_E^*$    & floor($2\pi R_\text{inner}^*/(N_\Phi C_\text{fw})$) \\
  Angular size of barrel segments in $\phi$                       & $d\phi_B$      & $d\Phi / N\phi_B$ \\
  Angular size of endcap segments in $\phi$                       & $d\phi_E^*$    & $d\Phi / N\phi_E^*$ \\\hline
  $*$ varies with $R_\text{inner}$ \\\hline
  \bottomrule
  \end{tabularx}
\end{table}

\begin{table}[h!]
\centering
\caption{Final crystal/readout counts and endcap dimensions from the parameterized geometry construction using input values from Table~\ref{scepcal_inputs}.}
\label{scepcal_final_parameters}
  \begin{tabularx}{\textwidth}{Xr}
  \toprule
  Quantity                                                               & Value  \\\hline
  \midrule
  Number of total barrel crystals                                        & 1,360,128 \\
  Number of barrel front crystal readout channels (1 SiPM per crystal)   & 680,064 \\
  Number of barrel rear crystal readout channels (2 SiPMs per crystal)   & 1,360,128 \\
  Number of total endcap crystals                                        & 251,136 \\
  Number of endcap front crystal readout channels (1 SiPM per crystal)   & 125,568 \\
  Number of endcap rear crystal readout channels (2 SiPMs per crystal)   & 251,136 \\
  Endcap innermost radius (center point of innermost front crystal face) & 360.2\,mm \\
  Endcap outermost radius (center point of outermost rear crystal face)  & 2386.8\,mm \\
  Endcap maximum pseudorapidity                                          & 2.595 \\\hline
  \bottomrule
\end{tabularx}
\end{table}

The top level loop of the construction routine loops over the number of global phi segmentations. A single global phi slice (Fig.~\ref{phisegment}) is then divided in theta into envelope volumes of asymmetrical trapezoids according to the nominal crystal tower face width along Z.

Crystal towers are then formed by dividing the XY extent of the chord formed by the global phi slice by the nominal tower face width to form quasi-square towers. Each tower is then split into front and rear crystals, also assymmetrical trapezoids, by the number of respective divisions. The rear vertices of the individual crystal trapezoids are additionally shifted by the ratio of the projective offset to the radial distance of the crystal from the IP to effectively point the crystal away from the IP by the offset distance, mitigating projective cracks.

\begin{figure}[h!]
  \centering
  \includegraphics[width=\textwidth]{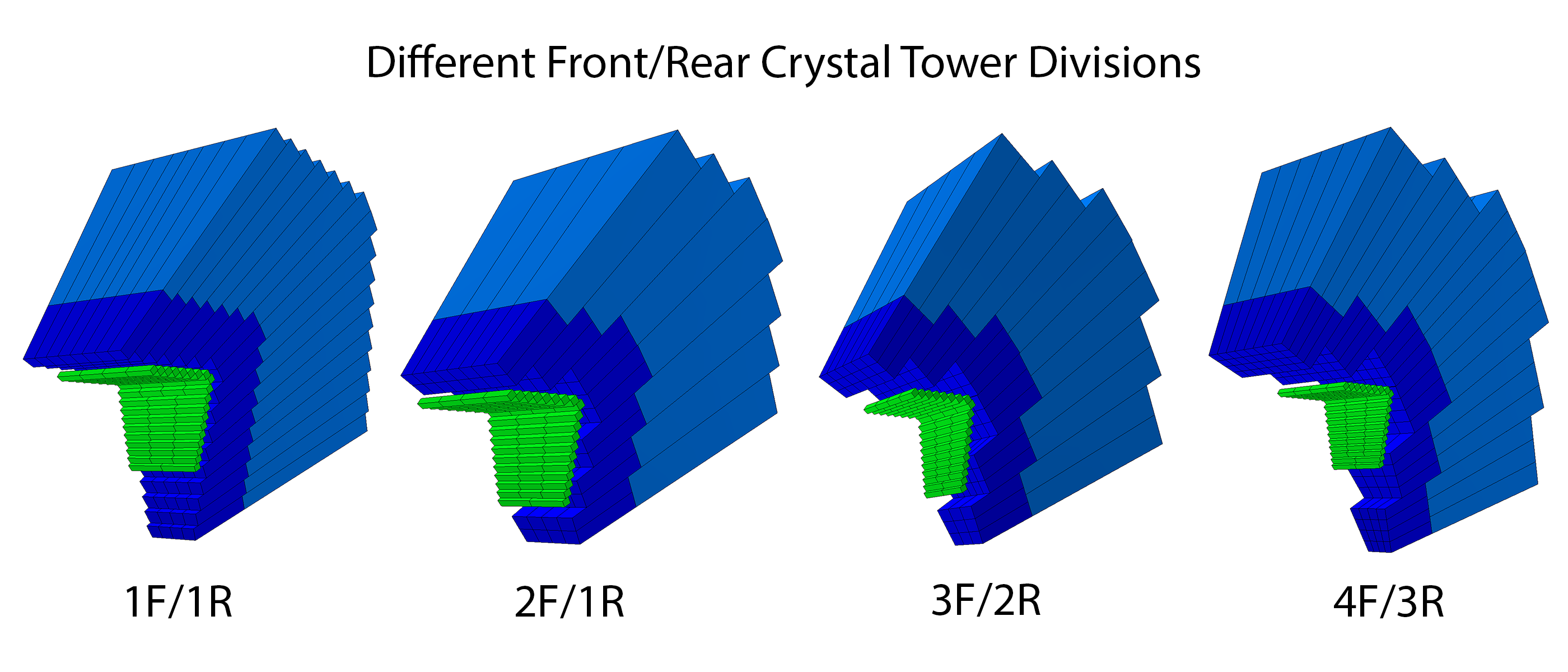}
  \caption{Barrel/endcap junction of a single global phi slice showing different front/rear crystal divisions. Crystal dimensions exaggerated for visibility. Dark/light blue: main layer front/rear quasi-projective crystals. Green: projective timing layer crystals.
  }
\label{tower_configs}
\end{figure}

Each crystal tower captures a constant solid angle and is therefore unique in the global phi slice. The basic assumptions for the calculation are that the center points of the tower front faces are collinear in Z along the inner radius of the barrel. The side lengths of the trapezoid faces are calculated by assuming the front face is normal to the IP, with an inner radius equal to that provided plus the projective offset, resulting in the crystals pointing at a point offset from the IP to mitigate projective cracks. The side lengths are then shrunk by the appropriate factor when placing the crystals along the true inner radius. The crystal lengths are fixed by the given inputs, resulting in the familiar teeth-like pattern of projective ECALs such as CMS and L3. 

Assembly volumes (dd4hep::Polyhedra) for the main and timing layers are made separately to avoid overlaps with the other subdetectors, matching the Z-extent, inner radius, and phi segmentation provided. The front, rear, and side protrusions due the tilted projective geometry are calculated to snugly fit all crstal towers in the polyhedra volumes. The construction routine in the endcap is similar as in the barrel, but the inner radius is updated for each theta.

The timing layer is a purely projective single layer of crystals. The nominal depth and length are used to calculate integer numbers of crystals for the array.

The global coordinates for each crystal center, as mapping from the cellID, are saved into the segmentation for later recall in the sensitive action to record the hit positions.

\subsection{Projective calculations}
The global dimensions $\EBz$ and $\Rin$ first determine the angular extent of the barrel and endcap regions. The inner surface of the detector is constructed as a $\PHISEGMENTS$-sided polygon instead of a circle to allow the timing layer segments to lie flush against the inner face of the barrel. 

The nominal face-width of a tower of projective crystals, $\nomfw$, is used to determine number of segments in $\theta$ for the barrel and endcap. Each phi segment of the barrel is further segmented in phi by $\nomfw$ into $\NPHIBARRELCRYSTAL$ towers in $\phi$ as shown in Figure~\ref{tower_configs}. The same is done for the endcap, but with $\Rin^*$ varying at each $\theta$ as opposed to a constant $\Rin$ for the barrel. Each crystal tower in a global phi segment is therefore a unique asymmetrical trapezoidal prism (or frustum). Barrel crystals capture a constant solid angle projection from the interaction point (IP), whereas the solid angle for the endcap crystals varies in $\theta$.

The dimensions of a single projective crystal tower are calculated as follows, assuming the crystal volume is centered at the origin. The towers are then rotated and placed in their individual positions, then further split into the number of crystal divisions specified.

First, the coordinates of the tower center in $\theta$, $\phi$, and the distances from the IP of the tower faces $\rinner$ and $\router$ are defined, where $n$ values denote elements of the total $N$ values in the tables above. $\gamm$ refers to the offset in $\phi$ from the center of the global phi segment. The expressions given below are for a front-segment tower; for a rear-segment tower, the distances $\rinner$ and $\router$ are changed accordingly.
\begin{align}
    \thC    & = \THETASIZEENDCAP + \DTHETABARREL/2 +(n\Theta_B \DTHETABARREL) \\
    \gamm   &= -\DPHIGLOBAL/2 + \DPHIBARRELCRYSTAL/2 + (n\phi_B \DPHIBARRELCRYSTAL) \\
    \rinner &= \Rin / \text{sin}(\thC) \\
    \router &= \rinner + \Fdz
\end{align}
The dimensions of the trapezoid face at $z=-\Fdz/2$ are then
\begin{align}
    \yinner    &= \rinner~\text{tan}(\DTHETABARREL/2) \\
    \xbotyinc  &= \text{tan}(\thC-\DTHETABARREL/2)[\rinner~\text{cos}(\thC) +\yinner~\text{sin}(\thC)] \\
    \xbotyinl  &= \xbotyinc~\text{tan}(\gamm-\DPHIBARRELCRYSTAL/2) \\
    \xbotyinr  &= \xbotyinc~\text{tan}(\gamm+\DPHIBARRELCRYSTAL/2) \\
    \xtopyinc  &= \text{tan}(\thC+\DTHETABARREL/2)[\rinner~\text{cos}(\thC) -\yinner~\text{sin}(\thC)] \\
    \xtopyinl  &= \xtopyinc~\text{tan}(\gamm-\DPHIBARRELCRYSTAL/2) \\
    \xtopyinr  &= \xtopyinc~\text{tan}(\gamm+\DPHIBARRELCRYSTAL/2) 
\end{align}
and similarly for the trapezoid face at $z=+\Fdz/2$, where the superscripts refer to the top (1) or bottom (0) edge in $y$ and the subscripts refer to the left (L) or right (R) edge in $x$ when offset by $\gamm$ from the center of the global phi segment $\gPhi=n\Phi~\DPHIGLOBAL$.
The position of the crystal center is then given by
\begin{align}
    \rSlice     &= \Rin +(\Fdz +\Rdz)/2 \\
    \dispSlice  &= [\rSlice~\text{cos}(\gPhi), \rSlice~\text{sin}(\gPhi), 0] \\
    \rlocal     &= \rinner +\Fdz/2 \\
    \dispLocal  &= [\rlocal~\text{sin}(\thC) - \rSlice,~\rlocal~\text{sin}(\thC)~\text{tan}(\gamm),\\
                &\ \rlocal~\text{cos}(\thC) ] \\
    \dispGlobal &= \dispSlice + \rotZ(\gPhi) \cdot \dispLocal
\end{align}
where $\Phi$ subscripts refer to the global phi segment and RotZ is the 3D rotation matrix about the Z axis for the specified angle. An additional rotation about the Y axis is applied for mirrored endcap crystals.

\subsection{Full simulation}
\begin{figure}[h]
    \centering
    \includegraphics[width=0.5\textwidth]{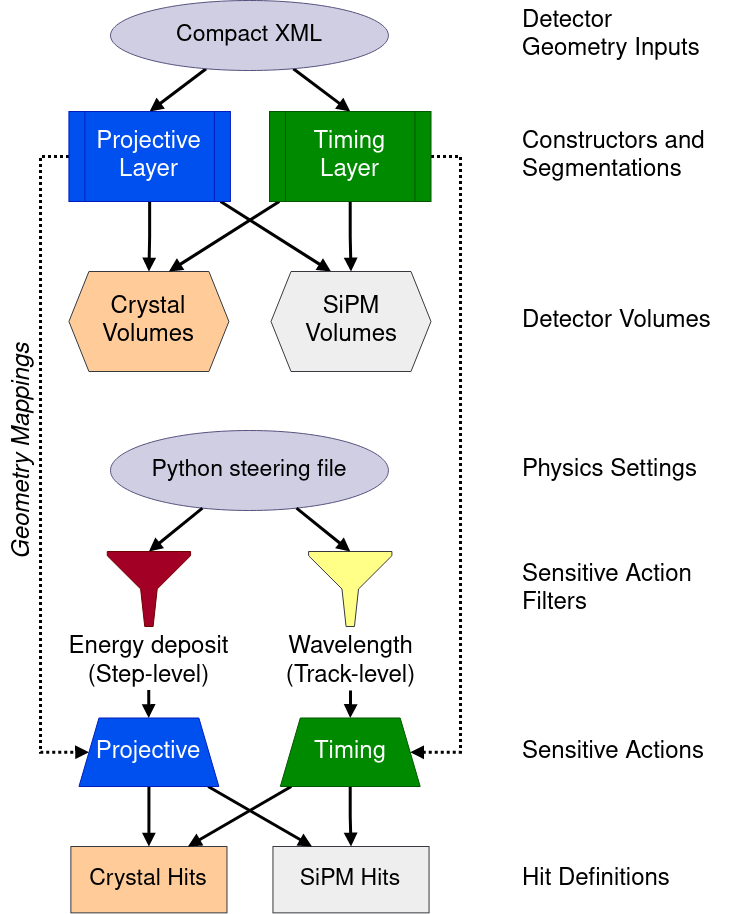}
    \caption{
    Schematic diagram of software components for a full detector simulation in dd4hep.
    }
    \label{software}
\end{figure}

The necessary software components for a new detector concept implementation in dd4hep are shown schematically in Figure~\ref{software}. The compact XML file specifies the subdetectors to be used and their parameters, as well as material properties and visualization attributes. The construction routines for each subdetector calculate, generate, and place the physical volume elements in Geant4~\cite{geant4}. Utility classes known as segmentations provide ancillary information such as global position coordinates for detector cells when writing out hits. A python steering file sets physics settings not directly related to the geometry, such as the physics process list, selection cuts, and event generation settings. The steering file also includes simulation-related settings such as logging, random number seeds, and threshold cuts for sensitive action filters, which determine the particles for which information is saved in the subdetector data writeout. Custom sensitive action classes and hits must be written for a new detector concept if existing ones, which record only positions and energy deposits, are not sufficient to capture information generated by the new detector, such as for SiPMs in dual-readout calorimeters, which record the digitized wavelength and time bins of incident scintillation and Cerenkov photons.

\subsection{Event display}
\begin{figure}[t]
    \centering
    \includegraphics[width=\textwidth]{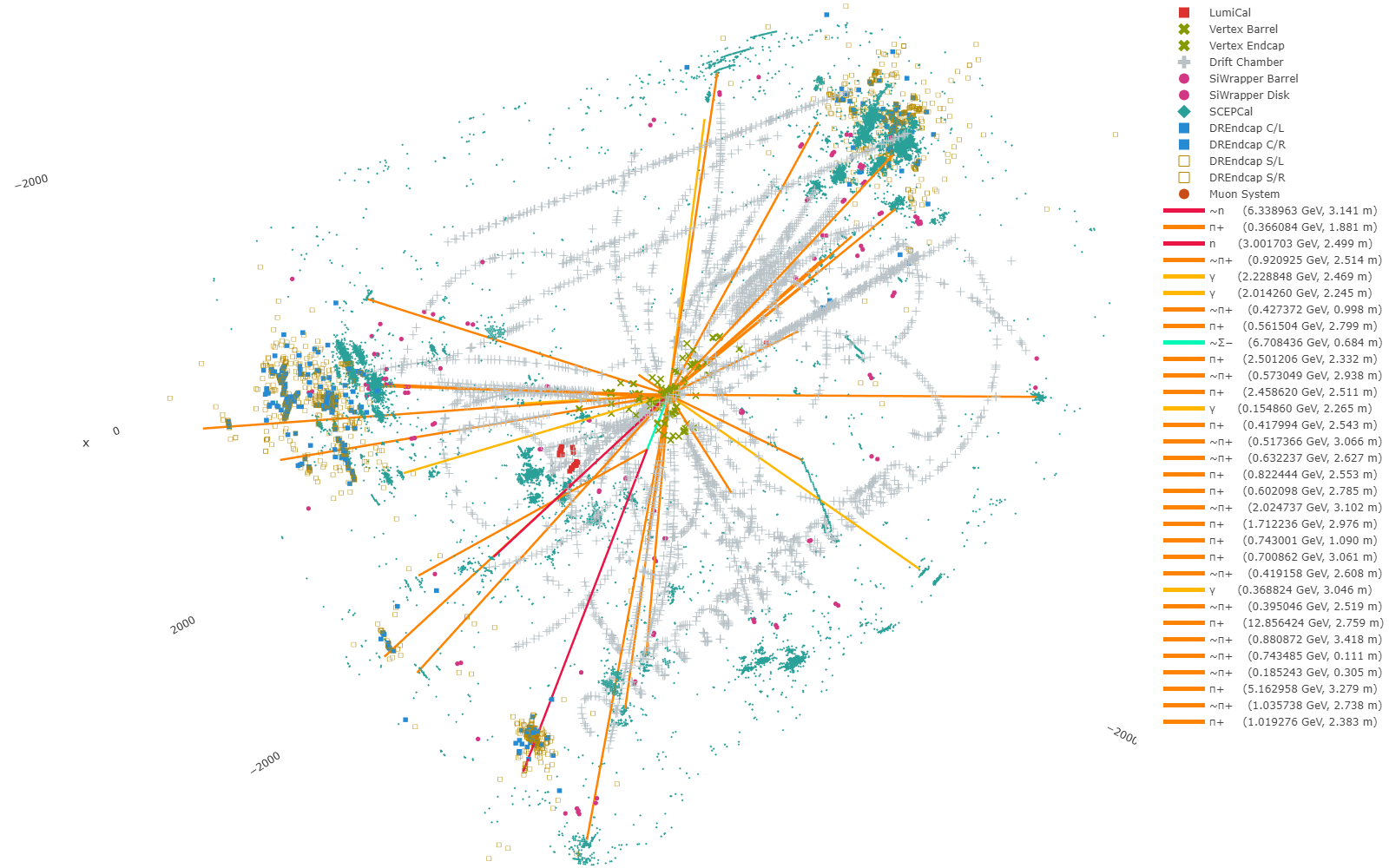}
    \includegraphics[width=\textwidth]{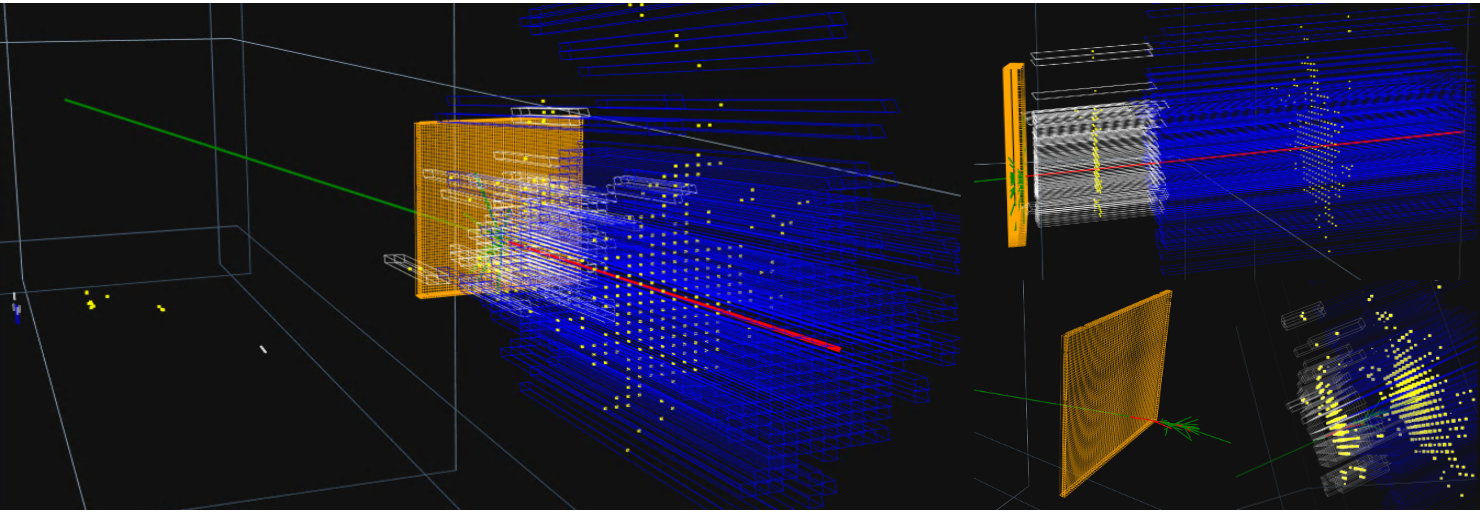}
    \caption{Event displays in plotly. (top) A 240\,GeV FCC-ee event in the full IDEA detector. (bottom) A 10\,GeV electron impinging on the SCEPCal.)}
    \label{eventdisplays}
\end{figure}

Event displays are in general difficult to produce for new detector designs because existing solutions typically require the export of the full ROOT geometry tree of the entire detector or have the detector description already built-in. The event display program itself must also be installed or hosted on a web server. Here we take advantage of modern JavaScript-based plotting libraries in python such as plotly to instead redraw the detector geometry elements for hits directly from the event readout file, circumventing the need for a separate display program and indeed able to run in an interactive python notebook. Since we have written the geometry calculations, we simply copy them and redraw the geometry element for a given hit, and place it in the correct position. The interactive nature of python also allows to select subsets of hits for a given event at will, such as selecting only timing layer hits, or front projective layer hits, etc. A selection of displays is shown in Figure ~\ref{eventdisplays}.

\section{Reconstruction in the AI/ML era}

Detectors at future colliders are widely expected to run real-time inference on front-end ASICs for so-called triggerless operation, and development of AI/ML reconstruction algorithms based on the latest neural network architectures is a fast-emerging area with wide interest~\cite{gravnet,mlpf}. In recent years the trend has been towards treating clusterization as an inner detail of overall reconstruction of unordered particle clouds~\cite{particlenet}, in contrast to the delineated approach towards jets used in cut- or rule-based Particle Flow algorithms (PFAs)~\cite{cmspfa}. 

A differentiable full simulation opens doors to a multitude of new types of studies in which granular components of the detector design can be optimized systematically against selected physics metrics. Optimizing geometry parameters against reconstruction algorithm parameters, which themselves are simultaneously being optimized, is surely a task for AI/ML. Even still, neural network architectures themselves are also the subject of intense research and are usually conceived and developed to solve specific classes of problems intuitively representing use-cases such as image recognition (convolutional neural networks) and natural language processing (transformers). In collider physics, recent work has tended toward graph neural networks for their ability to represent relationships between detector objects like calorimeter hits to physics objects and their kinematics. 

However, networks trained solely on static full simulations or parameterized fast-simulation models are inherently inefficient as these models lack insight into intrinsic detector characteristics that cannot be parameterized, such as angular resolution, material properties, and timing information. These approaches preclude bilevel optimization of reconstruction hyperparameters against geometry parameters.

\subsection{Picking the right neural network for the detector}

Particularly in calorimetry, the physical characteristics of a detector must be taken into account when considering the type of neural network architecture best suited for reconstruction of its readouts. Different detectors have different strengths, and hits from a silicon-tungsten sandwich calorimeter cannot be considered in the same class of objects as hits from a homogeneous, but slower and coarser, crystal apparatus, for example. The original class of problems for which a network architecture was designed should also be taken into account in view of its compatibility to the class of physics questions which are at hand in reconstruction.

A salient example is in the case of latent diffusion models~\cite{diffusion}, in which reconstruction from noise is a generative process differing fundamentally from the lineage of vector-token architectures like graph neural networks and transformers~\cite{transformers}. Recent extensions to diffusion models like ControlNet~\cite{controlnet}, which allows for spatial contextual-control of model outputs, evoke strong parallels to reconstruction tasks in pixelated sampling calorimeters, which have excellent mechanical resolution but high noise. One such implementation could use tracks as the spatial contextual-control against the stochastic structure of the noisy calorimeter hits, which the diffusion model has learned.

For calorimeters with granular longitudinal separation capabilities between EM showers and jets, such as with dual-readout crystals and fiber towers, additional handles such as precision timing information contribute to a long-range sequence of data suitable for use in a transformer architecture. Using tracks in the attention-step could potentially alleviate the short-range contextual limits that traditional graph neural networks are known to suffer from.

Different network architectures may also be combined within a single detector when applicable, such as in concepts like the Crilin detector~\cite{crilin}, which features both a sampling and homogeneous portion. A diffusion model could be used in the sampling portion, and a transformer model in the homogeneous section, and a generative adversarial network with a discriminator used to integrate the results of both networks into a final reconstructed object.

\section{Synthetic Training and Representation Bridging}

Reconstructing low-dimensional truth labels from high-dimensional experimental data is a central challenge in any scenario that relies on robust mappings across this so-called domain gap, from multi-particle final states in high-energy physics to large-scale early-universe structure in cosmological surveys.
We introduce a new method to bridge this domain gap with an intermediate, synthetic representation of truth that differs from methods operating purely in latent space, such as normalizing flows or invertible approaches, in that the synthetic data is specifically engineered to represent intrinsic detector hardware capabilities of the system at hand.
The hypothesis is that by encoding physical properties of the detector response available only in full simulation, such synthetic representations result in a less lossy compression and recovery than a direct mapping from truth to experimental data.

\begin{figure}[h!]
    \centering
    \includegraphics[width=\textwidth]{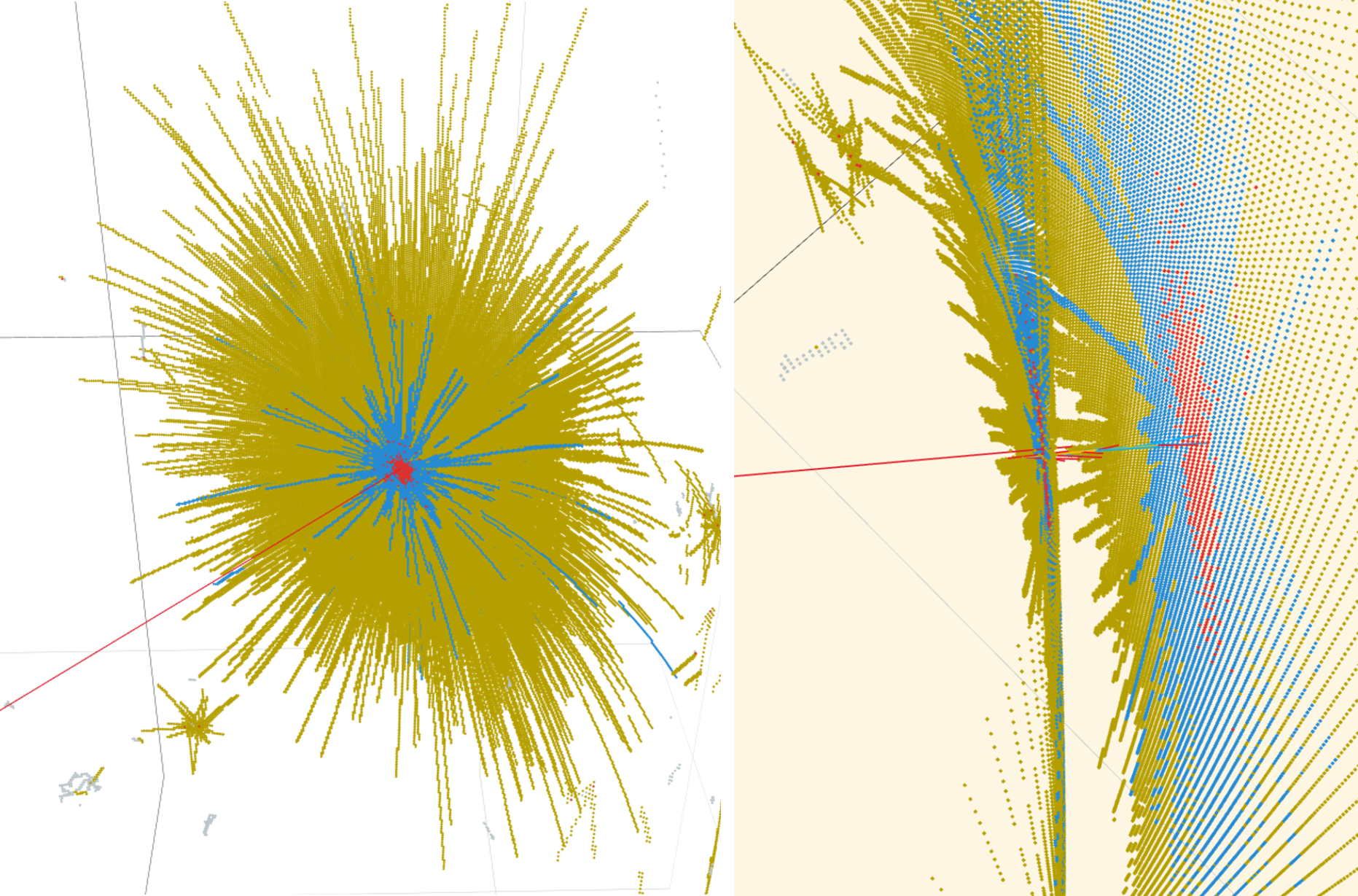}
    \caption{Example event display of the synthetic representation of a 50\,GeV electron event. Each dot is a single crystal hit. Yellow hits correspond to scintillation hit tracks, blue are Cerenkov. The red hits are those of a realistic detector hit, based on an energy deposit threshold of 1\,keV.}
    \label{fig:synthetic_hit}
\end{figure}

We demonstrate a first implementation of this concept with full simulation of a dual-readout crystal electromagnetic calorimeter for future collider detectors, in which the synthetic data is constructed to be the simulated detector hits corresponding to photon tracks of scintillation and Cerenkov photons.
We refer to these signals as simulated observables as they would not be physical observables in a real detector, but are nonetheless representations of a real physical process.
First results show that the synthetic representation naturally anchors the neural network architecture to a known physical method, in this case the dual-readout correction.
We believe this strategy opens new avenues for machinistic interpretability and explainability of ML-based reconstruction methods. In the case of anomalous signal detection, we hypothesize that anomalous signals detected in networks trained on synthetic data rooted in a physical process are more likely to be indicative of a genuinely physical anomaly.

\subsection{Introduction}
High-energy physics detectors produce vast, high-dimensional data from particle collisions.
In calorimeters alone, each event can generate hundreds to thousands of hit signals, encoding energy depositions in intricate spatial patterns.
Such complexity contrasts sharply with the relatively low-dimensional space of Monte Carlo (MC) truth labels—typically consisting of a particle type, momentum, and vertex—used for training reconstruction and classification algorithms.
This disparity in dimensionality often leads to degenerate mappings: multiple calorimeter hit patterns can correspond to the same nominal MC truth label.

While modern machine learning techniques excel at extracting robust features from high-dimensional data, their performance can still be limited by the fundamental mismatch between compact, idealized truth and the rich detector response.
Methods such as fast simulation with generative adversarial networks (GANs) or normalizing flows have attempted to learn transformations from particle-level truth to realistic calorimeter data, with varying degrees of success and fidelity to real detector effects~\cite{atlas_deepGAN,calogan,gan4hls}.
Likewise, invertible ML approaches (e.g., OmniFold)~\cite{omnifold,ml_landscape} aim to invert the detector response to recover underlying truth-level information.
Yet many of these frameworks implicitly treat MC truth as a low-dimensional label, leaving significant degeneracies when distinguishing topologically similar shower patterns.

We propose a new method to address the dimensionality gap and associated information loss.
Instead of directly learning the mapping between the low-dimensional truth space $\mathbf{T}$ and the high-dimensional detector space $\mathbf{D}$, we introduce a synthetic, intermediate representation of truth, denoted space $\mathbf{S}$.
This intermediate space is designed to be "closer" to the detector response space $\mathbf{D}$ in terms of dimensionality and information content.
Crucially, elements of $\mathbf{S}$ are constructed based on detector-specific hardware capabilities and are therefore representations of a real physical process that are visible only in full simulation.
The core hypothesis is that by factorizing the inference problem into two stages, $\mathbf{D}\rightarrow\mathbf{S}\rightarrow\mathbf{T}$, the overall process becomes less susceptible to information loss.
The mapping between the two high-dimensional spaces $\mathbf{D}\rightarrow\mathbf{S}$ is expected to be less degenerate and more information-preserving than the direct mapping $\mathbf{D}\rightarrow\mathbf{T}$.
The subsequent mapping $\mathbf{S}\rightarrow\mathbf{T}$ then operates on a richer, potentially better-conditioned feature space.

In this work, elements of $\mathbf{S}$ are instantiated as detector cell hits corresponding to the trajectories of optical Cerenkov (C) and scintillation (S) photons within longitudinally segmented dual-readout calorimeter crystals. These would not be real observables in a real detector, but here we refer to them as simulated observables, as they nonetheless represent a true physical process.
This approach encodes the intrinsic physics principles of dual-readout calorimetry as well as the segmented geometry of the detector into the space $\mathbf{S}$. The intent is to surface the hidden features of shower development in calorimetry which underpin reconstruction algorithms but are not directly observable, such skin-depth and fine-grained shower structure.

\begin{figure}[h!]
    \centering
    \includegraphics[width=0.6\textwidth]{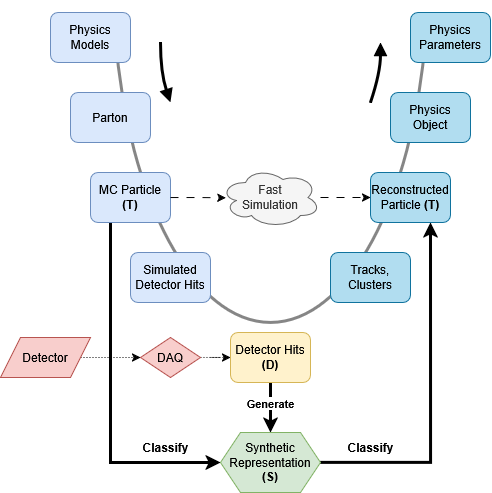}
    \caption{The U-Net structure of the traditional detector simulation chain is shown with the conceptual extension of the synthetic space $\mathbf{S}$. Starting from physics models, event generators and hadronization programs produce the MC particles ($\mathbf{T}$) to be simulated in a detector. Simulated detector hits are digitized to the space of realistic detector hits $\mathbf{D}$ produced by a real detector and DAQ. Raw detector hits typically undergo a process of track forming and clusterization for particle-flow algorithms to process them into reconstructed particles ($\mathbf{T}$) from which physics objects and parameters are extracted. The U-Net name comes from the process of folding dimensionality from one layer to the next, up to a so-called bottleneck layer (in this case, the detector hits $\mathbf{D}$), an arbitrary point at which global context is aggregated and skip-connections reconstruct features at matching scales on the way back up to the original domain (or equivalent). The synthetic representations $\mathbf{S}$ sit at a level below the detector hits and are engineered to encode additional information about the detector response available only in full simulation. The core hypothesis is that by basing the synthetic representation on intrinsic detector capabilities and known physics processes, the mapping $\mathbf{S}\rightarrow\mathbf{T}$, while still bridging a large dimensional-gap, is better-conditioned and less lossy than the direct connection $\mathbf{D}\rightarrow\mathbf{T}$.}
    \label{fig:detector_simulation_chain}
\end{figure}

The primary goal of this initial work is to establish that a machine learning model can successfully learn the mapping from realistic, observable detector hits ($\mathbf{D}$, comprising energy deposits above threshold and S/C photon counts) to the synthetic, intermediate representation based on S/C optical photon tracks. A U-Net architecture is employed for this image-to-image translation task. Qualitative results are presented, suggesting that the learned mapping preserves and reconstructs not only fine-grained topological information present in the synthetic representation, but does so according to the principles of dual-readout correction that underlie the construction of the synthetic representation. We interpret these results as being significant for machinistic interpretability of reconstruction algorithms and their physical context.

\subsection{Methodology}

Denote the three spaces as follows:
\begin{itemize}
    \item \( \mathbf{T} \in \mathbb{R}^{d} \): the traditional MC truth label space (e.g., particle type, momentum, vertex).
    \item \( \mathbf{S} \in \mathbb{R}^{m} \): the synthetic intermediate representation space, generated by a hardware-specific detector response function
    \item \( \mathbf{D} \in \mathbb{R}^{n} \): the space of realistic detector hits likely to be observed in a real experiment, with \( n \gg d \) and typically \( m \gtrsim n \) or at least \( m \gg d \).
\end{itemize}
 We train two neural networks:
 \begin{itemize}
    \item \textbf{NN1:}  \( f: \mathbf{D} \to \mathbf{S} \). This network is a forward mapping trained on pairs \( (D_i, S_i) \) to predict a synthetic intermediate representation from a given set of realistic hits. Architecturally, this can be implemented as an encoder-decoder model or a pair of generative/discriminative networks (e.g., using adversarial objectives).
    \item \textbf{NN2:}  \( g: \mathbf{S} \leftrightarrow \mathbf{T} \). This network is a forward-backward mapping trained on pairs \( (S_i, T_i) \) to perform classification and/or regression to learn and recover a traditional MC truth from a synthetic representation.
\end{itemize}
Given a new set of detector hits $D_j$, $\mathrm{NN}_1$ infers its synthetic representation $S_j$. $\mathrm{NN}_2$ then classifies $S_j$ to yield the final label $T_j$.
By factoring the mapping $\mathbf{D}\to \mathbf{T}$ through $\mathbf{S}$, we aim to anchor the convolutions in latent space to a detector-specific physical process.
Formally, given a new detector hit collection $D_1$, we compute
$$
 S_1 = f(D_1) \quad \text{and then} \quad T_1 = g(S_1), 
$$
so that the composite mapping $g\circ f: \mathbf{D} \to \mathbf{T}$ is used for particle identification or other reconstruction tasks.
 
Let $L_T$ and $L_S$ be appropriate loss functions (e.g., cross-entropy for classification, mean-squared error for regression). The two training objectives are:
$$
 \mathcal{L}_{S} = \mathbb{E}_{(D,S)}\left[ L_S\bigl(f(D),\,S\bigr) \right],
$$
$$
 \mathcal{L}_{T} = \mathbb{E}_{(S,T)}\left[ L_T\bigl(g(S),\,T\bigr) \right]. 
$$
The overall loss (when considering the full pipeline) is
$$
 \mathcal{L}_{\text{total}} = \mathbb{E}_{(D,S,T)}\left[ L_T\bigl(g(f(D)),\,T\bigr) \right] + \lambda\,\mathcal{L}_{S}, 
$$
where $\lambda$ is a hyperparameter balancing the two losses. Although we presently train the networks in two stages, an end-to-end training scheme is also conceivable.

\subsection{Detector Description and Differentiable Full Simulation}
The simulations in this study were performed using a custom full detector simulation implemented in the key4hep software ecosystem for future collider detectors.
The detector geometry and response functions are implemented in DD4hep, a modularized wrapper around the Geant4 simulation package for particle interactions with matter.
A custom edm4hep data class is used for the synthetic data.
This framework provides a flexible and comprehensive environment for simulating detector concepts proposed for future collider experiments.

\begin{figure}[h!]
    \centering
    \includegraphics[width=\textwidth]{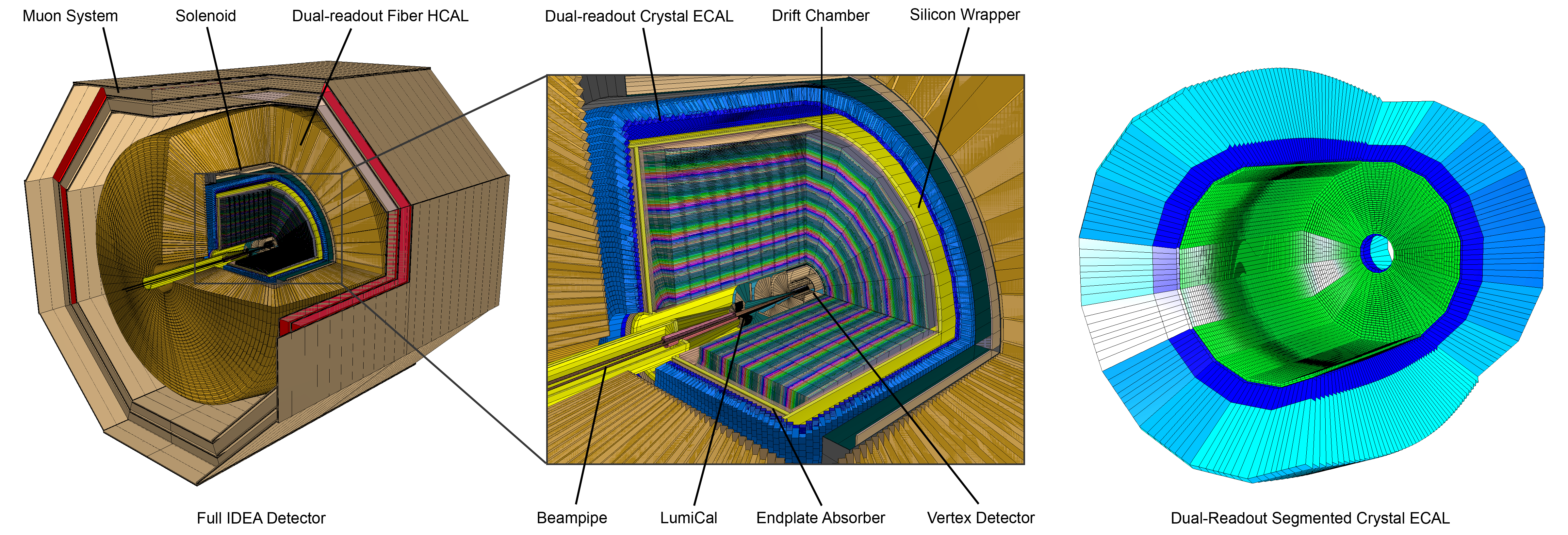}
    \caption{Left and middle panes: The full IDEA detector showing all subdetectors with labels. Right pane: isolated view of only the dual-readout segmented crystal electromagnetic calorimeter used for this study. Crystal dimensions are enlarged by a factor of 10x for visibility.}
    \label{fig:IDEA_SCEPCal}
\end{figure}

\begin{figure}[h!]
    \centering
    \includegraphics[width=\textwidth]{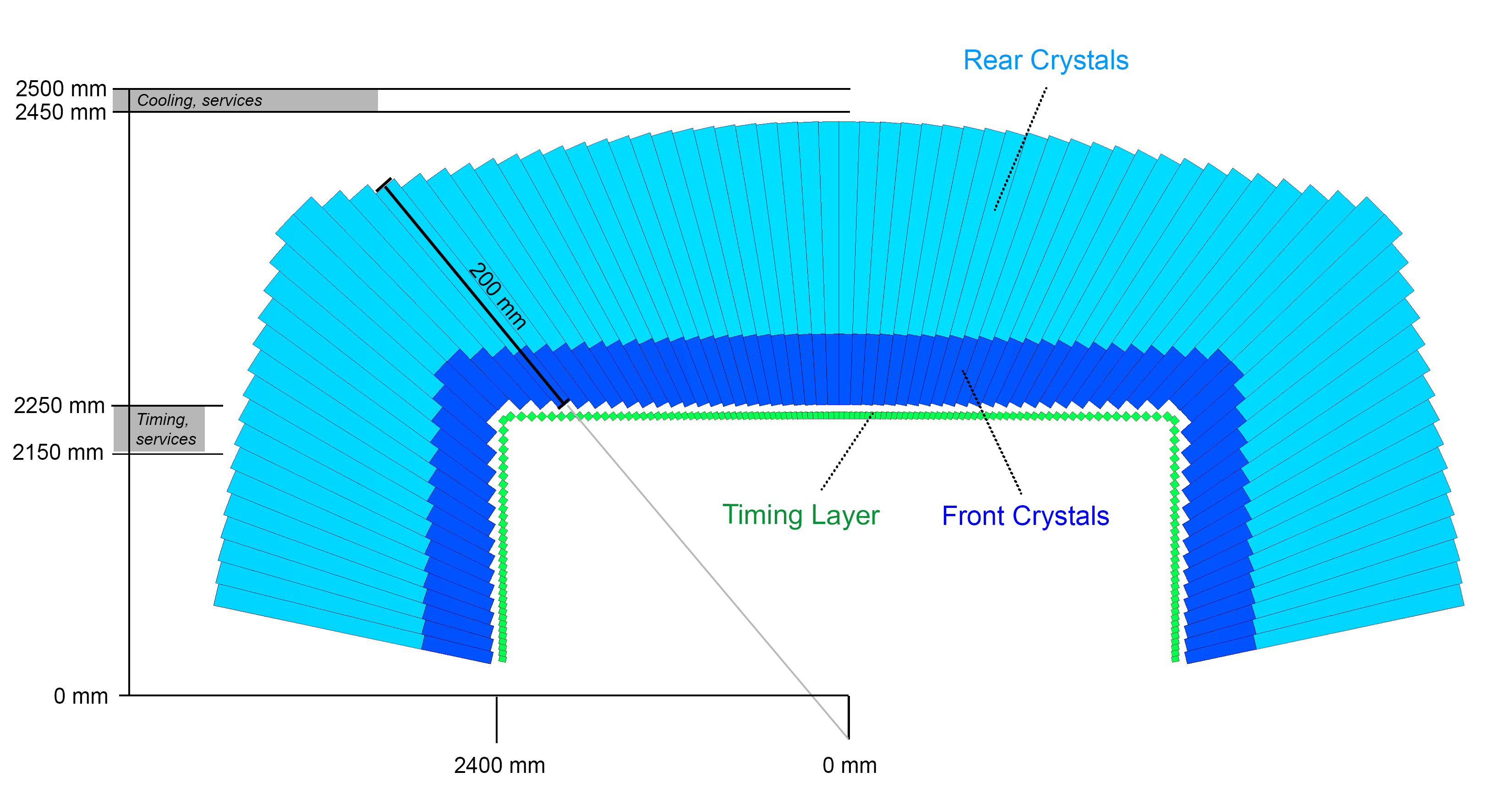}
    \caption{A single phi-slice of the dual-readout segmented crystal calorimeter used for this study. The main crystal layer (comprising front and rear crystals) is instrumented with a projective offset to point away from the interaction point as shown, to mitigate particle leakage through projective cracks. To maximize angular resolution, the timing layer has no projective offset. The timing layer is not used for this study.}
    \label{fig:scepcal_phislice}
\end{figure}

The detector is a dual-readout, homogeneous crystal electromagnetic calorimeter set in a projective geometry. A single phi-slice of the detector is shown in Figure~\ref{fig:scepcal_phislice}.
The crystals are longitudinally segmented into separate front and rear crystals with separate readout channels for scintillation and Cerenkov light. 
The nominal transverse granularity of the crystals is 1x1\,cm, with the front and rear crystal lengths in a ratio of 6:16 radiation lengths in depth for a total of 22 radiation lengths for full EM shower containment.
The crystal material for this study is lead tungstate, used in the CMS experiment for its short radiation length and moderate scintillation yield.
At a future $e^+e^-$ collider, this ECAL is designed to be paired with the IDEA dual-readout fiber calorimeter acting as the hadronic calorimeter (HCAL) and is part of the new IDEA baseline design shown in Figure~\ref{fig:IDEA_SCEPCal}.
When paired with the IDEA HCAL, the ECAL was shown in a previous work~\cite{Lucchini_2020} to significantly improve hadronic jet resolution, alongside providing excellent intrinsic EM resolution.
A precision timing layer providing picosecond time resolution is also envisioned to be instrumented in front of the main crystal layer, assumed to be made of a fast scintillating crystal such as LYSO, although LYSO is more suited for a proton-proton collider environment such as at the LHC.
The timing layer is not used for this study and the choice of material is under study.
The main projective layer mitigates projective cracks with a radial pointing offset around the interaction point similar to the CMS detector. The timing layer does not use a projective offset to maximize angular resolution.

The detector geometry is written to be dynamically reconfigurable for use in a differentiable full simulation pipeline, meaning the detector construction has no hard-coded parameters and is fully parameterized to a few input parameters such as the overall dimensions and nominal crystal granularities, including the option to bundle different numbers of front and rear crystals together in a single crystal tower.
The individual dimensions and projective placements of the crystals are then automatically calculated.
The number of front to rear crystals per tower is 1:1 in this study.

To our knowledge, this simulation is the first fully dynamic and reconfigurable geometry written for a collider detector and enables systematic studies of detector geometry optimizations alongside reconstruction algorith parameters in a process called bilevel optimization.
More details of the detector construction can be found in~\cite{calor2024}.

Optical surface definitions are implemented on the crystals and amount to defining the reflection and transmission coefficients of the material.
However, the effect on the simulated outputs was found to be minimal at the expense of a significant increase in compute, so for this study the optical surface effects were turned off.

The standard output of a DD4hep simulation is edm4hep calorimeter hits, which store accumulated energy deposit per event in each detector cell, or crystal.
For dual-readout calorimetery, we are interested in optical photons, so we additionally implement a custom readout class to save the number of scintillation and Cerenkov photons produced per event in each detector cell, representing the realistic detector response in the spacee $D$.

\subsection{Synthetic Representation of Detector Response}
An important capability of full simulation which has traditionally been overlooked is the possibility to customize the detector response function, which runs at every step of every particle in the simulation and contains the logic of when to create and save a detector hit.
The typical approach is to save hits based on the energy deposited in the material at each step of the simulation, subject to a threshold value, usually 1\,keV, reflecting the finite sensitivity and noise thresholds of real sensors.

In addition to the counts of S/C photons produced per crystal, we implement a custom detector response function to generate a synthetic detector response in the space $S$.

Instead of applying an energy-deposit cut at the step-level, we apply a cut at the track-level for optical photons based on the wavelength of the photon, registering a detector hit for photons with a wavelength within the specified range, even if the energy deposited in a particular step is zero, effectively saving the tracks of optical photons through the detector as blank, zero-energy hits.
In this study we use the wavelength range 200-600\,nm in accordance with the scintillation and cerenkov production wavelengths of lead tungstate.

\begin{figure}[h!]
    \centering
    \includegraphics[width=\textwidth]{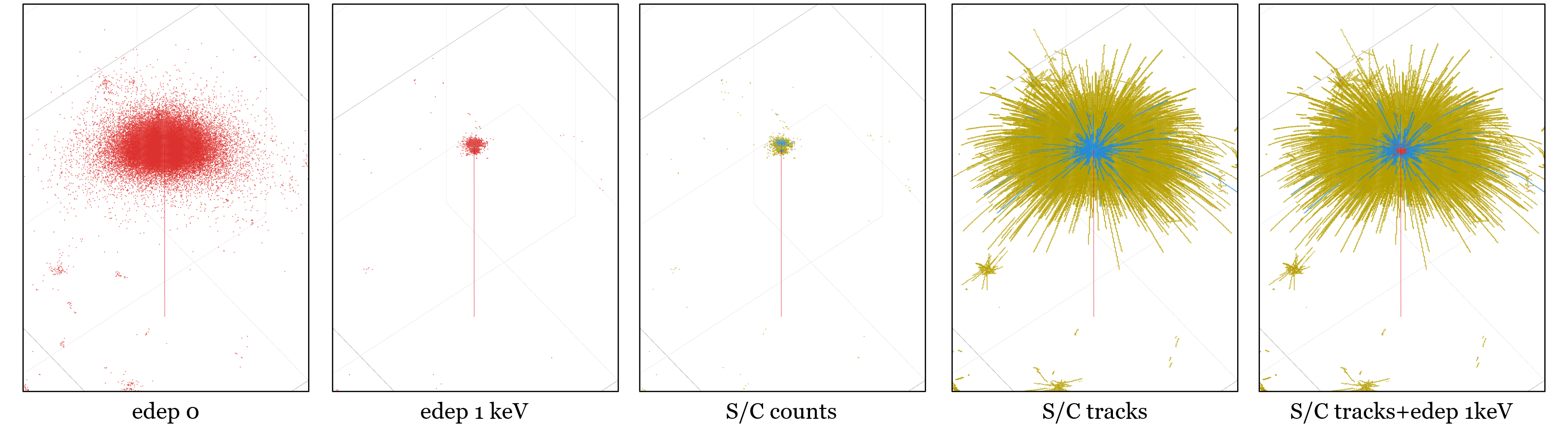}
    \caption{Event displays for a single 50\,GeV electron showing detector hits for realistic and synthetic detector response functions. Each point is a single crystal hit. From left to right: edep0 results in a very dense cloud of hits as every step with non-zero energy deposit is saved. Using edep 1\,keV or non-zero S/C photon counts results in a much more concentrated cloud of hits and is most representative of a realistic detector response. Next, saving all hits belonging to a S/C photon track, even if the energy deposit is zero, produces an image with rich geometrical structure encoding properties of the detector interaction visible only in full simulation. Finally, superimposing edep 1\,keV on top of S/C tracks illustrates the dimensional-gap we propose to bridge.}
    \label{fig:detector_responses}
\end{figure}

The event displays in Figure~\ref{fig:detector_responses} illustrate the core idea of this work, which is that the synthetic representations offer a rich source of structural information which are unphysical in the sense that one would not ever see them in a real detector, but they are not entirely unphysical in that they are representations of a true physical process tied to the physical capabilities of the detector hardware, in this case, dual-readout and longitudinal segmentation. 

The hypothesis is that because this synthetic representation is grounded in the detector hardware capabilities rather than existing solely in abstract latent space, it can act as a less lossy bridge between the realistic detector hits $D$ and the truth labels $T$. The mapping $T\rightarrow S$ is effectively a transformation of the labels into the space of pure detector response, where particles are no longer described by a particle type and momentum, but rather as rich geometrical structures of detector hits in a synthetic space only available in full simulation.

\subsection{Event Generation and Image Encoding}

Single particle events are run using the built-in DD4hep/Geant4 particle gun. We simulate 10,000 events each of electrons, photons (gammas), $\pi^0$, $\pi^+$, $\pi^-$, and neutrons, all emitted at a fixed momentum of 50 GeV from the interaction point with no smearing, in a uniform angular distribution in the barrel of the detector only. We determine 10,000 events to be a sufficient sample size for this study because single-particle events have a high degree of self-similarity and the intent is to first establish a handle in hyperparameter space and perform qualitative examinations of the method.

\begin{figure}[h!]
    \centering
    \includegraphics[width=\textwidth]{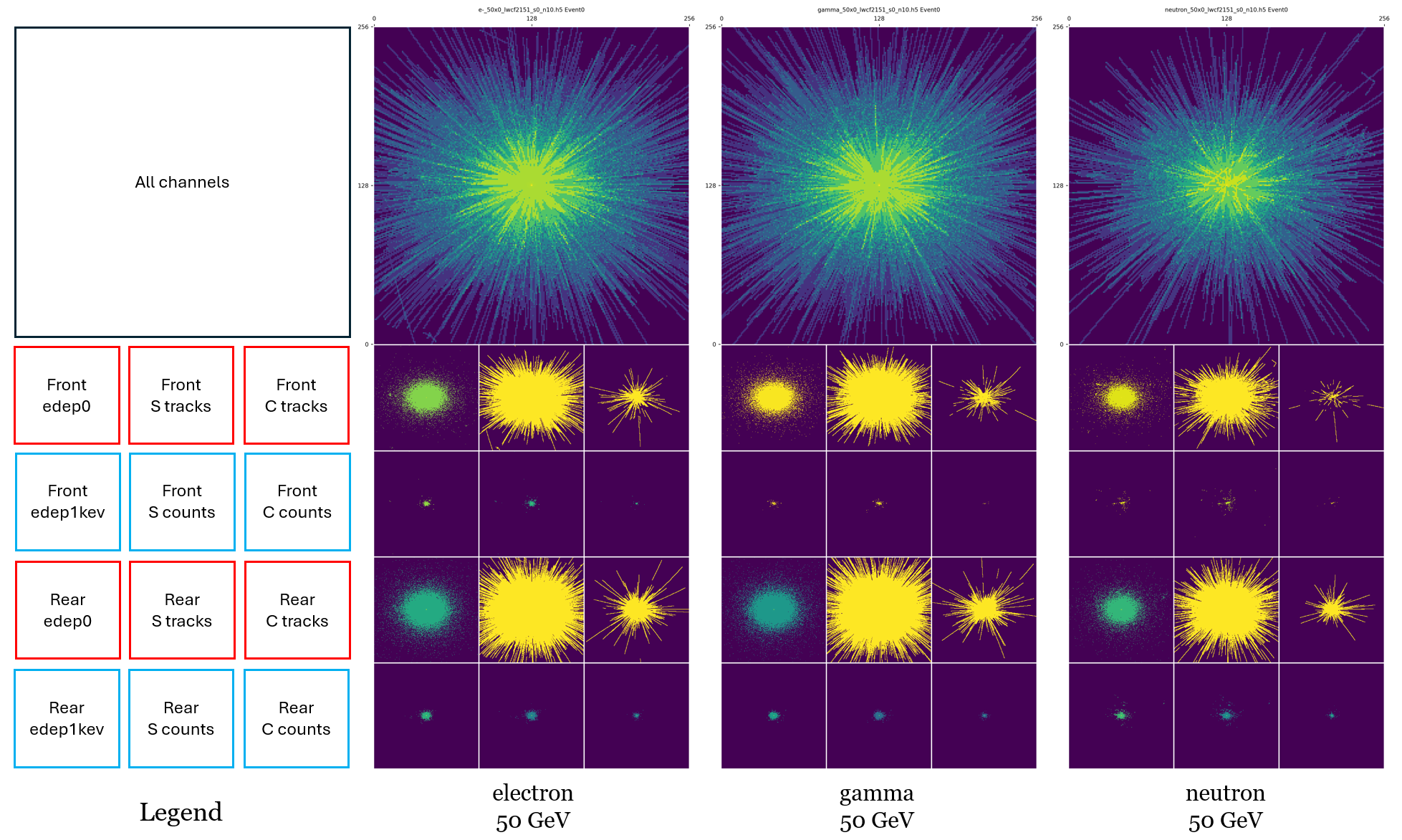}
    \caption{Representative samples of constructed images for 50\,GeV electrons, gammas, and neutrons. The sum of all channels per pixel is displayed in the top pane for each particle type, and the 12 channels shown separately below, corresponding to the legend at left. The channels boxed in red in the legend are those considered synthetic, and blue channels represent realistic detector responses. Compared to the realistic channels, the synthetic channels, S/C tracks in particular, offer a rich geometrical structure which encode aspects of detector response visible only in full simulation.}
    \label{fig:image_gen_egn}
\end{figure}

\begin{figure}[h!]
    \centering
    \includegraphics[width=\textwidth]{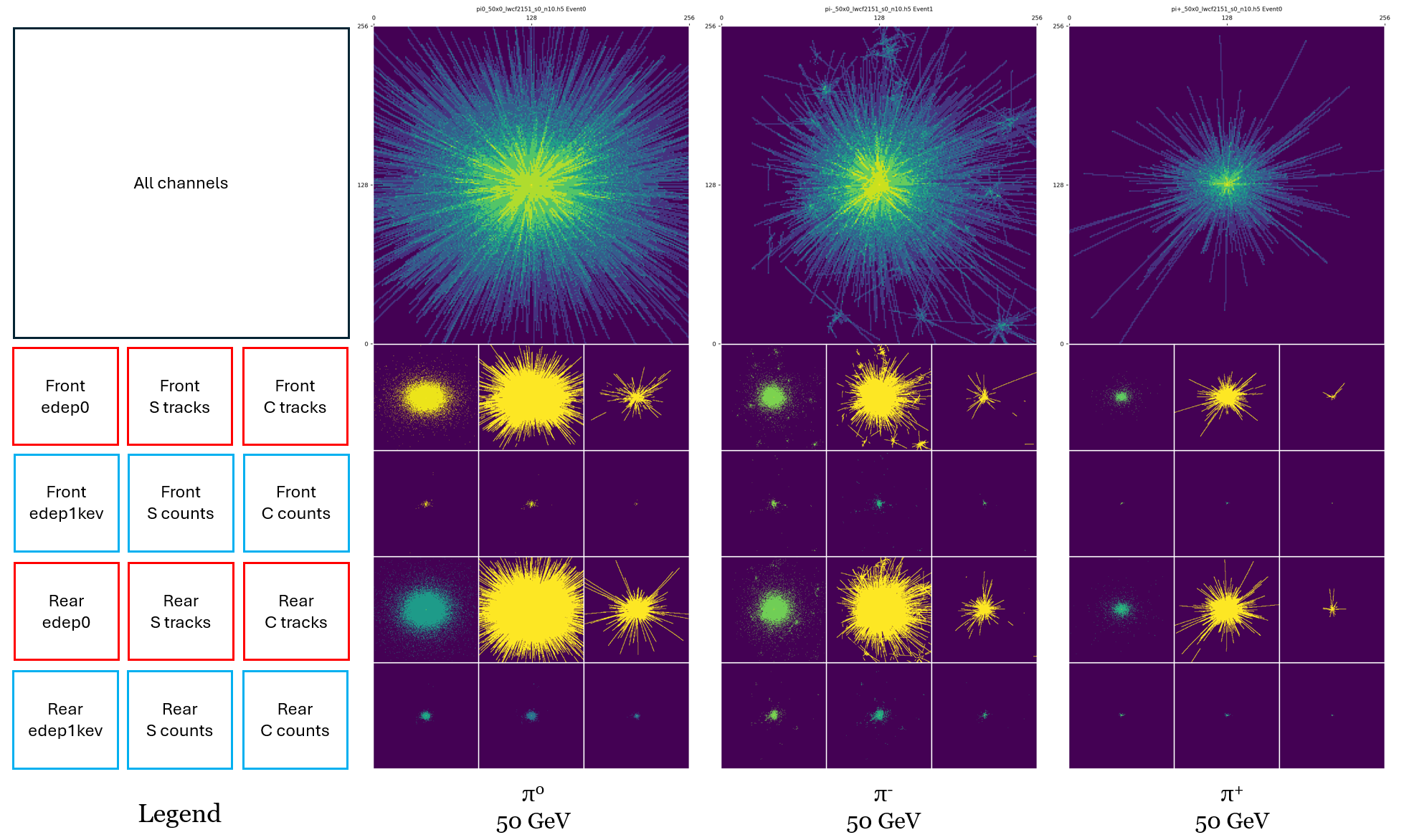}
    \caption{Representative samples of constructed images for 50\,GeV neutral and charged pions. Neutral pions typically decay almost immediately into two photons, and resolving the signature of two overlapping photons underpins $\pi^0$/gamma separation. In contrast, charged pions may shower early or late in the ECAL. In these samples, the $\pi^-$, middle, exhibits a typical pattern of early-showing, with multiple secondaries around the main island clearly visible. The $\pi^+$, right, is a typical late-shower, with a noticeably sparse scintillation signal.}
    \label{fig:image_gen_pions}
\end{figure}

A 12-channel image is constructed for each event using the following procedure.
First, the hit with the maximum energy deposit in the event is identified and a breadth-first-search starting from this hit is conducted to collect the set of directly neighboring detector hits, including depth, comprising the "island" of hits originating from the maximum-energy hit.
A fixed-size 256\,x\,256 pixel image is then constructed for the island of hits, centered around the maximum-energy hit, with each pixel representing a single crystal tower of one front and one rear crystal. Each pixel is encoded with 12 channels of data, 6 for the front crystal and 6 for the rear crystal. The channel descriptions and values are listed in Table~\ref{tab:image_channels}. Representative samples of images constructed for electrons, gammas, and neutrons are shown in Figure~\ref{fig:image_gen_egn}, and neutral and charged pions in Figure~\ref{fig:image_gen_pions}.

\begin{table}[h!]
\centering
\caption{Channel encoding scheme for each event image. $E$, $S$, and $C$ are the sum of the energy deposits and S/C photon counts for the island of hits. $e_i^0$, $e_i^{\text{1 keV}}$, $s_i$, and $c_i$ are the energy deposits (using edep0 and edep1keV) and S/C photon counts for the single crystal at each pixel. A log-weighted formula, which compresses small signals, is used to normalize the dynamic range of the energy deposits and S/C counts, with $s_e$, $s_p$, $w_e$, $w_p$ as scaling and offset factors. In this study $s_e=2$, $s_p=5$, and $w_e=w_p=1$. An alternative weighting scheme such as a square-root function or arcsine may be used to preserve signal linearity, however to first-order, the choice of weighting function is not expected to be significant for a qualitative assessment of the method.}
\label{tab:image_channels}

\begin{tabularx}{\textwidth}{lll}
\toprule
    Channel & Description & Value \\ 
    \hline 
    \midrule
    1  & Front crystal edep0 & $s_e\,\log(e_i^0/E+1)+w_e$ \\
    2  & Front S track hits  & 1 if present, 0 otherwise \\
    3  & Front C track hits  & 1 if present, 0 otherwise \\
    4  & Front edep1 keV     & $s_e\,\log(e_i^{\text{1 keV}}/E+1)+w_e$ \\
    5  & Front S counts      & $s_p\,\log(s_i/S+1)+w_p$ \\
    6  & Front C counts      & $s_p\,\log(c_i/C+1)+w_p$ \\
    7  & Rear crystal edep0  & $s_e\,\log(e_i^0/E+1)+w_e$ \\
    8  & Rear S track hits   & 1 if present, 0 otherwise \\
    9  & Rear C track hits   & 1 if present, 0 otherwise \\
    10 & Rear edep1 keV      & $s_e\,\log(e_i^{\text{1 keV}}/E+1)+w_e$ \\
    11 & Rear S counts       & $s_p\,\log(s_i/S+1)+w_p$ \\
    12 & Rear C counts       & $s_p\,\log(c_i/C+1)+w_p$ \\
    \hline 
    \bottomrule
\end{tabularx}
\end{table}

\subsection{First Implementation: A 3-level U-Net}

For a first test of this idea we choose a U-Net architecture to implement the image-to-image translation $D\rightarrow S$, extending the traditional structure of the detector simulation chain as shown in Figure~\ref{fig:detector_simulation_chain}.

We implement a 3-level U-Net to train over the 10,000 images for each particle in various combinations (just electrons, or electrons and gammas, etc.). The training loop makes 2 copies of each image: full and masked. The masked image zeros-out the synthetic channels and the model is trained to infer the missing synthetic channels when given an image with only the realistic channels present. The loss function used is a weighted L1+SSIM loss, which, concisely, considers a combination of absolute pixel difference (L1) and structural correlations (SSIM) within the image. We train over a modest hyperparameter space with batch sizes 2, 4, 8, or 16, with an initial learning rate of $1e-3$ with a scheduler halving the rate to a minimum of $1e-7$ with a patience of 5. The training loop is run for 500 epochs. Then we run inference to generate images for a test set of 1000 images for each particle type and classify them.

\subsection{Inference: First Look}
\begin{figure}[h!]
    \centering
    \includegraphics[width=\textwidth]{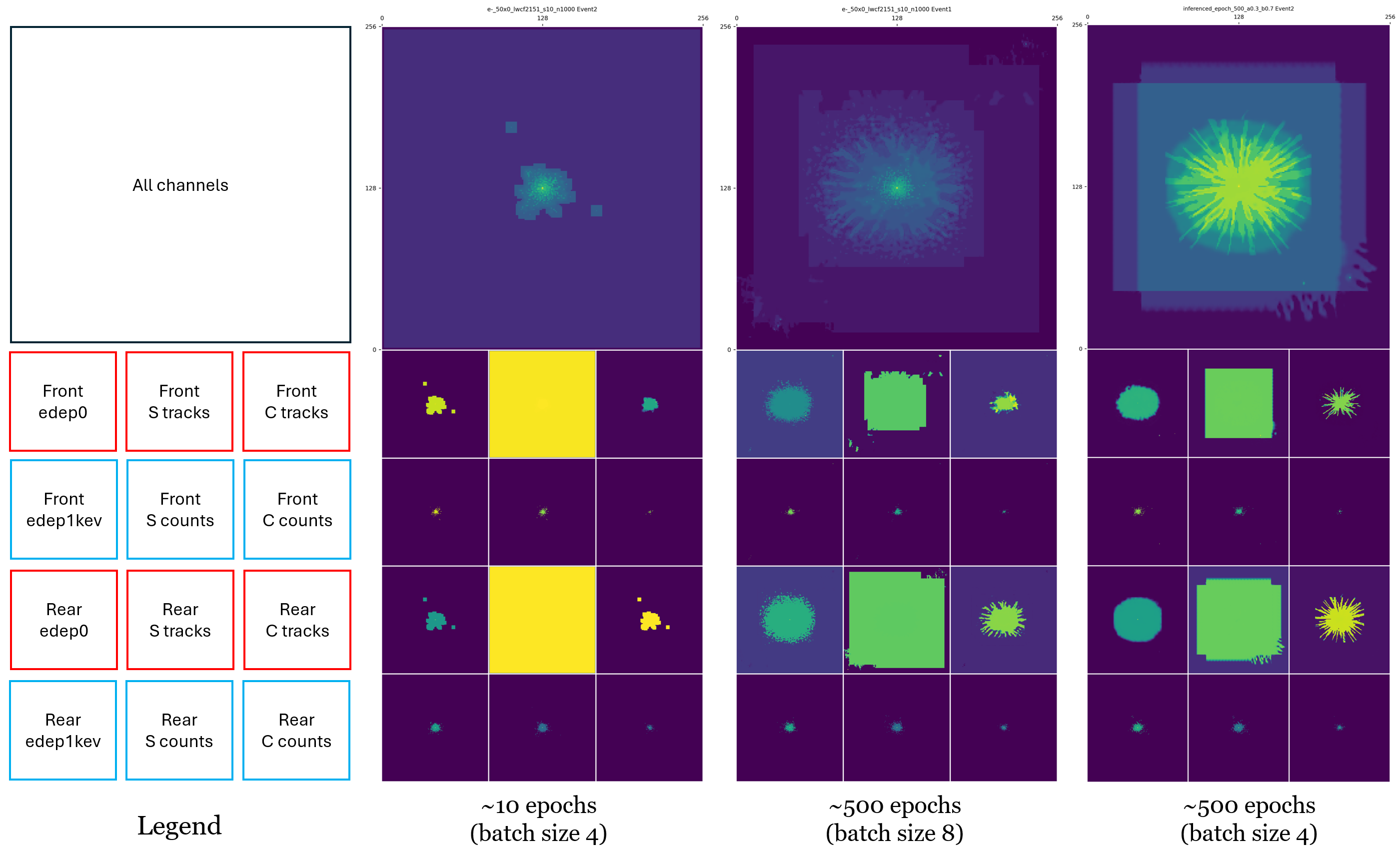}
    \caption{Inference images of a 50\,GeV electron after various training epochs and batch sizes. See text.}
    \label{fig:inference_firstlook}
\end{figure}

A sample of inferenced images for various hyperparameters are shown in Figure~\ref{fig:inference_firstlook}.
While tuning hyperparameters is clearly the challenge for a production-grade model, here we discuss qualitative assesments of the model's progression and interpret its output.

The performance of the model appears to depend most significantly on the batch size, i.e. the number of simultaneous images over which the model attempts to calculate a gradient and apply the loss function.
As noted previously, the constructed images have an inherently high degree of self-similarly because they effectively encode the stochastic differences in detector response from the same type of single-particle event.
The small structural differences we are interested in discriminating are likely to be washed out if a large batch size is used. Conversely, using the minimum batch size of 2 would in principle be most effective at discerning fine, pixel-level differences, however the tradeoff here is that the model may become stuck in the local extrema of these features and never converge to the global context.
Using a modest batch size of 2-4 with a more robust scheduler is likely to be the most direct path to improvement.

Nonetheless, we observe that the cerenkov signal begins to be resolved first across all variations, a robust indication of the model's explainability.
In the context of image-to-image translation, this is also an expected result because the Cerenkov signal is in general much more sparse than the scintillation signal and therefore, assuming all channels of the image are weighted equally in the loss, one would expect the model to pick up on the gradient landscape of the Cerenkov signal during feedforward and backpropagation more quickly than the scintillation signal.

Conversely, the scintillation signals already nearly saturate the available canvas before being zeroed-out, and subsequently, in early epochs, appear as a large uniform square, reflecting the difficulty of feature resolution in this regime.
In later epochs we observe structure beginning to emerge in the scintillation signal from the outside-in, indicating that the radial structure of the tracks, in particular outlier "hair"-like protrusions, offers the most discriminating power.
Given the negative effects of a saturated signal, we note that it may be advantageous to attentuate the scintillation signal for this type of analysis, contrary to the traditional wisdom in non-dual-readout systems that that more signal is always better. 

Our interpretation of these results is that the model is effectively machine-learning the dual-readout correction factor of the detector.
In a more traditional analysis, one would calibrate the detector on known electromagnetic and hadronic processes separately to obtain the respective S/C scaling response factors, then use these to determine the EM fraction of the shower, mitigating the event-by-event hadronic shower fluctuations seen in non-dual-readout calorimeters.
By locking on to the sparse Cerenkov signal first while the more degenerate scintillation signal is resolved slowly, we interpret the model to be implicitly performing the dual-readout correction.
Because we have deliberately engineered the synthetic data to represent the known physical process of dual-readout rooted in the specific detector geometry, we show for the first time that a simulated observables can augment real observables in an interpretable and explainable way. We posit that in systems engineering in such a fashion, anomalous ML signals may be more likely to genuinely represent physically anomalous signals.

\subsection{Outlook}
We have introduced a new method of encoding simulated observables into a space of synthetic representations based on intrinsic hardware capabilities and real physical processes.
The concept is demonstrated to be interpretable in machine learning models with respect to an established instrumentation technique, in this study, dual-readout calorimetry.
We expect the core idea introduced here to be applicable to a wide range of experiments that rely on high-statistics Monte Carlo simulations mapping low-dimensional truth-labels to high-dimensional, complex detector responses.

For the image-to-image translation task of this specific work we have used a U-Net model as a natural and easy conceptual extension of the detector simulation chain.
In general, any suitable architecture may be used as long as the motivations underlying the synthetic response are grounded in real physical processes and detector capabilities.
For image generation, denoising latent diffusion models are an attractive next avenue for testing and offer distinct advantages with respect to the possibility of encoding detector systematics directly into the model.
Although timing information was not used in this study, vector sequencing methods such as GNNs or transformers may be used in a similar way in studies using timing information or other dimensional information specific to the use case.

\printbibliography

@article{25years,
  author  = {Wigmans, Richard},
  title   = {{25 Years of Dual-Readout Calorimetry}},
  doi     = {10.3390/instruments6030036},
  journal = {Instruments},
  volume  = {6},
  number  = {3},
  pages   = {36},
  year    = {2022}
}

@article{Alfven1940,
  title   = {{On the motion of a charged particle in a magnetic field}},
  author  = {Alfv\'en, Hannes},
  journal = {Ark. Mat. Astron. Fys.},
  volume  = {25B},
  pages   = {29},
  number  = {1--20},
  year    = {1940}
}

@article{angelo2022,
  title         = {Heisenberg's uncertainty principle in the PTOLEMY project: A theory update},
  author        = {Apponi, A. and Betti, M. G. and Borghesi, M. and Boyarsky, A. and Canci, N. and Cavoto, G. and Chang, C. and Cheianov, V. and Cheipesh, Y. and Chung, W. and Cocco, A. G. and Colijn, A. P. and D'Ambrosio, N. and de Groot, N. and Esposito, A. and Faverzani, M. and Ferella, A. and Ferri, E. and Ficcadenti, L. and Frederico, T. and Gariazzo, S. and Gatti, F. and Gentile, C. and Giachero, A. and Hochberg, Y. and Kahn, Y. and Lisanti, M. and Mangano, G. and Marcucci, L. E. and Mariani, C. and Marques, M. and Menichetti, G. and Messina, M. and Mikulenko, O. and Monticone, E. and Nucciotti, A. and Orlandi, D. and Pandolfi, F. and Parlati, S. and Pepe, C. and P\'erez de los Heros, C. and Pisanti, O. and Polini, M. and Polosa, A. D. and Puiu, A. and Rago, I. and Raitses, Y. and Rajteri, M. and Rossi, N. and Rozwadowska, K. and Rucandio, I. and Ruocco, A. and Strid, C. F. and Tan, A. and Teles, L. K. and Tozzini, V. and Tully, C. G. and Viviani, M. and Zeitler, U. and Zhao, F.},
  collaboration = {PTOLEMY Collaboration},
  journal       = {Phys. Rev. D},
  volume        = {106},
  issue         = {5},
  pages         = {053002},
  numpages      = {14},
  year          = {2022},
  month         = {Sep},
  publisher     = {American Physical Society},
  doi           = {10.1103/PhysRevD.106.053002},
  url           = {https://link.aps.org/doi/10.1103/PhysRevD.106.053002}
}

@article{atlas_deepGAN,
  collaboration = {ATLAS Collaboration},
  title         = {{Deep Generative Models for Fast Photon Shower Simulation in ATLAS}},
  archiveprefix = {arXiv},
  eprint        = {2210.06204},
  reportnumber  = {CERN-EP-2022-185},
  journal       = {Comput. Softw. Big Sci.},
  volume        = {8},
  pages         = {7},
  year          = {2024},
  url           = {https://cds.cern.ch/record/2836604},
  doi           = {10.1007/s41781-023-00106-9}
}

@article{Beamson1980Collimating,
  author               = {Beamson, G. and Porter, H. Q. and Turner, D. W.},
  day                  = {01},
  doi                  = {10.1088/0022-3735/13/1/018},
  issn                 = {0022-3735},
  journal              = {Journal of Physics E: Scientific Instruments},
  month                = dec,
  number               = {1},
  pages                = {64},
  title                = {{The collimating and magnifying properties of a superconducting field photoelectron spectrometer}},
  url                  = {http://dx.doi.org/10.1088/0022-3735/13/1/018},
  volume               = {13},
  year                 = {1980}
}

@article{betti2019design,
  title     = {A design for an electromagnetic filter for precision energy measurements at the tritium endpoint},
  author    = {Betti, MG and Biasotti, M and Bosc{\'a}, A and Calle, F and Carabe-Lopez, J and Cavoto, G and Chang, C and Chung, W and Cocco, AG and Colijn, AP and others},
  journal   = {Progress in Particle and Nuclear Physics},
  volume    = {106},
  pages     = {120--131},
  year      = {2019},
  publisher = {Elsevier}
}

@article{calogan,
  title     = {CaloGAN: Simulating 3D high energy particle showers in multilayer electromagnetic calorimeters with generative adversarial networks},
  author    = {Paganini, Michela and de Oliveira, Luke and Nachman, Benjamin},
  journal   = {Phys. Rev. D},
  volume    = {97},
  issue     = {1},
  pages     = {014021},
  numpages  = {12},
  year      = {2018},
  month     = {Jan},
  publisher = {American Physical Society},
  doi       = {10.1103/PhysRevD.97.014021},
  url       = {https://link.aps.org/doi/10.1103/PhysRevD.97.014021}
}

@article{calor2024,
  author  = {{Chung, Wonyong}},
  title   = {Differentiable Full Detector Simulation of a Projective Dual-Readout Crystal Electromagnetic Calorimeter with Longitudinal Segmentation and Precision Timing},
  doi     = {10.1051/epjconf/202532000052},
  url     = {https://doi.org/10.1051/epjconf/202532000052},
  journal = {EPJ Web Conf.},
  year    = 2025,
  volume  = 320,
  pages   = {00052}
}

@article{calorimetryatfccee,
  author        = {Aleksa, Martin and Bedeschi, Franco and Ferrari, Roberto and Sefkow, Felix and Tully, Christopher G.},
  title         = {{Calorimetry at FCC-ee}},
  eprint        = {2109.00391},
  archiveprefix = {arXiv},
  primaryclass  = {hep-ex},
  doi           = {10.1140/epjp/s13360-021-02034-2},
  journal       = {Eur. Phys. J. Plus},
  volume        = {136},
  number        = {10},
  pages         = {1066},
  year          = {2021}
}

@article{cary2009hamiltonian,
  title     = {Hamiltonian theory of guiding-center motion},
  author    = {Cary, John R and Brizard, Alain J},
  journal   = {Reviews of Modern Physics},
  volume    = {81},
  number    = {2},
  pages     = {693},
  year      = {2009},
  publisher = {APS}
}

@article{cmb_discovery1965,
  author   = {{Penzias}, A.~A. and {Wilson}, R.~W.},
  title    = {{A Measurement of Excess Antenna Temperature at 4080 Mc/s.}},
  journal  = {Astrophysical Journal (U.S.)},
  year     = {1965},
  month    = {07},
  volume   = {142},
  pages    = {419-421},
  doi      = {10.1086/148307},
  adsurl   = {https://ui.adsabs.harvard.edu/abs/1965ApJ...142..419P}
}

@article{cmb1965,
  author  = {Dicke, R H and Peebles, R J.E. and Roll, P G and Wilkinson, D T},
  title   = {COSMIC BLACK-BODY RADIATION},
  doi     = {10.1086/148306},
  url     = {https://www.osti.gov/biblio/4587058},
  journal = {Astrophysical Journal (U.S.)},
  volume  = {Vol: 142},
  place   = {Country unknown/Code not available},
  year    = {1965},
  month   = {07}
}

@techreport{cmsmtdtdr,
  author       = {CMS, Collaboration},
  title        = {{A MIP Timing Detector for the CMS Phase-2 Upgrade}},
  institution  = {CERN},
  reportnumber = {CERN-LHCC-2019-003, CMS-TDR-020},
  address      = {Geneva},
  year         = {2019},
  url          = {https://cds.cern.ch/record/2667167}
}

@article{cmspfa,
  title     = {Particle-flow reconstruction and global event description with the CMS detector},
  volume    = {12},
  issn      = {1748-0221},
  url       = {http://dx.doi.org/10.1088/1748-0221/12/10/P10003},
  doi       = {10.1088/1748-0221/12/10/p10003},
  number    = {10},
  journal   = {Journal of Instrumentation},
  publisher = {IOP Publishing},
  author    = {CMS Collaboration},
  year      = {2017},
  month     = oct,
  pages     = {P10003–P10003}
}

@article{Cocco_2007,
  doi       = {10.1088/1475-7516/2007/06/015},
  url       = {https://dx.doi.org/10.1088/1475-7516/2007/06/015},
  year      = {2007},
  month     = {jun},
  publisher = {},
  volume    = {2007},
  number    = {06},
  pages     = {015},
  author    = {Cocco, Alfredo G and Mangano, Gianpiero and Messina, Marcello},
  title     = {Probing low energy neutrino backgrounds with neutrino capture on beta decaying
               nuclei},
  journal   = {Journal of Cosmology and Astroparticle Physics}
}

@misc{controlnet,
  title         = {Adding Conditional Control to Text-to-Image Diffusion Models},
  author        = {Lvmin Zhang and Anyi Rao and Maneesh Agrawala},
  year          = {2023},
  eprint        = {2302.05543},
  archiveprefix = {arXiv},
  primaryclass  = {cs.CV},
  url           = {https://arxiv.org/abs/2302.05543}
}

@article{crilin,
  title    = {Crilin: A CRystal calorImeter with Longitudinal InformatioN for a future Muon Collider},
  journal  = {Nucl. Instrum. Meth. A},
  volume   = {1047},
  pages    = {167817},
  year     = {2023},
  issn     = {0168-9002},
  doi      = {https://doi.org/10.1016/j.nima.2022.167817},
  url      = {https://www.sciencedirect.com/science/article/pii/S0168900222011093},
  author   = {S. Ceravolo and F. Colao and C. Curatolo and E. {Di Meco} and E. Diociaiuti and D. Lucchesi and S. Martellotti and M. Moulson and D. Paesani and N. Pastrone and A. Saputi and I. Sarra and L. Sestini and D. Tagnani}
}

@software{dd4hep,
  author    = {Frank, Markus and
               Gaede, Frank and
               Petric, Marko and
               Sailer, Andre},
  title     = {AIDASoft/DD4hep: v01-25-01},
  month     = feb,
  year      = 2023,
  publisher = {Zenodo},
  version   = {v01-25-01},
  doi       = {10.5281/zenodo.7673417},
  url       = {https://doi.org/10.5281/zenodo.7673417}
}

@misc{diffusion,
  title         = {Denoising Diffusion Probabilistic Models},
  author        = {Jonathan Ho and Ajay Jain and Pieter Abbeel},
  year          = {2020},
  eprint        = {2006.11239},
  archiveprefix = {arXiv},
  primaryclass  = {cs.LG},
  url           = {https://arxiv.org/abs/2006.11239}
}

@misc{fcc-fsr,
  doi       = {10.17181/CERN.9DKX.TDH9},
  url       = {http://cds.cern.ch/record/2928193},
  author    = {Bartmann, Wolfgang and Burnet, Jean-Paul and Carli, Christian and Chance, Antoine and Craievich, Paolo and Giovannozzi, Massimo and Grojean, Christophe and Gutleber, Johannes and Hanke, Klaus and Henriques, Andre and Janot, Patrick and Lourenco, Carlos and Mangano, Michelangelo and Otto, Thomas and Poole, John Howard and Rajagopalan, Srini and Raubenheimer, Tor and Todesco, Ezio and Watson, Timothy Paul and Wilkinson, Guy},
  title     = {Future Circular Collider Feasibility Study Report Volume 1: Physics and Experiments},
  publisher = {CERN Document Server},
  year      = {2025}
}

@article{filter_paper1,
  title   = {A design for an electromagnetic filter for precision energy measurements at the tritium endpoint},
  journal = {Progress in Particle and Nuclear Physics},
  volume  = {106},
  pages   = {120-131},
  year    = {2019},
  issn    = {0146-6410},
  doi     = {https://doi.org/10.1016/j.ppnp.2019.02.004},
  url     = {https://www.sciencedirect.com/science/article/pii/S0146641019300080},
  author  = {M.G. Betti and M. Biasotti and A. Boscá and F. Calle and J. Carabe-Lopez and G. Cavoto and C. Chang and W. Chung and A.G. Cocco and A.P. Colijn and J. Conrad and N. D’Ambrosio and P.F. {de Salas} and M. Faverzani and A. Ferella and E. Ferri and P. Garcia-Abia and G. Garcia Gomez-Tejedor and S. Gariazzo and F. Gatti and C. Gentile and A. Giachero and J.E. Gudmundsson and Y. Hochberg and Y. Kahn and M. Lisanti and C. Mancini-Terracciano and G. Mangano and L.E. Marcucci and C. Mariani and J. Martínez and M. Messina and A. Molinero-Vela and E. Monticone and A. Nucciotti and F. Pandolfi and S. Pastor and J. Pedrós and C. Pérez {de los Heros} and O. Pisanti and A.D. Polosa and A. Puiu and Y. Raitses and M. Rajteri and N. Rossi and R. Santorelli and K. Schaeffner and C.F. Strid and C.G. Tully and F. Zhao and K.M. Zurek}
}

@article{filter_paper2,
  doi       = {10.1088/1748-0221/17/05/P05021},
  url       = {https://dx.doi.org/10.1088/1748-0221/17/05/P05021},
  year      = {2022},
  month     = {5},
  publisher = {IOP Publishing},
  volume    = {17},
  number    = {05},
  pages     = {P05021},
  author    = {Apponi, A. and Betti, M.G. and Borghesi, M. and Canci, N. and Cavoto, G. and Chang, C. and Chung, W. and Cocco, A.G. and Colijn, A.P. and D'Ambrosio, N. and de Groot, N. and Faverzani, M. and Ferella, A. and Ferri, E. and Ficcadenti, L. and Gariazzo, S. and Gatti, F. and Gentile, C. and Giachero, A. and Hochberg, Y. and Kahn, Y. and Kievsky, A. and Lisanti, M. and Mangano, G. and Marcucci, L.E. and Mariani, C. and Messina, M. and Monticone, E. and Nucciotti, A. and Orlandi, D. and Pandolfi, F. and Parlati, S. and Pérez de los Heros, C. and Pisanti, O. and Polosa, A.D. and Puiu, A. and Rago, I. and Raitses, Y. and Rajteri, M. and Rossi, N. and Rozwadowska, K. and Ruocco, A. and Strid, C.F. and Tan, A. and Tully, C.G. and Viviani, M. and Zeitler, U. and Zhao, F.},
  title     = {Implementation and optimization of the PTOLEMY transverse drift electromagnetic filter},
  journal   = {Journal of Instrumentation}
}

@incollection{fusion_bookchapter,
  author    = {Wonyong Chung, Andi Tan, and Christopher Tully},
  title     = {Charged Particle Beam Injection into Magnetically Confined Plasmas},
  booktitle = {Advances in Fusion Energy Research},
  publisher = {IntechOpen},
  address   = {Rijeka},
  year      = {2022},
  editor    = {Bruno Carpentieri and Aamir Shahzad},
  chapter   = {8},
  doi       = {10.5772/intechopen.106037},
  url       = {https://doi.org/10.5772/intechopen.106037}
}

@article{gan4hls,
  author        = {Musella, Pasquale and Pandolfi, Francesco},
  title         = {{Fast and Accurate Simulation of Particle Detectors Using Generative Adversarial Networks}},
  eprint        = {1805.00850},
  archiveprefix = {arXiv},
  primaryclass  = {hep-ex},
  doi           = {10.1007/s41781-018-0015-y},
  journal       = {Comput. Softw. Big Sci.},
  volume        = {2},
  number        = {1},
  pages         = {8},
  year          = {2018}
}

@article{geant4,
  title   = {GEANT4},
  journal = {Nuclear Instruments and Methods in Physics Research},
  volume  = {A835},
  pages   = {186-225},
  year    = {2016}
}

@article{gravnet,
  title     = {Learning representations of irregular particle-detector geometry with distance-weighted graph networks},
  volume    = {79},
  issn      = {1434-6052},
  url       = {http://dx.doi.org/10.1140/epjc/s10052-019-7113-9},
  doi       = {10.1140/epjc/s10052-019-7113-9},
  number    = {7},
  journal   = {The European Physical Journal C},
  publisher = {Springer Science and Business Media LLC},
  author    = {Qasim, Shah Rukh and Kieseler, Jan and Iiyama, Yutaro and Pierini, Maurizio},
  year      = {2019},
  month     = jul
}

@article{idea,
  author        = {Antonello, M.},
  editor        = {D'Ambrosio, G. and De Nardo, G.},
  collaboration = {RD-FA},
  title         = {{IDEA: A detector concept for future leptonic colliders}},
  doi           = {10.1393/ncc/i2020-20027-2},
  journal       = {Nuovo Cim. C},
  volume        = {43},
  number        = {2-3},
  pages         = {27},
  year          = {2020}
}

@misc{ideadetectorpaper,
  title         = {The IDEA detector concept for FCC-ee},
  author        = {The IDEA Study Group},
  year          = {2025},
  eprint        = {2502.21223},
  archiveprefix = {arXiv},
  primaryclass  = {physics.ins-det},
  url           = {https://arxiv.org/abs/2502.21223}
}

@article{katrin2025,
  author  = {KATRIN Collaboration† and Max Aker  and Dominic Batzler  and Armen Beglarian  and Jan Behrens  and Justus Beisenkötter  and Matteo Biassoni  and Benedikt Bieringer  and Yanina Biondi  and Fabian Block  and Steffen Bobien  and Matthias Böttcher  and Beate Bornschein  and Lutz Bornschein  and Tom S. Caldwell  and Marco Carminati  and Auttakit Chatrabhuti  and Suren Chilingaryan  and Byron A. Daniel  and Karol Debowski  and Martin Descher  and Deseada Díaz Barrero  and Peter J. Doe  and Otokar Dragoun  and Guido Drexlin  and Frank Edzards  and Klaus Eitel  and Enrico Ellinger  and Ralph Engel  and Sanshiro Enomoto  and Arne Felden  and Caroline Fengler  and Carlo Fiorini  and Joseph A. Formaggio  and Christian Forstner  and Florian M. Fränkle  and Kevin Gauda  and Andrew S. Gavin  and Woosik Gil  and Ferenc Glück  and Steffen Grohmann  and Robin Grössle  and Rainer Gumbsheimer  and Nathanael Gutknecht  and Volker Hannen  and Leonard Hasselmann  and Norman Haußmann  and Klaus Helbing  and Hanna Henke  and Svenja Heyns  and Stephanie Hickford  and Roman Hiller  and David Hillesheimer  and Dominic Hinz  and Thomas Höhn  and Anton Huber  and Alexander Jansen  and Christian Karl  and Jonas Kellerer  and Khanchai Khosonthongkee  and Matthias Kleifges  and Manuel Klein  and Joshua Kohpeiß  and Christoph Köhler  and Leonard Köllenberger  and Andreas Kopmann  and Neven Kovač  and Alojz Kovalík  and Holger Krause  and Luisa La Cascio  and Thierry Lasserre  and Joscha Lauer  and Thanh-Long Le  and Ondřej Lebeda  and Bjoern Lehnert  and Gen Li  and Alexey Lokhov  and Moritz Machatschek  and Martin Mark  and Alexander Marsteller  and Eric L. Martin  and Christin Melzer  and Susanne Mertens  and Shailaja Mohanty  and Jalal Mostafa  and Klaus Müller  and Andrea Nava  and Holger Neumann  and Simon Niemes  and Anthony Onillon  and Diana S. Parno  and Maura Pavan  and Udomsilp Pinsook  and Alan W. P. Poon  and Jose Manuel Lopez Poyato  and Stefano Pozzi  and Florian Priester  and Jan Ráliš  and Shivani Ramachandran  and R. G. Hamish Robertson  and Caroline Rodenbeck  and Marco Röllig  and Carsten Röttele  and Milos Ryšavý  and Rudolf Sack  and Alejandro Saenz  and Richard Salomon  and Peter Schäfer  and Magnus Schlösser  and Klaus Schlösser  and Lisa Schlüter  and Sonja Schneidewind  and Ulrich Schnurr  and Michael Schrank  and Jannis Schürmann  and Ann-Kathrin Schütz  and Alessandro Schwemmer  and Adrian Schwenck  and Michal Šefčík  and Daniel Siegmann  and Frank Simon  and Felix Spanier  and Daniela Spreng  and Warintorn Sreethawong  and Markus Steidl  and Jaroslav Štorek  and Xaver Stribl  and Michael Sturm  and Narumon Suwonjandee  and Nicholas Tan Jerome  and Helmut H. Telle  and Larisa A. Thorne  and Thomas Thümmler  and Simon Tirolf  and Nikita Titov  and Igor Tkachev  and Korbinian Urban  and Kathrin Valerius  and Drahoslav Vénos  and Christian Weinheimer  and Stefan Welte  and Jürgen Wendel  and Christoph Wiesinger  and John F. Wilkerson  and Joachim Wolf  and Sascha Wüstling  and Johanna Wydra  and Weiran Xu  and Sergey Zadorozhny  and Genrich Zeller },
  title   = {Direct neutrino-mass measurement based on 259 days of KATRIN data},
  journal = {Science},
  volume  = {388},
  number  = {6743},
  pages   = {180-185},
  year    = {2025},
  doi     = {10.1126/science.adq9592},
  url     = {https://www.science.org/doi/abs/10.1126/science.adq9592},
  eprint  = {https://www.science.org/doi/pdf/10.1126/science.adq9592}
}

@misc{katrindesign2005,
  author        = {Angrik, J. and others},
  collaboration = {KATRIN},
  title         = {{KATRIN design report 2004}},
  reportnumber  = {FZKA-7090},
  month         = {2},
  year          = {2005},
  url           = {https://publikationen.bibliothek.kit.edu/270060419}
}

@misc{key4hep,
  title         = {Key4hep, a framework for future HEP experiments and its use in FCC},
  author        = {Gerardo Ganis and Clément Helsens and Valentin Völkl},
  year          = {2021},
  eprint        = {2111.09874},
  archiveprefix = {arXiv},
  primaryclass  = {hep-ex},
  url           = {https://arxiv.org/abs/2111.09874}
}

@article{Kraus:2004zw,
  author        = {Kraus, Ch. and Otten, E. W. and Bonn, J. and Bornschein, B. and Bornschein, L. and Flatt, B. and Kazachenko, O. and Kovalik, A. and Weinheimer, Ch.},
  collaboration = {Mainz},
  title         = {{Final results from phase II of the Mainz neutrino mass search in tritium beta decay}},
  eprint        = {hep-ex/0412056},
  archiveprefix = {arXiv},
  doi           = {10.1140/epjc/s2005-02164-5},
  journal       = {Eur. Phys. J. C},
  volume        = {40},
  pages         = {447-468},
  year          = {2005}
}

@article{Lobashev:1999tp,
  author        = {Lobashev, V. M. and others},
  collaboration = {Troitsk},
  title         = {{Direct search for mass of neutrino and anomaly in the tritium beta spectrum}},
  eprint        = {hep-ex/9902018},
  archiveprefix = {arXiv},
  reportnumber  = {INR-1011-99},
  doi           = {10.1016/S0370-2693(99)00489-0},
  journal       = {Phys. Lett. B},
  volume        = {460},
  pages         = {227-235},
  year          = {1999}
}

@article{Lucchini_2020,
  url       = {https://doi.org/10.1088/1748-0221/15/11/p11005},
  year      = 2020,
  month     = {nov},
  publisher = {{IOP} Publishing},
  volume    = {15},
  number    = {\textbf{11}},
  pages     = {P11005--P11005},
  author    = {M.T. Lucchini and W. Chung and S.C. Eno and Y. Lai and L. Lucchini and M. Nguyen and C.G. Tully},
  title     = {New perspectives on segmented crystal calorimeters for future colliders},
  journal   = {Journal of Instrumentation}
}

@article{ml_landscape,
  title     = {{The landscape of unfolding with machine learning}},
  author    = {Nathan Huetsch and Javier Mariño Villadamigo and Alexander Shmakov and Sascha Diefenbacher and Vinicius Mikuni and Theo Heimel and Michael Fenton and Kevin Greif and Benjamin Nachman and Daniel Whiteson and Anja Butter and Tilman Plehn},
  journal   = {SciPost Phys.},
  volume    = {18},
  pages     = {070},
  year      = {2025},
  publisher = {SciPost},
  doi       = {10.21468/SciPostPhys.18.2.070},
  url       = {https://scipost.org/10.21468/SciPostPhys.18.2.070}
}

@article{mlpf,
  title     = {MLPF: efficient machine-learned particle-flow reconstruction using graph neural networks},
  volume    = {81},
  issn      = {1434-6052},
  url       = {http://dx.doi.org/10.1140/epjc/s10052-021-09158-w},
  doi       = {10.1140/epjc/s10052-021-09158-w},
  number    = {5},
  journal   = {The European Physical Journal C},
  publisher = {Springer Science and Business Media LLC},
  author    = {Pata, Joosep and Duarte, Javier and Vlimant, Jean-Roch and Pierini, Maurizio and Spiropulu, Maria},
  year      = {2021},
  month     = may
}

@article{neutrino_oscillations1998,
  title         = {Evidence for Oscillation of Atmospheric Neutrinos},
  author        = {Fukuda, Y. and Hayakawa, T. and Ichihara, E. and Inoue, K. and Ishihara, K. and Ishino, H. and Itow, Y. and Kajita, T. and Kameda, J. and Kasuga, S. and Kobayashi, K. and Kobayashi, Y. and Koshio, Y. and Miura, M. and Nakahata, M. and Nakayama, S. and Okada, A. and Okumura, K. and Sakurai, N. and Shiozawa, M. and Suzuki, Y. and Takeuchi, Y. and Totsuka, Y. and Yamada, S. and Earl, M. and Habig, A. and Kearns, E. and Messier, M. D. and Scholberg, K. and Stone, J. L. and Sulak, L. R. and Walter, C. W. and Goldhaber, M. and Barszczxak, T. and Casper, D. and Gajewski, W. and Halverson, P. G. and Hsu, J. and Kropp, W. R. and Price, L. R. and Reines, F. and Smy, M. and Sobel, H. W. and Vagins, M. R. and Ganezer, K. S. and Keig, W. E. and Ellsworth, R. W. and Tasaka, S. and Flanagan, J. W. and Kibayashi, A. and Learned, J. G. and Matsuno, S. and Stenger, V. J. and Takemori, D. and Ishii, T. and Kanzaki, J. and Kobayashi, T. and Mine, S. and Nakamura, K. and Nishikawa, K. and Oyama, Y. and Sakai, A. and Sakuda, M. and Sasaki, O. and Echigo, S. and Kohama, M. and Suzuki, A. T. and Haines, T. J. and Blaufuss, E. and Kim, B. K. and Sanford, R. and Svoboda, R. and Chen, M. L. and Conner, Z. and Goodman, J. A. and Sullivan, G. W. and Hill, J. and Jung, C. K. and Martens, K. and Mauger, C. and McGrew, C. and Sharkey, E. and Viren, B. and Yanagisawa, C. and Doki, W. and Miyano, K. and Okazawa, H. and Saji, C. and Takahata, M. and Nagashima, Y. and Takita, M. and Yamaguchi, T. and Yoshida, M. and Kim, S. B. and Etoh, M. and Fujita, K. and Hasegawa, A. and Hasegawa, T. and Hatakeyama, S. and Iwamoto, T. and Koga, M. and Maruyama, T. and Ogawa, H. and Shirai, J. and Suzuki, A. and Tsushima, F. and Koshiba, M. and Nemoto, M. and Nishijima, K. and Futagami, T. and Hayato, Y. and Kanaya, Y. and Kaneyuki, K. and Watanabe, Y. and Kielczewska, D. and Doyle, R. A. and George, J. S. and Stachyra, A. L. and Wai, L. L. and Wilkes, R. J. and Young, K. K.},
  collaboration = {Super-Kamiokande Collaboration},
  journal       = {Phys. Rev. Lett.},
  volume        = {81},
  issue         = {8},
  pages         = {1562--1567},
  numpages      = {0},
  year          = {1998},
  month         = {Aug},
  publisher     = {American Physical Society},
  doi           = {10.1103/PhysRevLett.81.1562},
  url           = {https://link.aps.org/doi/10.1103/PhysRevLett.81.1562}
}

@article{omnifold,
  title     = {OmniFold: A Method to Simultaneously Unfold All Observables},
  author    = {Andreassen, Anders and Komiske, Patrick T. and Metodiev, Eric M. and Nachman, Benjamin and Thaler, Jesse},
  journal   = {Phys. Rev. Lett.},
  volume    = {124},
  issue     = {18},
  pages     = {182001},
  numpages  = {7},
  year      = {2020},
  month     = {May},
  publisher = {American Physical Society},
  doi       = {10.1103/PhysRevLett.124.182001},
  url       = {https://link.aps.org/doi/10.1103/PhysRevLett.124.182001}
}

@article{particlenet,
  title     = {Jet tagging via particle clouds},
  volume    = {101},
  issn      = {2470-0029},
  url       = {http://dx.doi.org/10.1103/PhysRevD.101.056019},
  doi       = {10.1103/physrevd.101.056019},
  number    = {5},
  journal   = {Physical Review D},
  publisher = {American Physical Society (APS)},
  author    = {Qu, Huilin and Gouskos, Loukas},
  year      = {2020},
  month     = mar
}

@article{pdg2024,
  title         = {Review of Particle Physics},
  author        = {Navas, S. and Amsler, C. and Gutsche, T. and Hanhart, C. and Hern\'andez-Rey, J. J. and Louren\ifmmode \mbox{\c{c}}\else \c{c}\fi{}o, C. and Masoni, A. and Mikhasenko, M. and Mitchell, R. E. and Patrignani, C. and Schwanda, C. and Spanier, S. and Venanzoni, G. and Yuan, C. Z. and Agashe, K. and Aielli, G. and Allanach, B. C. and Alvarez-Mu\~niz, J. and Antonelli, M. and Aschenauer, E. C. and Asner, D. M. and Assamagan, K. and Baer, H. and Banerjee, Sw. and Barnett, R. M. and Baudis, L. and Bauer, C. W. and Beatty, J. J. and Beringer, J. and Bettini, A. and Biebel, O. and Black, K. M. and Blucher, E. and Bonventre, R. and Briere, R. A. and Buckley, A. and Burkert, V. D. and Bychkov, M. A. and Cahn, R. N. and Cao, Z. and Carena, M. and Casarosa, G. and Ceccucci, A. and Cerri, A. and Chivukula, R. S. and Cowan, G. and Cranmer, K. and Crede, V. and Cremonesi, O. and D'Ambrosio, G. and Damour, T. and de Florian, D. and de Gouv\^ea, A. and DeGrand, T. and Demers, S. and Demiragli, Z. and Dobrescu, B. A. and D'Onofrio, M. and Doser, M. and Dreiner, H. K. and Eerola, P. and Egede, U. and Eidelman, S. and El-Khadra, A. X. and Ellis, J. and Eno, S. C. and Erler, J. and Ezhela, V. V. and Fava, A. and Fetscher, W. and Fields, B. D. and Freitas, A. and Gallagher, H. and Gershon, T. and Gershtein, Y. and Gherghetta, T. and Gonzalez-Garcia, M. C. and Goodman, M. and Grab, C. and Gritsan, A. V. and Grojean, C. and Groom, D. E. and Gr\"unewald, M. and Gurtu, A. and Haber, H. E. and Hamel, M. and Hashimoto, S. and Hayato, Y. and Hebecker, A. and Heinemeyer, S. and Hikasa, K. and Hisano, J. and H\"ocker, A. and Holder, J. and Hsu, L. and Huston, J. and Hyodo, T. and Ianni, Al. and Kado, M. and Karliner, M. and Katz, U. F. and Kenzie, M. and Khoze, V. A. and Klein, S. R. and Krauss, F. and Kreps, M. and Kri\ifmmode \check{z}\else \v{z}\fi{}an, P. and Krusche, B. and Kwon, Y. and Lahav, O. and Lellouch, L. P. and Lesgourgues, J. and Liddle, A. R. and Ligeti, Z. and Lin, C.-J. and Lippmann, C. and Liss, T. M. and Lister, A. and Littenberg, L. and Lugovsky, K. S. and Lugovsky, S. B. and Lusiani, A. and Makida, Y. and Maltoni, F. and Manohar, A. V. and Marciano, W. J. and Matthews, J. and Mei\ss{}ner, U.-G. and Melzer-Pellmann, I.-A. and Mertsch, P. and Miller, D. J. and Milstead, D. and M\"onig, K. and Molaro, P. and Moortgat, F. and Moskovic, M. and Nagata, N. and Nakamura, K. and Narain, M. and Nason, P. and Nelles, A. and Neubert, M. and Nir, Y. and O'Connell, H. B. and O'Hare, C. A. J. and Olive, K. A. and Peacock, J. A. and Pianori, E. and Pich, A. and Piepke, A. and Pietropaolo, F. and Pomarol, A. and Pordes, S. and Profumo, S. and Quadt, A. and Rabbertz, K. and Rademacker, J. and Raffelt, G. and Ramsey-Musolf, M. and Richardson, P. and Ringwald, A. and Robinson, D. J. and Roesler, S. and Rolli, S. and Romaniouk, A. and Rosenberg, L. J and Rosner, J. L. and Rybka, G. and Ryskin, M. G. and Ryutin, R. A. and Safdi, B. and Sakai, Y. and Sarkar, S. and Sauli, F. and Schneider, O. and Sch\"onert, S. and Scholberg, K. and Schwartz, A. J. and Schwiening, J. and Scott, D. and Sefkow, F. and Seljak, U. and Sharma, V. and Sharpe, S. R. and Shiltsev, V. and Signorelli, G. and Silari, M. and Simon, F. and Sj\"ostrand, T. and Skands, P. and Skwarnicki, T. and Smoot, G. F. and Soffer, A. and Sozzi, M. S. and Spiering, C. and Stahl, A. and Sumino, Y. and Takahashi, F. and Tanabashi, M. and Tanaka, J. and Ta\ifmmode \check{s}\else \v{s}\fi{}evsk\'y, M. and Terao, K. and Terashi, K. and Terning, J. and Thoma, U. and Thorne, R. S. and Tiator, L. and Titov, M. and Tovey, D. R. and Trabelsi, K. and Urquijo, P. and Valencia, G. and Van de Water, R. and Varelas, N. and Verde, L. and Vivarelli, I. and Vogel, P. and Vogelsang, W. and Vorobyev, V. and Wakely, S. P. and Walkowiak, W. and Walter, C. W. and Wands, D. and Weinberg, D. H. and Weinberg, E. J. and Wermes, N. and White, M. and Wiencke, L. R. and Willocq, S. and Woody, C. L. and Workman, R. L. and Yao, W.-M. and Yokoyama, M. and Yoshida, R. and Zanderighi, G. and Zeller, G. P. and Zhu, R.-Y. and Zhu, S.-L. and Zimmermann, F. and Zyla, P. A. and Anderson, J. and Kramer, M. and Schaffner, P. and Zheng, W.},
  collaboration = {Particle Data Group Collaboration},
  journal       = {Phys. Rev. D},
  volume        = {110},
  issue         = {3},
  pages         = {030001},
  numpages      = {5},
  year          = {2024},
  month         = {Aug},
  publisher     = {American Physical Society},
  doi           = {10.1103/PhysRevD.110.030001},
  url           = {https://link.aps.org/doi/10.1103/PhysRevD.110.030001}
}

@article{Planck2018,
  author  = {Aghanim, N. and Akrami, Y. and Ashdown, M. and {\it et al.} (Planck Collaboration)},
  title   = {Planck 2018 results. VI. Cosmological parameters},
  journal = {Astron. Astrophys.},
  volume  = {641},
  pages   = {A6},
  year    = {2020},
  doi     = {10.1051/0004-6361/201833910}
}

@article{rajteri2020tes,
  title     = {TES Microcalorimeters for PTOLEMY},
  author    = {Rajteri, M and Biasotti, M and Faverzani, M and Ferri, E and Filippo, R and Gatti, F and Giachero, A and Monticone, E and Nucciotti, A and Puiu, A},
  journal   = {Journal of Low Temperature Physics},
  volume    = {199},
  number    = {1},
  pages     = {138--142},
  year      = {2020},
  publisher = {Springer}
}

@incollection{roederer2014particle,
  title     = {Particle fluxes, distribution functions and violation of invariants},
  author    = {Roederer, Juan and Zhang, Hui},
  booktitle = {Dynamics of Magnetically Trapped Particles},
  pages     = {89-122},
  year      = {2014},
  publisher = {Springer}
}

@misc{transformers,
  title         = {Attention Is All You Need},
  author        = {Ashish Vaswani and Noam Shazeer and Niki Parmar and Jakob Uszkoreit and Llion Jones and Aidan N. Gomez and Lukasz Kaiser and Illia Polosukhin},
  year          = {2023},
  eprint        = {1706.03762},
  archiveprefix = {arXiv},
  primaryclass  = {cs.CL},
  url           = {https://arxiv.org/abs/1706.03762}
}

@article{weinberg1962,
  title     = {Universal Neutrino Degeneracy},
  author    = {Weinberg, Steven},
  journal   = {Phys. Rev.},
  volume    = {128},
  issue     = {3},
  pages     = {1457--1473},
  numpages  = {0},
  year      = {1962},
  month     = {Nov},
  publisher = {American Physical Society},
  doi       = {10.1103/PhysRev.128.1457},
  url       = {https://link.aps.org/doi/10.1103/PhysRev.128.1457}
}

\end{document}